\documentclass[aps,prd,notitlepage,superscriptaddress,nofootinbib,showkeys]{revtex4-1}

\usepackage{graphicx}
\usepackage{grffile}
\usepackage{amsmath}
\usepackage{amsfonts}
\usepackage{todonotes}
\usepackage{amssymb}
\usepackage{bm}
\usepackage{epstopdf}
\usepackage{float}
\usepackage{subfigure}
\usepackage{capt-of}
\allowdisplaybreaks

\usepackage{hyperref}

\hypersetup{
     colorlinks   = true,
     citecolor    = red,
     linkcolor    = blue,
     urlcolor     = blue,
}
\definecolor{deeppink}{rgb}{0.9, 0.17, 0.31}

\def\beq{\begin{equation}}
\def\eeq{\end{equation}}
\def\bea{\begin{eqnarray}}
\def\eea{\end{eqnarray}}
\def\be{\begin{equation}}
\def\ee{\end{equation}}

\def\bse{\begin{subequations}}
\def\ese{\end{subequations}}

\def\nn{\nonumber} 
\def\pa{\partial}
\def\f{\frac}

\def\l{\left}
\def\r{\right}
\def\d{{\rm d}}
\def\Mpl{M_{_{\mathrm{Pl}}}}

\def\ee{\eta_{\mathrm{e}}}

\def\ef{\eta_{\mathrm{f}}}

\def\are{a_{\mathrm{re}}}
\def\af{a_{\mathrm{f}}}
\def\Are{A_{\mathrm{re}}}

\def\Tre{T_{\mathrm{re}}}
\def\kre{k_{\mathrm{re}}}

\def\Nre{N_{\mathrm{re}}}

\def\Hre{H_{\mathrm{re}}}

\def\rhoR{\rho_{_{\mathrm{R}}}}

\def\phif{\phi_{\mathrm{f}}}

\def\ngw{n_{_\mathrm{GW}}}

\def\af{a_{\mathrm{f}}}

\def\Are{A_{\mathrm{re}}}
\def\kf{k_{\mathrm{f}}}

\def\kre{k_{\mathrm{re}}}
\def\fre{f_{\mathrm{re}}}

\def\pt{\mathcal{P}_{_{\mathrm{T}}}}
\def\ns{n_{_{\mathrm{S}}}}
\def\cAs{\mathcal{A}_{_{\mathrm{S}}}}
\def\ogw{\Omega_{_{\mathrm{GW}}}}
\def\rhogw{\rho_{_{\mathrm{GW}}}}
\def\HI{H_{_{\mathrm I}}}

\graphicspath{{./figs/}}
\begin{document}

\title{Decoding the phases of early and late time reheating\\ 
through imprints on primordial gravitational waves}
\author{Md Riajul Haque}
\email{E-mail: riaju176121018@iitg.ac.in}
\author{Debaprasad Maity}
\email{E-mail: debu@iitg.ac.in}
\affiliation{Department of Physics, Indian Institute of Technology, 
Guwahati~781039, India}
\author{Tanmoy Paul}
\email{E-mail: pul.tnmy9@gmail.com}
\affiliation{Department of Physics, Chandernagore College, Hooghly~712136, India}
\affiliation{International Laboratory for Theoretical Cosmology, TUSUR, 
634050 Tomsk, Russia}
\author{L. Sriramkumar}
\email{E-mail: sriram@physics.iitm.ac.in}
\affiliation{Department of Physics, Indian Institute of Technology Madras,
Chennai 600036, India}
\begin{abstract}
Primordial gravitational waves (GWs) carry the imprints of the dynamics of the 
universe during its earliest stages.
With a variety of GW detectors being proposed to operate over a wide range of 
frequencies, there is great expectation that observations of primordial GWs
can provide us with an unprecedented window to the physics operating during
inflation and reheating.
In this work, we closely examine the effects of the regime of reheating on 
the spectrum of primordial GWs observed today.
We consider a scenario wherein the phase of reheating is described by an averaged 
equation of state (EoS) parameter with an abrupt transition to radiation domination
as well as a scenario wherein there is a gradual change in the effective EoS 
parameter to that of radiation due to the perturbative decay of the inflaton.
We show that the perturbative decay of the inflaton leads to oscillations in 
the spectrum of GWs, which, if observed, can possibly help us decipher finer 
aspects of the reheating mechanism.
We also examine the effects of a secondary phase of reheating arising due 
to a brief epoch driven possibly by an exotic, non-canonical, scalar field.
Interestingly, we find that, for suitable values of the EoS parameter 
governing the secondary phase of reheating, the GWs can be of the strength
as suggested by the recent NANOGrav observations.
We conclude with a discussion of the wider implications of our analysis.
\end{abstract}
\maketitle


\section{Introduction}

The inflationary scenario offers the most attractive mechanism for 
the generation of the primordial perturbations (for the original discussions,
see Refs.~\cite{Hawking:1982cz,Guth:1982ec,Starobinsky:1982ee,Bardeen:1983qw};
for reviews, see, for example, Refs.~\cite{Mukhanov:1990me,Martin:2003bt,
Martin:2004um,Bassett:2005xm,Sriramkumar:2009kg,Baumann:2008bn,Baumann:2009ds,
Sriramkumar:2012mik,Linde:2014nna,Martin:2015dha}).
The existence of primordial gravitational waves (GWs) is one of the profound 
predictions of inflationary dynamics (for the initial discussions, see, for 
example, Refs.~\cite{Grishchuk:1974ny,Starobinsky:1979ty}; for recent reviews 
on the generation of primary and secondary GWs, see, for instance,
Refs.~\cite{Guzzetti:2016mkm,Caprini:2018mtu}).
If the primordial GWs or their imprints are detected, it will not only prove 
the quantum origin of the perturbations, it can also, in principle, provide 
us with insights into the fundamental nature of gravitation.
The primordial GWs provide a unique window to probe the dynamics of our 
universe during its very early stages, which seems difficult to observe 
by any other known means.
However, the extremely weak nature of the gravitational force makes the 
detection of GWs rather challenging.
Decades of effort towards detecting GWs have finally achieved success only 
in the last few years with the observations of GWs from merging binary black 
holes and neutron stars by LIGO~\cite{TheLIGOScientific:2016agk,TheLIGOScientific:2016qqj,TheLIGOScientific:2016wfe,Abbott:2016blz,Abbott:2016nmj,Abbott:2017vtc,Abbott:2017gyy,Abbott:2017oio,TheLIGOScientific:2017qsa,LIGOScientific:2020stg,Abbott:2020uma,Abbott:2020khf}.
These observations have led to a surge of experimental proposals across the 
globe to observe GWs over a wide range of frequencies.
The proposed GW observatories include 
advanced LIGO ($10$--$10^3\, \mathrm{Hz}$)~\cite{TheLIGOScientific:2016dpb}, 
ET ($1$--$10^4\, \mathrm{Hz}$)~\cite{Punturo:2010zz,Sathyaprakash:2012jk},
BBO ($10^{-3}$--$10\, 
\mathrm{Hz}$)~\cite{Crowder:2005nr,Corbin:2005ny,Baker:2019pnp},
DECIGO ($10^{-3}$--$1\, \mathrm{Hz}$)~\cite{Seto:2001qf,Kawamura:2011zz,Sato:2017dkf,Kawamura:2019jqt},
eLISA ($10^{-5}$--$1\,\mathrm{Hz}$)~\cite{AmaroSeoane:2012km,Audley:2017drz,
Barausse:2020rsu}, and 
SKA ($10^{-9}$--$10^{-6}\, \mathrm{Hz}$)~\cite{Janssen:2014dka}. 

Apart from various astrophysical mechanisms that can generate GWs, as 
we mentioned, inflation provides an exclusive mechanism to produce GWs 
of quantum mechanical origin (for discussions on the generation of 
primary and secondary GWs, see, for instance, 
Refs.~\cite{Easther:2006gt,Ananda:2006af,Baumann:2007zm,
Saito:2008jc,Saito:2009jt,Kuroyanagi:2018csn,Espinosa:2018eve,Kohri:2018awv,
Inomata:2019zqy,Inomata:2019ivs,Braglia:2020eai,Ragavendra:2020sop,
Ragavendra:2020vud,Bhattacharya:2020lhc} and references therein).  
The tensor perturbations generated from the quantum vacuum are amplified
during inflation, which subsequently evolve through the various phases of
the universe until they reach the GW detectors today. 
Therefore, the spectrum of primordial GWs today is a convolution of their
origin as well as dynamics. 
On the one hand, they contain the signatures of the mechanism that generates
them, viz. the specific model that drives inflation as well as the initial 
conditions from which the perturbations emerge. 
On the other, they also carry the imprints of the dynamics of the subsequent
cosmological phases as the GWs propagate through them. 
As is well known, immediately after inflation, the universe is expected to be 
reheated through the decay of the inflaton into radiation, which eventually 
leads to the epoch of radiation domination. 
In this work, we shall examine the evolution of primordial GWs with special 
emphasis on the effects due to the epoch of reheating. 
Specifically, our aim is to decode the mechanism of reheating from the 
spectrum of GWs today.  

Over the years, major cosmological observations have considerably 
improved the theoretical understanding of the various epochs of our 
universe~\cite{Akrami:2018odb,Aubourg:2014yra,Vazquez:2018qdg}. 
However, due to the lack of direct observational signatures, the phase of 
reheating remains poorly understood. 
The effects of reheating on the dynamics of GWs have already been 
examined in the standard cosmological scenario (see, for instance,
Refs.~\cite{Turner:1993vb,Boyle:2005se,Sa:2007pc,Nakayama:2008wy,
Nakayama:2008ip,Sa:2010qw,Kuroyanagi:2011fy,Kuroyanagi:2014nba})
as well as in certain non-standard scenarios (see, for example, 
Refs.~\cite{Assadullahi:2009nf,Nakayama:2009ce,Durrer:2011bi,Alabidi:2013lya,
DEramo:2019tit,Ricciardone:2016lym,Koh:2018qcy,Fujita:2018ehq,Bernal:2019lpc,
Bernal:2020ywq,Mishra:2021wkm}). 
Moreover, the imprints of specific microscopic physical effects 
on the spectrum of GWs~---~such as decoupling 
neutrinos~\cite{Weinberg:2003ur,Mangilli:2008bw}
or the variation in the number of relativistic degrees of freedom in the early 
universe~\cite{Watanabe:2006qe,Kuroyanagi:2008ye,Caldwell:2018giq}~---~have 
also been explored.
In this work, we shall study the effects of reheating on the spectrum of primordial GWs over a wide range of scales and illustrate the manner in 
which the spectrum captures specific aspects of the different phases. 
We shall consider two of the simplest reheating mechanisms. 
We shall first consider a scenario wherein the reheating phase is described 
by an averaged equation of state (EoS) parameter, with reheating ending
instantaneously~\cite{Martin:2010kz,Dai:2014jja,Cook:2015vqa}.
We shall then consider a scenario wherein there is a gradual change in  
the EoS parameter from its initial value during the phase of coherent 
oscillations to its eventual value during radiation domination achieved 
through the perturbative decay of the
inflaton~\cite{Chung:1998rq,Giudice:2000ex,Maity:2018dgy,Haque:2019prw,Haque:2020zco}. 
We shall explicitly illustrate the effects of the reheating dynamics on
the spectrum of GWs.
It seems worthwhile to highlight here that the following aspects of the 
reheating dynamics can, in principle, be decoded from the spectrum of 
primordial GWs: (i)~the shape of the inflaton potential near its minimum 
which is responsible for the end of inflation and the dynamics during reheating, 
(ii)~the decay width of the inflaton, which governs the entire process of 
reheating and therefore determines the reheating temperature, and 
(iii)~the thermalization time scale over which the EoS parameter during 
the period of coherent oscillations of the inflaton say, $w_{\phi}$, 
is modified to the EoS parameter corresponding to radiation. 
Further, it is expected that determining the inflaton decay width and the
thermalization time scale would permit us not only to arrive at the form of 
the coupling between the inflaton and radiation, but also help us understand
the nature of the coupling amongst all the relativistic degrees of freedom. 
We shall briefly discuss these possibilities in this work, and we shall return 
to examine the issue in greater detail in a future effort. 

Finally, we shall also consider the implications of our analysis on the 
recent observations involving the pulsar-timing data by the North American 
Nanohertz Observatory for Gravitational Waves (NANOGrav), which has been 
attributed to stochastic GWs~\cite{Arzoumanian:2020vkk,Pol:2020igl}. 
A variety of mechanisms that can possibly occur in the early universe 
have been explored in the literature to explain this interesting 
observation~\cite{Kuroyanagi:2020sfw,Inomata:2020xad,Tahara:2020fmn,
Kitajima:2020rpm,Bhattacharya:2020lhc,Domenech:2020ers,Li:2020cjj,
Bian:2020bps,Blasi:2020mfx,DeLuca:2020agl,Vaskonen:2020lbd,
Ellis:2020ena,Buchmuller:2020lbh,Kohri:2020qqd,Vagnozzi:2020gtf,
Bhattacharya:2020lhc}. 
We find that introducing a secondary phase of reheating~---~apart 
from the original, inflaton driven, primary reheating phase~---~can 
account for the NANOGrav observations. 
We introduce an exotic, non-canonical scalar field to drive such a phase 
and show that a suitable EoS for the non-canonical field can 
lead to primordial GWs of strength as observed by NANOGrav.

This paper is structured as follows. 
In Sec.~\ref{sec:pgw}, we shall briefly sketch the arguments leading 
to the standard scale-invariant spectrum of GWs generated in de 
Sitter inflation.
We shall also discuss the typical inflationary model that we shall have 
in mind when we later discuss the effects due to reheating. 
Moreover, we shall introduce the dimensionless density parameter~$\ogw$
characterizing the spectrum of GWs.
In Sec.~\ref{sec:egw-drh}, we shall discuss the evolution of the tensor
perturbations during the epoch of reheating.
We shall first consider the scenario wherein the epoch of reheating is 
described by an averaged EoS parameter associated with the inflaton.
Such a description allows us to arrive at analytic solutions for the
tensor perturbations during the epoch.
We shall also consider the scenario of perturbative reheating wherein
we take into account the continuous decay of the inflaton into radiation.
As it proves to be involved in constructing analytical solutions for the 
background as well as the tensor perturbations in such a case, we shall 
resort to numerics. 
In Sec.~\ref{sec:egw-drd}, we shall briefly discuss the evolution of 
the tensor perturbations during the epoch of radiation domination and 
arrive at the spectrum of GWs today by comparing the behavior of the 
the energy density of GWs in the sub-Hubble domain with that of radiation.
In Secs.~\ref{sec:averaged} and~\ref{sec:actual}, we shall evaluate the 
dimensionless energy density of GWs today that arise in the two types
of reheating scenarios mentioned above.
We shall focus on the spectrum of primordial GWs over wave numbers (or, 
equivalently, frequencies) that correspond to small scales which reenter
the Hubble radius either during the epochs of reheating or radiation 
domination. 
In Sec.~\ref{sec:actualinflation}, we shall numerically evaluate the 
inflationary tensor power spectrum and discuss the behavior of the 
spectrum of GWs today close to the scale that leaves the Hubble 
radius at the end of inflation.
In Sec.~\ref{sec:probe}, we shall outline the manner in which we should 
be able to decode various information concerning the epochs of inflation 
and reheating from the observations of the anisotropies in the cosmic
microwave background (CMB) and the spectrum of GWs today. 
In Sec.~\ref{sec:NANOGrav}, we shall evaluate the spectrum of GWs in 
a scenario involving late time production of entropy and discuss the 
implications for the recent observations by NANOGrav.
Lastly, in Sec.~\ref{sec:c}, we shall conclude with a summary of the 
main results.

Before we proceed further, a few clarifications concerning the conventions 
and notations that we shall adopt are in order. 
We shall work with natural units such that $\hbar=c=1$, and set the reduced 
Planck mass to be $\Mpl=\l(8\,\pi\, G\r)^{-1/2}$.
We shall adopt the signature of the metric to be~$(-,+,+,+)$.
Note that Latin indices shall represent the spatial coordinates, except 
for~$k$, which shall be reserved for denoting the wavenumber of the tensor 
perturbations, i.e.~GWs. 
We shall assume the background to be the spatially flat
Friedmann-Lema\^itre-Robertson-Walker (FLRW) line element described by the 
scale factor~$a$ and the Hubble parameter~$H$.
Also, an overprime shall denote differentiation with respect to the 
conformal time~$\eta$.
We should mention that the frequency, say, $f$, is related to the wave 
number~$k$ of the tensor perturbations through the relation
\begin{equation}
f = \frac{k}{2\,\pi} 
= 1.55\times10^{-15}\,\l(\f{k}{1\, \mathrm{Mpc}^{-1}}\r)\,\mathrm{Hz}
\label{eq:f}
\end{equation}
and, as convenient, we shall refer to the spectrum of GWs either in 
terms of wave numbers or frequencies.

\section{Spectrum of GWs generated during inflation}\label{sec:pgw}

In this section, we shall briefly recall the equation governing the tensor
perturbations and arrive at the spectrum of GWs generated in de Sitter 
inflation.
We shall also introduce the dimensionless energy density of GWs, which is 
the primary observational quantity of interest in this work.


\subsection{Generation of GWs during inflation}

Let $h_{ij}$ denote the tensor perturbations characterizing the GWs in a
FLRW universe.
When these tensor perturbations are taken into account, the line-element 
describing the spatially flat, FLRW universe can be expressed 
as~\cite{Maggiore:1999vm}
\begin{equation}
\d s^2
= a^2(\eta)\,\biggl\{-\d\eta^2+\l[\delta_{ij}+h_{ij}(\eta,{\bm x})\r]\,
\d x^i \d x^j\biggr\}.
\end{equation}
Since the tensor perturbations are transverse and traceless, they satisfy 
the conditions $\pa^i\,h_{ij}=0$ and $h^i_i=0$.
We shall assume that no anisotropic stresses are present.
In such a case, the first order Einstein's equations then lead to the following 
equation of motion for the tensor perturbations~$h_{ij}$:
\begin{equation}
h_{ij}'' + 2\,\f{a'}{a}\,h_{ij}'
-{\bm \nabla^2}\,h_{ij}=0.
\end{equation}

On quantization, the tensor perturbations can be decomposed in terms of the 
Fourier modes~$h_k$ as follows~\cite{Mukhanov:1990me,Martin:2003bt,
Martin:2004um,Bassett:2005xm,Sriramkumar:2009kg,Baumann:2008bn,Baumann:2009ds,
Sriramkumar:2012mik,Linde:2014nna,Martin:2015dha}:
\begin{eqnarray}
\hat{h}_{ij}(\eta, {\bm x}) 
&=& \int \f{\d^{3}{\bm k}}{\l(2\,\pi\r)^{3/2}}\,
\hat{h}_{ij}^{\bm k}(\eta)\, \mathrm{e}^{i\,{\bm k}\cdot{\bm x}}\nn\\
&=& \sum_{\lambda={+,\times}}\int \f{\d^{3}{\bm k}}{(2\,\pi)^{3/2}}\,
\l[\hat{a}^{\lambda}_{\bm k}\, \varepsilon^{\lambda}_{ij}({\bm k})\,
h_{k}(\eta)\, \mathrm{e}^{i\,{\bm k}\cdot{\bm x}}
+\hat{a}^{\lambda\dag}_{\bf k}\,\varepsilon^{\lambda\ast}_{ij}({\bm k})\, 
h^{\ast}_{k}(\eta)\,\mathrm{e}^{-i\,{\bm k}\cdot{\bm x}}\r],\label{eq:tp-m-dc}
\end{eqnarray}
where the quantity $\varepsilon^{\lambda}_{ij}({\bm k})$ represents the 
polarization tensor, with the index~$\lambda$ denoting the 
polarization~$+$ or~$\times$ of the GWs.
The polarization tensor obeys the relations 
$\delta^{ij}\,\varepsilon^{\lambda}_{ij}({\bm k})
=k^{i}\,\varepsilon_{ij}^\lambda({\bm k})=0$, 
and we shall work with the normalization such 
that $\varepsilon^{ij\,\lambda}({\bm k})\,
\varepsilon_{ij}^{\lambda'\ast}({\bm k})=2\,\delta^{\lambda\lambda'}$.
In the above decomposition, the operators $(\hat{a}_{\bm k}^{\lambda},
\hat{a}^{\lambda\dagger}_{\bm k})$ denote the annihilation and creation 
operators corresponding to the tensor modes associated with the wave 
vector~${\bm k}$.
They obey the following commutation 
relations:~$[\hat{a}^{\lambda}_{\bm k},\hat{a}_{\bm k'}^{\lambda'}]
=[\hat{a}^{\lambda\dag}_{\bm k},\hat{a}_{\bm k'}^{\lambda'\dagger}]=0$ and
$[\hat{a}^{\lambda}_{\bm k},\hat{a}_{\bm k'}^{\lambda'\dagger}]
=\delta^{(3)}({\bm k}-{\bm k}')\, \delta^{\lambda\lambda'}$. 
In the absence of sources with anisotropic stresses, the Fourier mode $h_k$ 
satisfies the differential equation 
\begin{equation}
h_k''+2\,\f{a'}{a}\,h_k'+k^2\,h_k=0.\label{eq:em-hk}    
\end{equation}
The tensor power spectrum $\pt(k)$ is defined through the relation
\begin{equation}
\langle 0 \vert {\hat h}^{ij}_{\bm k}(\eta)\,
{\hat h}_{ij}^{\bm k'}(\eta)\vert 0\rangle
=\f{(2\,\pi)^2}{2\, k^3}\, \pt(k)\;
\delta^{(3)}({\bm k}+{\bm k'}),\label{eq:tps-d}
\end{equation}
where the vacuum state $\vert 0\rangle$ is defined as 
${\hat a}_{\bm k}^{\lambda}\vert 0\rangle=0$ for all ${\bm k}$ and $\lambda$.
On utilizing the above decomposition in terms of the Fourier modes, we 
obtain that 
\begin{equation}
\pt(k)=4\,\f{k^3}{2\, \pi^2}\, \vert h_k\vert^2,\label{eq:tps}
\end{equation}
and it is often assumed that the spectrum is evaluated on super-Hubble
scales during inflation.

Motivated by form of the second order action governing the tensor 
perturbation~$h_{ij}$, the Fourier mode $h_k$ is usually written in 
terms of the Mukhanov-Sasaki variable $u_k$ as $h_k=(\sqrt{2}/\Mpl)\, (u_k/a)$.
The  Mukhanov-Sasaki variable $u_k$ satisfies the differential 
equation~\cite{Mukhanov:1990me,Martin:2003bt,Martin:2004um,
Bassett:2005xm,Sriramkumar:2009kg,Baumann:2008bn,Baumann:2009ds,
Sriramkumar:2012mik,Linde:2014nna,Martin:2015dha}
\begin{equation}
u_k''+ \l(k^2 - \frac{a''}{a}\r)\, u_k = 0.\label{inflation eom1}
\end{equation}
We shall focus on the simple case of slow roll inflation and work in the de Sitter
approximation wherein the scale factor is given by $a(\eta)=(1-\HI\,\eta)^{-1}$, 
with $\HI$ denoting the constant Hubble scale during inflation.
In such a case, $a''/a=2\,\HI^2/\l(1-\HI\,\eta\r)^2$, and the 
solution to Eq.~\eqref{inflation eom1} corresponding to the 
Bunch-Davies initial condition is given by
\begin{equation}
u_k(\eta) 
= \f{1}{\sqrt{2\,k}}\,
\l[1 + \f{i\,\HI\, a(\eta)}{k}\r]\,
\mathrm{e}^{-i\,k\,\eta}.\label{inflation solution1}
\end{equation}
Or, equivalently, we can write that
\begin{eqnarray}
h_k(\eta) = h_k(a)
= \f{\sqrt{2}}{\Mpl}\,\f{i\,\HI}{\sqrt{2\,k^3}}\,
\l[1-\f{i\,k}{\HI\,a(\eta)}\r]\,\mathrm{e}^{-i\,k/\HI}\;
\mathrm{e}^{i\,k/[\HI\,a(\eta)]}\label{inflation solution2}
\end{eqnarray}
with $a(\eta)$ being given by the de Sitter form mentioned above.
Let us assume that inflation ends at the conformal time~$\ef$ such 
that $0< \ef <\HI^{-1}$, and let $\af=a(\ef)$.
Under these conditions, upon using the above solution, the tensor 
power spectrum at~$\af$ can be obtained to be
\begin{equation}
\pt(k) = \frac{2\,\HI^2}{\pi^2\,\Mpl^2}\,\l(1+\f{k^2}{\kf^2}\r),
\end{equation}
where $\kf =\af\,\HI$ is the mode that leaves the Hubble radius at 
the end of inflation.
For $k \ll \kf$, the above spectrum reduces to 
\begin{equation}
\pt(k) \simeq \frac{2\,\HI^2}{\pi^2\,\Mpl^2}\,\label{eq:pt-i},  
\end{equation}
which is the well known scale invariant spectrum often discussed in the context
of de Sitter inflation~\cite{Mukhanov:1990me,Martin:2003bt,
Martin:2004um,Bassett:2005xm,Sriramkumar:2009kg,Baumann:2008bn,Baumann:2009ds,
Sriramkumar:2012mik,Linde:2014nna,Martin:2015dha}.
Actually, in slow roll inflation, the tensor power spectrum will contain a 
small spectral tilt, which we shall choose to ignore in our discussion.
We should emphasize the point that the above scale invariant spectrum is valid 
only for $k \ll \kf$ since the de Sitter form for the scale factor would not 
hold true close to the end of inflation.
Therefore, in our discussion below, we shall mostly restrict ourselves to wave 
numbers such that $k < 10^{-2}\,\kf$. 
In Sec.~\ref{sec:actualinflation}, we shall evaluate tensor power spectrum 
numerically in the inflationary model of interest to arrive at the present 
day spectrum of GWs near~$\kf$.
 

\subsection{A typical inflationary model of interest}

In order to illustrate the results later, we shall focus on a specific 
inflationary model that permits slow roll inflation.
We shall consider the so-called $\alpha$-attractor model, which unifies 
a large number of inflationary potentials (in this context, see
Refs.~\cite{Kallosh:2013hoa,Kallosh:2013yoa}). 
If $\phi$ is the canonical scalar field driving inflation, the model is 
described by the potential~$V(\phi)$ of the form
\begin{equation}
V(\phi) = \Lambda^4\,
\l[1 - \mathrm{exp}\l(-\sqrt{\f{2}{3\,\alpha}}\,\f{\phi}{\Mpl}\r)\r]^{2\,n},
\label{eq:aap}
\end{equation}
where, as we shall soon discuss, the scale $\Lambda$ can be determined 
using the constraints from the observations of the anisotropies in the 
CMB. 
It is worth pointing out here that, for $n = 1$ and $\alpha = 1$,
the above potential reduces to the Higgs-Starobinsky 
model~\cite{Starobinsky:1980te,Bezrukov:2007ep}.
We should also mention that the potential~\eqref{eq:aap} contains a 
plateau at suitably large values of the field, which is favored by the 
CMB data~\cite{Akrami:2018odb}.

Let~$N_k$ denote the e-fold at which the wave number $k$ leaves the Hubble 
radius, \textit{when counted from the end of inflation}.\/
For the potential~\eqref{eq:aap}, the quantity $N_k$ can be easily determined 
in the slow roll approximation to be (see, for instance, Ref.~\cite{Drewes:2017fmn})
\begin{equation}
N_k \simeq \f{3\,\alpha}{4\,n}\,
\l[\mathrm{exp}\l({\sqrt{\f{2}{3\,\alpha}}}\,\f{\phi_k}{\Mpl}\r) 
- \mathrm{exp}\l({\sqrt{\f{2}{3\,\alpha}}}\,\f{\phif}{\Mpl}\r) 
- \sqrt{\f{2}{3\,\alpha}}\,\l(\f{\phi_k}{\Mpl} - \f{\phif}{\Mpl}\r)\r],
\label{e-folding-alpha-attractor}
\end{equation}
where $\phi_k$ and $\phif$ denote the values of the field when the mode with wave
number~$k$ crosses the Hubble radius and at the end of inflation, respectively.
It is easy to show that $\phif$ can be expressed in terms of the parameters~$\alpha$ 
and~$n$ as
\begin{eqnarray}
\f{\phif}{\Mpl}
\simeq \sqrt{\f{3\,\alpha}{2}}\,\mathrm{ln}\l(\f{2\,n}{\sqrt{3\,\alpha}} + 1\r).
\label{end of phi}
\end{eqnarray}
As we mentioned, the parameter~$\Lambda$ can be determined by utilizing the 
constraints from the CMB.
One finds that the parameter can be expressed in terms of the scalar 
amplitude~$\cAs$, the scalar spectral index~$\ns$ and the tensor-to-scalar 
ratio~$r$ as follows:
\begin{equation}
\Lambda = \Mpl\, \l(\f{3\,\pi^2\, r\,\cAs}{2}\r)^{1/4}\,
\l[\f{2\,n\,(1+2\,n) 
+ \sqrt{4\,n^2 
+ 6\,\alpha\,(1+n)\,\l(1-\ns\r)}}{4\,n\,(1+n)}\r]^{n/2}.\label{lambda}
\end{equation}
Given the best-fit values for the inflationary parameters~$\cAs$ and $\ns$ as 
well as the upper bound on~$r$ from Planck~\cite{Akrami:2018odb}, evidently, 
using the above relation, we can choose a set of values for the parameters~$\Lambda$,
$\alpha$ and~$n$ describing the potential~\eqref{eq:aap} that are compatible 
with the CMB data.

As we shall see, to evolve the background beyond inflation, we shall require 
the value of the energy density of the scalar field at the end of inflation, 
say, $\rho_\mathrm{f}$.
One finds that the quantity $\rho_\mathrm{f}$ can be expressed in terms of
the corresponding value of the potential, say,  $V_\mathrm{f}$, as 
\begin{equation}
\rho_\mathrm{f}= \f{3}{2}\,V_\mathrm{f}.\label{eq:rhof}
\end{equation}
It is useful to note here that the value of the potential~\eqref{eq:aap} at 
the end of inflation is given by
\begin{eqnarray}
V_\mathrm{f} = V(\phif)
=\Lambda^4\,\l(\f{2\,n}{2\,n 
+ \sqrt{3\,\alpha}}\r)^{2\,n}.\label{end of potential}
\end{eqnarray}
We should mention that, hereafter, we shall set the parameter $\alpha$ to be
unity.


\subsection{The dimensionless energy density of GWs}

Let us turn to now discuss the observable quantity of our interest, viz.
the dimensionless energy density of GWs.
We shall be interested in evaluating the quantity over the domain of wave 
numbers which reenter the Hubble radius during the epochs of reheating and
radiation domination.

The energy density of GWs at any given time, say, $\rhogw(\eta)$, is given 
by (see, for example, Refs.~\cite{Boyle:2005se,Maggiore:1999vm})
\begin{equation}
\rhogw(\eta)=\f{\Mpl^2}{4\, a^2}\, \l(\f{1}{2}\,\langle {\hat h}_{ij}^{\prime 2}\rangle    
+\f{1}{2}\,\langle \vert{\bm \nabla} {\hat h}_{ij}\vert^2\rangle\r),
\end{equation}
where the expectation values are to be evaluated in the initial Bunch-Davies 
vacuum imposed in the sub-Hubble regime during inflation.
The energy density per logarithimic interval, say, $\rhogw(k,\eta)$, is
defined through the relation
\begin{equation}
\rhogw(\eta)=\int_{0}^{\infty}\d\, \mathrm{ln}\, k\; \rhogw(k,\eta).
\end{equation}
Upon using the mode decomposition~\eqref{eq:tp-m-dc} and the above expressions
for $\rhogw(\eta)$, we obtain the quantity~$\rhogw(k,\eta)$ to be
\begin{equation}
\rhogw(k,\eta)=\f{\Mpl^2}{a^2}\, \f{k^3}{2\,\pi^2}\,
\l(\f{1}{2}\,\vert h_k'(\eta)\vert^2+\f{k^2}{2}\,\vert h_k(\eta)\vert^2\r).
\label{eq:rho-gw-at}
\end{equation}
The corresponding dimensionless energy density $\ogw(k,\eta)$ is defined as 
\begin{equation}
\ogw(k,\eta)=\f{\rhogw(k,\eta)}{\rho_\mathrm{c}(\eta)}
=\f{\rhogw(k,\eta)}{3\,H^2\,\Mpl^2},
\end{equation}
where $\rho_\mathrm{c}=3\,H^2\,\Mpl^2$ is the critical density at time~$\eta$.

The observable quantity of interest is the dimensionless energy 
density~$\ogw(k,\eta)$ evaluated \textit{today},\/ which we shall 
denote as $\ogw(k)$.
We shall often refer to $\ogw(k)$ or, equivalently, $\ogw(f)$ [recall
that $f$ is the frequency associated with the wave number $k$, cf.
Eq.~\eqref{eq:f}], as the spectrum of GWs today.
In the following sections, we shall evolve the tensor perturbations
through the epochs of reheating and radiation domination.
As we shall see, at late times during radiation domination, once all
the wave numbers of interest are well inside the Hubble radius, the 
energy density of GWs behave in a manner similar to that of the 
energy density of radiation.
We shall utilize this behavior to arrive at the spectrum of GWs today.


\section{Evolution of GWs during reheating}\label{sec:egw-drh}

In order to follow the evolution of the tensor perturbations post 
inflation, it proves to be convenient to write (in this context, 
see, for instance, Ref.~\cite{Boyle:2005se})
\begin{equation}
h_k(\eta) = h_k^{^{_{\mathrm{P}}}}\,\chi_k(\eta),\label{transfer function}
\end{equation}
where $h_k^{^{_{\mathrm{P}}}}$ denotes the primordial value evaluated 
at the end of inflation, and is given by [cf. Eq.~\eqref{inflation solution2}]
\begin{equation}
h_k^{^{_{\mathrm{P}}}}
=h_k(\af) 
=\f{\sqrt{2}}{\Mpl}\,\f{i\,\HI}{\sqrt{2\,k^3}}\,
\l(1-\f{i\,k}{\kf}\r)\,\mathrm{e}^{-i\,k/\HI}\;
\mathrm{e}^{i\,k/\kf}.\label{eq:hkp}
\end{equation}
The quantity $\chi_k$ is often referred to as the tensor transfer function, 
which obeys same equation of motion as $h_k$, viz. Eq.~\eqref{eq:em-hk}.
Clearly, the strength of the primordial GWs observed today will not only 
depend on the amplitude of the tensor perturbations generated during 
inflation, but also on their evolution during the subsequent epochs. 
In this section, we shall discuss the evolution of the transfer function $\chi_k$
during the epoch of reheating.
As we mentioned earlier, we shall consider two types of scenarios for reheating.
We shall first consider the case wherein the period of reheating is described by
the constant EoS parameter, say, $w_\phi$, often associated with the coherent
oscillations of the scalar field around the minimum of the inflationary potential.
We shall assume that the transition to radiation domination occurs instantaneously
after a certain duration of time.
In such a case, as we shall see, the transfer function $\chi_k$ can be arrived 
at analytically.
We shall then consider the scenario of perturbative reheating wherein there is a 
gradual transfer of energy from the inflaton to radiation.
It seems difficult to treat such a situation analytically and, hence, we shall 
examine the problem numerically.

To follow the evolution of GWs in the post-inflationary regime, we shall 
choose to work with rescaled scale factor as the independent variable rather 
than the conformal time coordinate.
If we define $A=a/\af$, where $\af=a(\ef)$ is the scale factor at the end 
of inflation, then we find that the equation governing the transfer function 
$\chi_k$ is given by 
\begin{equation}
\f{\d^2\chi_k}{\d A^2} + \l(\f{4}{A}+\f{1}{H}\,\f{\d H}{\d A}\r)\,
\f{\d\chi_k}{\d A} 
+ \f{(k/\kf)^2}{(H/\HI)^2\,A^4}\,\chi_k=0,\label{eq:de-tf}
\end{equation}
where, recall that, $\kf=\af\,\HI$ denotes the wave number that leaves the 
Hubble radius at the end of inflation. 
Our aim now is to solve the above equation for the transfer function during 
the epoch of reheating.
As is evident from the equation, the evolution of GWs is dictated by the 
behavior of the Hubble parameter.
Therefore, if we can first determine the behavior of the Hubble parameter during 
reheating, we can solve the above equation to understand the evolution of 
transfer function~$\chi_k$ during the epoch.
We shall also require the initial conditions for the transfer function~$\chi_k$
and its derivative $\d \chi_k/\d A$ at the end of inflation, i.e. when $a=\af$
or, equivalently, when $A=1$.
Since we have introduced the transfer function through 
the relation~\eqref{transfer function}, clearly, 
\begin{equation}
\chi_k^{_\mathrm{I}}(A=1)=1.\label{boundary condition inflation1}
\end{equation}
Also, on using the solution~\eqref{inflation solution2} and the 
expression~\eqref{eq:hkp} for $h_k^{^{_{\mathrm{P}}}}$, we find that 
\begin{equation}
\f{\d \chi_k^{_\mathrm{I}}(A=1)}{\d A} =-\f{(k/\kf)^2}{1-i\,(\,k/\kf)}
\simeq 0,\label{boundary condition inflation2}
\end{equation}
where the final equality is applicable when $k \ll \kf$.
We shall make use of these initial conditions to determine the transfer 
function during the epoch of reheating.

In the following two sub-sections, we shall discuss the solutions in the two 
scenarios involving instantaneous and gradual transfer of energy from the
inflaton to radiation.


\subsection{Reheating described by an averaged EoS parameter}\label{subsec:averaged}

Let us first consider the scenario wherein the epoch of reheating is dominated 
by the dynamics of the scalar field as it oscillates at the bottom of an 
inflationary potential, such as the $\alpha$-attractor model~\eqref{eq:aap}
we had introduced earlier.
In such a case, the evolution of the scalar field can be described by an 
averaged EoS parameter, say,~$w_\phi$~\cite{Martin:2010kz,Dai:2014jja,Cook:2015vqa}.
The conservation of energy implies that the energy density of the inflaton 
behaves as $\rho_\phi \propto a^{-3\,(1+w_\phi)}$, which, in turn,
implies that the Hubble parameter behaves as 
$H^2 =\HI^2\, A^{-3\,(1 + w_\phi)}$. 
Note that, $H=\HI$ when $A=1$, as required.
As a result, during such a phase, equation~\eqref{eq:de-tf} 
governing the tensor transfer function reduces to the form
\begin{equation}
\f{\d^2\chi_k}{\d A^2} 
+ (5-3\,w_\phi)\,\f{1}{2\,A}\,\f{\d\chi_k}{\d A} 
+ \f{(k/\kf)^2}{A^{1-3\,w_\phi}}\,\chi_k = 0.
\label{transfer equation reheating case1}
\end{equation}
The general solution to this differential equation can be expressed as
\begin{equation}
\chi_k^{_\mathrm{RH}}(A)
=A^{-\nu}\, \l[C_k\, J_{-\nu/\gamma}\l(\f{k}{\gamma\, \kf}\,A^{\gamma}\r) 
+ D_k\,J_{\nu/\gamma}\l(\f{k}{\gamma\,\kf}\,A^{\gamma}\r)\r],
\label{transfer solution reheating case1}
\end{equation}
where $J_{\alpha}(z)$ denote Bessel functions of order $\alpha$, while 
the quantities $\nu$ and $\gamma$ are given by
\begin{equation}
\nu=\f{3}{4}\,(1-w_\phi),\quad 
\gamma=\f{1}{2}\,\l(1+3\,w_\phi\r).\label{eq:n-g}
\end{equation}
The coefficients $C_k$ and $D_k$ can be arrived at by using the 
conditions~\eqref{boundary condition inflation1} 
and~\eqref{boundary condition inflation2} for the transfer function
$\chi_k$ and its derivative $\d \chi_k/\d A$ at end of inflation, i.e. when $A=1$.
We find that the coefficients $C_k$ and $D_k$ can be expressed as follows:
\begin{subequations}\label{eq:Ck-Dk}
\begin{eqnarray}
C_k &=&\f{\pi\,k}{2\,\gamma\,\kf}\, \l[\f{1}{1-i\,(k/\kf)}\r]\,
\l[\f{k}{\kf}\;J_{\nu/\gamma}\l(\f{k}{\gamma\, \kf}\r) 
- \l(1-\f{i\, k}{\kf}\r)\,J_{(\nu/\gamma)+1}\l(\f{k}{\gamma\, \kf}\r) \r]\,
\csc\l(\frac{\pi\,\nu}{\gamma}\r),\\
D_k &=& -\f{\pi\,k}{2\,\gamma\,\kf}\, \l[\f{1}{1-i\,(k/\kf)}\r]\,
\l[\f{k}{\kf}\;J_{-\nu/\gamma}\l(\f{k}{\gamma\, \kf}\r) 
+ \l(1-\f{i\,k}{\kf}\r)\,J_{-(\nu/\gamma)-1}\l(\f{k}{\gamma\, \kf}\r) \r]\,
\csc\l(\frac{\pi\,\nu}{\gamma}\r).
\end{eqnarray}
\end{subequations}
We shall later make use of these coefficients and the 
solution~\eqref{transfer solution reheating case1} to eventually arrive at 
the spectrum of GWs today.
It is useful to note here that the quantity $\d \chi_k^{_\mathrm{RH}}/\d A$ 
is given by
\begin{equation}
\f{\d \chi_k^{_\mathrm{RH}}}{\d A}
=\f{k}{\kf}\, A^{-1+\gamma-\nu}\, 
\l[C_k\, J_{-(\nu/\gamma)-1}\l(\f{k}{\gamma\, \kf}\,A^{\gamma}\r) 
- D_k\,J_{(\nu/\gamma)+1}\l(\f{k}{\gamma\,\kf}\,A^{\gamma}\r)\r].
\label{eq:dchi-dA}
\end{equation}
\color{black}

We should mention here that the duration of the reheating phase characterized by 
the number of e-folds~$\Nre$ and the reheating temperature~$\Tre$ can be expressed 
in terms of the equation of state parameter~$w_\phi$ and the inflationary 
parameters as follows (in this context, see, for example, 
Refs.~\cite{Dai:2014jja,Cook:2015vqa}):
\begin{subequations}
\begin{eqnarray}
\Nre &=& \f{4}{(3\,w_\phi-1)}
\l[N_\ast
+\f{1}{4}\,\ln \l(\f{30}{\pi^2\,g_\mathrm{r, re}}\r)
+ \f{1}{3}\ln \l(\f{11\,g_{s,\mathrm{re}}}{43}\r) + \ln \l(\f{k_\ast}{a_0\,T_0}\r) 
+ \ln \l(\f{\rho_\mathrm{f}^{1/4}}{\HI}\r)\r],\label{reheating e-folding}\\
\Tre &=& \l(\f{43}{11\,g_{s,\mathrm{re}}}\r)^{1/3}
\l(\f{a_0\,\HI}{k_\ast}\r)\,
\mathrm{e}^{-(N_\ast + \Nre)}\;
T_0,\label{eq:Tre}
\end{eqnarray}
\end{subequations}
where $T_0 = 2.725\,\mathrm{K}$ is the present temperature of the CMB and 
$H_0$~denotes the current value of the Hubble parameter.
Moreover, note that, $k_\ast/a_0$ represents the CMB pivot scale, with $a_0$ 
denoting the scale factor today.
We shall assume that $k_\ast/a_0 \simeq 0.05\,\mathrm{Mpc}^{-1}$.
We should also point out that $N_\ast$ denotes the number of e-folds prior to 
the end of inflation when the pivot scale~$k_\ast$ leaves the Hubble radius.


\subsection{The case of perturbative reheating}\label{subsec:actual}

As a second possibility, we shall consider the perturbative 
reheating scenario (for recent discussions, see, for instance, 
Refs.~\cite{Maity:2018dgy,Haque:2020zco}).
In such a  case, after inflation, the inflaton energy density, say,
$\rho_\phi$, gradually decays into the radiation energy density, say, 
$\rhoR$, with the decay process being governed by the Boltzmann equations. 
As a result, the effective EoS parameter during the reheating phase becomes 
time dependent.
In our analysis below, for simplicity, we shall assume that the EoS
parameter~$w_\phi$ of the scalar field during the reheating phase 
remains constant. 
Such an assumption is valid as far as the oscillation time scale of the 
inflaton is much smaller than the Hubble time scale.
This turns out to be generally true when the field is oscillating immediately 
after the end of inflation, and we should point out that such a behavior has 
also been seen in lattice 
simulations~\cite{Podolsky:2005bw,Figueroa:2016wxr,Maity:2018qhi}. 
Hence, for a wide class of inflationary potentials which behave as $V(\phi) 
\propto \phi^{2\,n}$ near the minimum, the time averaged EoS parameter of the 
inflaton can be expressed as $w_\phi = (n-1)/(n+1)$~\cite{Mukhanov:2005sc}.
We should emphasize that this scenario is different from the one considered
in the previous section wherein the explicit decay of the inflaton field was 
not taken into account.
Specifically, in the earlier case, the energy density of the inflaton was 
supposed to be converted instantaneously into the energy density of radiation 
at a given time, leading to the end of the phase of reheating. 
In due course, we shall demonstrate the manner in which the detailed mechanism 
of reheating leaves specific imprints on the spectrum of primordial GWs observed 
today.

Let us define the following dimensionless variables to describe the comoving 
energy densities of the scalar field and radiation 
\begin{equation}\label{rescale}
\Phi(A) = \f{\rho_\phi}{m_\phi^4}\, A^{3\,(1 + w_\phi)}, \quad  
R(A) = \f{\rhoR}{m_\phi^4}\, A^4,
\end{equation}
where $m_\phi$ denotes the mass of the inflaton.
Also, let $\Gamma_\phi$ represent the decay rate of the
inflaton to radiation.
In such a case, the Boltzmann equations governing the evolution of the energy 
densities $\rho_\phi$ and $\rhoR$ can be 
expressed as~\cite{Maity:2018dgy,Haque:2020zco,Haque:2020bip}
\begin{subequations}\label{eq:be}
\begin{eqnarray}
\f{\d\Phi}{\d A} + \f{\sqrt{3}\, \Mpl\, \Gamma_\phi}{m_\phi^2}\, (1 + w_\phi)\;
\f{A^{1/2}\,\Phi}{(\Phi/A^{3\, w_\phi}) + (R/A)} &=& 0,\label{Boltzmann equation1}\\
\f{\d R}{\d A} - \f{\sqrt{3}\, \Mpl\, \Gamma_\phi}{m_\phi^2}\, (1 + w_\phi)\;
\f{A^{3\, (1 - 2\,w_\phi)/2}\,\Phi}{(\Phi/A^{3\, w_\phi})
+ (R/A)} &=& 0.\label{Boltzmann equation2}
\end{eqnarray}
\end{subequations}
Note that the tensor transfer function $\chi_k$ is essentially governed by 
the behavior of the Hubble parameter~$H$ [cf. Eq.~\eqref{eq:de-tf}].
It proves to be difficult to solve the above set of equations analytically.
Therefore, to arrive at the Hubble parameter during the epoch of reheating, we 
shall  solve the Boltzmann equations~\eqref{eq:be} numerically with the following 
conditions imposed at the end of inflation:
\begin{equation}
\rho_\phi(A=1) = \rho_\mathrm{f}=\f{3}{2}\,V_\mathrm{f},\quad
\rhoR(A=1)= 0,\label{boundary Boltzmann equation1}
\end{equation}
where, recall that, $V_\mathrm{f}$ is the value of the inflationary potential 
at the end of inflation.
We should point out that, for the inflationary potential~\eqref{eq:aap} 
of our interest, $V_\mathrm{f}$ is given by Eq.~\eqref{end of potential} .

We should also clarify a couple of points in this regard.
In the case of perturbative reheating, the phase of reheating is assumed to 
be complete when $H = \Gamma_\mathrm{\phi}$.
In other words, we require
\begin{equation} \label{reheating 1}
H^2(\Are)
= \f{1}{3\,\Mpl^2}\,
\l[\rho_\mathrm{\phi}(\Are,\ns,\Gamma_\mathrm{\phi})
+ \rhoR(\Are,\ns,\Gamma_\phi)\r]= \Gamma_\phi^2,
\end{equation}
where $\Are=a_\mathrm{re}/\af$, with $a_\mathrm{re}$ denoting the scale factor 
when reheating has been achieved.
In fact, this corresponds to the point in time when the rate of transfer of 
the energy from the inflaton to radiation is the maximum.
The associated reheating temperature can then be determined from the energy 
density of radiation~$\rhoR$ through the relation
\begin{equation} \label{reheating 2}
\Tre= \l(\f{30}{\pi^2\, g_{r, \mathrm{re}}}\r)^{1/4}\,
\rhoR^{1/4}(\Are,\ns,\Gamma_\phi).
\end{equation}
In fact, later, to arrive at the results on the spectrum of GWs today, along 
with specific values for the parameters describing the inflationary potential 
(that are consistent with the CMB data), we shall also choose a value of~$\Tre$.
Having fixed the value of $\Tre$, we shall make use of Eq.~\eqref{eq:Tre} to 
arrive at $\Nre$ and thereby determine the value $\Gamma_\phi$ using the 
condition~$H(\Are)=\Gamma_\phi$.

\begin{figure}[t!]
\centering
\includegraphics[width=8.25cm]{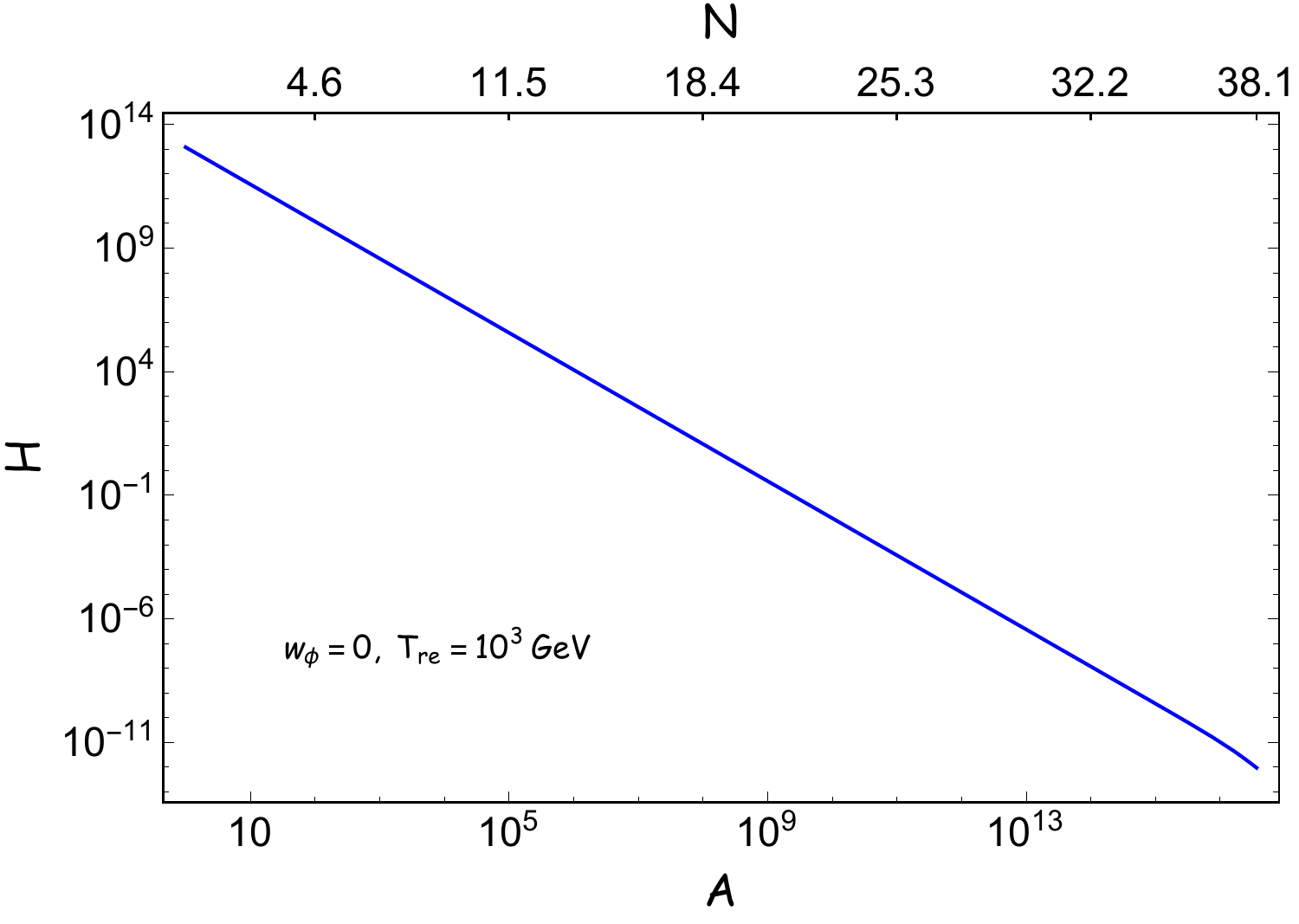}
\hskip 5pt
\includegraphics[width=8.00cm]{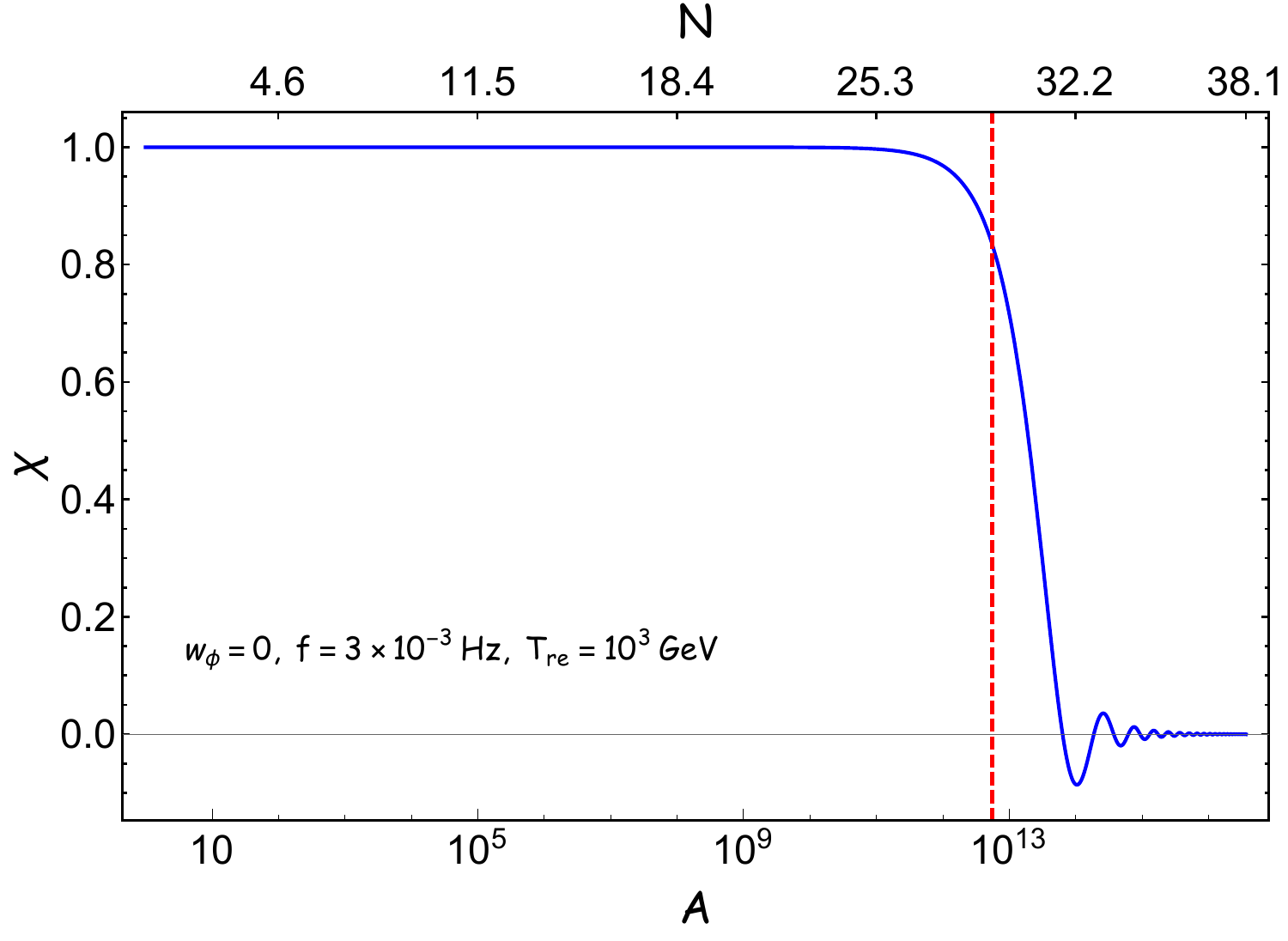}
\caption{The evolution of the Hubble parameter~$H$ (in blue, on the left) 
and the amplitude of the tensor transfer function~$\chi_k$ for a given wave 
number (in blue, on the right), obtained numerically in the case of the 
perturbative reheating scenario, have been plotted as a function of $A=a/\af$ 
over the domain $1 \leq A \leq \Are$, which corresponds to the period of 
reheating. 
We have also indicated the lapse in time in terms of e-folds (counted from the 
end of inflation) on the top of the two figures.
We have assumed that $w_{\phi} = 0$ and have set $\Tre = 10^{3}\,\mathrm{GeV}$
in plotting the quantities.
We have chosen the wave number to be $k\simeq 2\times 10^{12}\,\mathrm{Mpc}^{-1}$, 
which corresponds to the frequency of $f\simeq 3\times10^{-3}\,\mathrm{Hz}$ [cf.
Eq.~\eqref{eq:f}].
The wave number has been chosen so that it renters the Hubble radius during the 
epoch of reheating.
The slope of the straight line describing $H(A)$ (on the left) is $-3/2$, which is
consistent with $w_\phi=0$.
We find that the slope changes as $A$ approaches $\Are$ indicating the beginning of
the transition to the radiation dominated epoch.
The vertical line (in red, on the right) indicates the time when the mode reenters 
the Hubble radius.
As expected, the transfer function proves to be constant on super-Hubble scales and 
it oscillates once the mode is inside the Hubble radius.} 
\label{plot_solution_reheating}
\end{figure}

With the solutions to the coupled background equations~\eqref{eq:be} at hand, we 
shall proceed to solve the differential equation~\eqref{eq:de-tf} for the transfer
function~$\chi_k$ during reheating.
To illustrate the nature of the solutions, we have plotted the behavior of the Hubble 
parameter~$H$ and the transfer function~$\chi_k$ in Fig.~\ref{plot_solution_reheating}
during the period of perturbative reheating.
We have chosen specific values for the EoS parameter~$w_\phi$, the reheating 
temperature~$\Tre$, and wave number~$k$ in plotting the figures.
We have considered a wave number so that the mode reenters the Hubble radius 
during the period of reheating.
As one would have expected, while the amplitude of the transfer function is a 
constant on super-Hubble scales, it decreases as the mode reenters the Hubble 
radius and begins to oscillate once inside.


\section{Evolution during radiation domination and the spectrum 
of GWs today}\label{sec:egw-drd}

The Hubble parameter during the radiation dominated epoch evolves as
\begin{eqnarray}
H^2=\Hre^2\, \f{A_\mathrm{re}^4}{A^4}\label{connection1}
\end{eqnarray}
with $\Hre$ and $\Are$ denoting the Hubble parameter and the rescaled 
scale factor at the end of reheating, respectively. 
During radiation domination, the transfer function is governed by the
equation
\begin{equation}
\frac{\d^2 \chi_k}{\d A^2} 
+ \f{2}{A}\,\f{\d\chi_k}{\d A} 
+ \f{(k/\kre)^2}{\Are^2}\,\chi_k = 0,\label{transfer equation radiation}
\end{equation}
where $\kre = a_\mathrm{re}\,\Hre$ is the mode which reenters the Hubble
radius at the end of the reheating era. 
The above differential equation can be immediately solved to arrive at 
the following general solution:
\begin{equation}
\chi_k^{_\mathrm{RD}}(A)
= \f{1}{A}\,\biggl\{E_k\, \mathrm{e}^{-i\,(k/\kre)\,\l[(A/\Are)-1\r]} 
+ F_k\,\mathrm{e}^{i\,(k/\kre)\,\l[(A/\Are)-1\r]}\biggr\}.
\label{transfer solution radiation 1}
\end{equation}
The coefficients $E_k$ and $F_k$ need to be determined by matching this solution 
and its derivative with the solution~\eqref{transfer solution reheating case1} 
during the epoch of reheating and its derivative at~$A=\Are$. 
Upon carrying out the matching at the junction between the eras of reheating 
and the radiation domination, we find that the coefficients $E_k$ and $F_k$ 
can be expressed as
\begin{subequations}\label{eq:Ek-Fk}
\begin{eqnarray}
E_k &=&\f{\Are}{2}\, \l[\l(1+ \f{i\,\kre}{k}\r)\,\chi_k^{_\mathrm{RH}}(\Are)
+ \f{i\,\kre}{k}\,\Are\,\f{\d\chi_k^{_\mathrm{RH}}(\Are)}{\d A}\r]
=i\,\f{\Are}{2}\ \f{\kre}{k}\,\mathcal{E}_k,\\
F_k &=& \f{\Are}{2}\, \l[\l(1- \f{i\,\kre}{k}\r)\,\chi_k^{_\mathrm{RH}}(\Are)
- \f{i\,\kre}{k}\,\Are\,\f{\d\chi_k^{_\mathrm{RH}}(\Are)}{\d A}\r]
=-i\,\f{\Are}{2}\ \f{\kre}{k}\,\mathcal{F}_k,
\end{eqnarray}
\end{subequations}
where we have also introduced the quantities $\mathcal{E}_k$ and 
$\mathcal{F}_k$ which we shall make use of later.
At this stage, we should mention that the quantities $\chi_k^{_\mathrm{RH}}(\Are)$
and $\d\chi_k^{_\mathrm{RH}}(\Are)/\d A$ and hence the coefficients $E_k$ and~$F_k$ 
will depend on the details of the reheating mechanism. 
In particular, they will be different for two the types of reheating mechanisms under 
consideration.

Since we have set $h_k=h_k^{^{_{\mathrm{P}}}}\,\chi_k$, where $h_k^{^{_{\mathrm{P}}}}$ 
denotes the primordial amplitude of the tensor perturbations 
[cf. Eq.~\eqref{eq:rho-gw-at}], the energy density of GWs $\rhogw(k,\eta)$ is 
given by  [cf. Eq.~\eqref{transfer function}]
\begin{equation}
\rhogw(k,\eta)=\f{\Mpl^2}{a^2}\, 
\f{k^3}{2\,\pi^2}\, \vert h_k^{^{_{\mathrm{P}}}}\vert^2\,
\l(\f{1}{2}\,\vert {\chi_k^{_\mathrm{RD}}}'(\eta)\vert^2
+\f{k^2}{2}\,\vert \chi_k^{_\mathrm{RD}}(\eta)\vert^2\r).
\end{equation}
On using the definition~\eqref{eq:tps} of the inflationary tensor power 
spectrum~$\pt(k)$, this energy density of GWs can be expressed as
\begin{equation}
\rhogw(k,\eta)=\f{\Mpl^2}{4\,a^2}\, \pt(k)\,
\l(\f{1}{2}\,\vert {\chi_k^{_\mathrm{RD}}}'(\eta)\vert^2
+\f{k^2}{2}\,\vert \chi_k^{_\mathrm{RD}}(\eta)\vert^2\r).
\end{equation}
Upon substituting the solution~\eqref{transfer solution radiation 1} in the
above expression, we obtain $\rhogw(k,\eta)$ to be
\begin{eqnarray}
\rhogw(k,\eta)
&=& \f{\Mpl^2\,k^2}{8\,\af^2\,A^4}\,\pt(k)\,
\biggl\{\l(\vert E_k\vert^2 +\vert F_k \vert^2\r)\,
\l[2+\l(\f{\kre\,\Are}{k\,A}\r)^2\r]\nn\\
& &+\,E_k\,F_k^\ast\,\l(\f{\kre\,\Are}{k\,A}\r)^2\,
\l(1+\f{2\,i\,k\,A}{\kre\,\Are}\r)\,
\mathrm{e}^{-2\,i\,(k/\kre)\,\l[(A/\Are)-1\r]}\nn\\
& &+\,E_k^\ast\,F_k\, \l(\f{\kre\,\Are}{k\,A}\r)^2\,
\l(1-\f{2\,i\,k\,A}{\kre\,\Are}\r)\,
\mathrm{e}^{2\,i\,(k/\kre)\,\l[(A/\Are)-1\r]}\biggr\}.
\end{eqnarray}
We had mentioned earlier that we shall be interested in the range of wave numbers 
which reenter the Hubble radius during the epochs of reheating and radiation 
domination.
At late times during radiation domination such that $A/\Are\gg 1$, all the modes
of our interest would be well inside the Hubble radius, i.e. $k\, A \gg 1$.
In such a case, we find that, the above expression for $\rhogw(k,\eta)$
simplifies to be 
\begin{equation}
\rhogw(k,\eta)
= \f{\Mpl^2\,k^2}{4\,\af^2\,A^4}\,\pt(k)\,
\l(\vert E_k\vert^2 +\vert F_k \vert^2\r).\label{eq:rho-gw-rd}
\end{equation}
Hence, the corresponding dimensionless parameter describing the energy 
density of GWs, viz. $\ogw(k,\eta)$, is given by
\begin{equation}
\ogw(k,\eta)
= \f{k^2}{12\,\af^2\,H^2\,A^4}\,\pt(k)\,
\l(\vert E_k\vert^2 +\vert F_k \vert^2\r)
= \f{\kre^2\,\Are^2}{48\,\af^2\,H^2\,A^4}\,\pt(k)\,
\l(\vert \mathcal{E}_k\vert^2 
+\vert \mathcal{F}_k \vert^2\r),\label{PGW_spectrum_main}
\end{equation}
where we have made use of the expressions~\eqref{eq:Ek-Fk} relating 
the coefficients $E_k$ and $F_k$ to the quantities $\mathcal{E}_k$ 
and  $\mathcal{F}_k$.
During radiation domination, $H^2\,A^4=\Hre^2\,\Are^4$.
Also, recall that, $\kre=\are\, \Hre$.
On using these relations, at late times during radiation domination,
we obtain that
\begin{equation}
\ogw(k,\eta)
= \f{\pt(k)}{48}\, \l(\vert \mathcal{E}_k\vert^2 
+\vert \mathcal{F}_k \vert^2\r).\label{PGW_spectrum_main2}
\end{equation}
The task that remains is to explicitly determine the quantities
$\mathcal{E}_k$ and $\mathcal{F}_k$.
In the special case of instantaneous reheating, $\Are=1$ and $\kre=\kf$.  
Therefore, on using the conditions~\eqref{boundary condition inflation1}
and~\eqref{boundary condition inflation2}, one can readily show that
\begin{equation}
\mathcal{E}_k
=\f{1-2\,i\,(k/\kf)-2\,(k^2/\kf^2)}{1-i\,(k/\kf)},\quad  
\mathcal{F}_k=\f{1}{1-i\,(k/\kf)}.
\end{equation}
For $k\ll \kf$, we find that, $\mathcal{E}_k \simeq F_k \simeq 1$,
which lead to
\begin{equation}
\ogw(k,\eta)= \f{\pt(k)}{24}=\f{\HI^2}{12\,\pi^2\,\Mpl^2},
\end{equation}
where, in arriving at the final expression, we have made use of the scale invariant
inflationary tensor power spectrum~\eqref{eq:pt-i}. 
In other words, the dimensionless density parameter $\ogw(k,\eta)$ is strictly scale 
invariant over all wave numbers in the instantaneous reheating scenario.
Evidently, such a behavior can be expected to be hold true even when we have an 
epoch of reheating with $w_\phi=1/3$, a result we shall encounter in due course.

Note that the energy density of GWs behaves as $a^{-4}$ 
[cf. Eq.~\eqref{eq:rho-gw-rd}], exactly as the energy density of radiation does.
Such a behavior should not come as a surprise and it arises due to the fact that
the modes of interest are well inside the Hubble radius at late times (say,
close to the epoch of radiation-matter equality) during radiation domination.
On utilizing this property, the dimensionless energy density
parameter~$\ogw(k)$ \textit{today}\/ can be expressed in terms 
of $\ogw(k,\eta)$ as follows:
\begin{equation}
\ogw(k)\,h^2
= \l(\f{g_{r,\mathrm{eq}}}{g_{r,0}}\r)\, 
\l(\f{g_{s,0}}{g_{s,\mathrm{eq}}}\r)^{4/3}\,
\Omega_{_{\mathrm{R}}}\, h^2\; \ogw(k,\eta)
\simeq  \l(\f{g_{r,0}}{g_{r,\mathrm{eq}}}\r)^{1/3}\,
\Omega_{_{\mathrm{R}}}\, h^2\; \ogw(k,\eta),\label{presentrelic}
\end{equation}
where $\Omega_{_\mathrm{R}}$ denotes the present day dimensionless 
energy density of radiation.
We should mention that, while $g_{r,\mathrm{eq}}$ and $g_{r,0}$ represent 
the number of relativistic degrees of freedom at equality and today, 
respectively, $g_{s,\mathrm{eq}}$ and $g_{s,0}$ represent the number of 
such degrees of freedom that contribute to the entropy at these epochs.
Further, the Hubble parameter today, as usual, has been expressed as 
$H_0=100\,h\,\mathrm{km}\,\mathrm{sec}^{-1}\,\mathrm{Mpc}^{-1}$.

The spectrum of primordial GWs in the reheating scenario with an averaged EoS 
parameter can be expected to be different when compared to the one arising in 
the perturbative reheating scenario, 
In the following two sections, we shall derive the spectrum of primordial GWs 
at the present epoch in the two cases. 


\section{Spectrum of GWs in reheating described by an averaged EoS
parameter}\label{sec:averaged}

Before we go on to discuss the results, we should mention that the 
spectrum of GWs arising in the scenario wherein the epoch of reheating 
is described by an averaged EoS parameter and the transition to radiation
domination is assumed to occur instantaneously at a given time has been 
evaluated earlier in the literature (see, for instance, 
Refs.~\cite{Turner:1993vb,Nakayama:2008wy,Nakayama:2009ce}; for a recent 
discussion, see Ref.~\cite{Mishra:2021wkm}). 
However, we find that, in the earlier investigations, the initial conditions 
that determine the dynamics during reheating have not always been chosen to 
be consistent with the dynamics during inflation. 
In this work, we shall consider a specific model of inflation and we 
shall show that model dependent initial conditions play a primary role 
in determining the range of frequencies which reenter the Hubble radius 
during reheating.
Therefore, in this section, we shall reanalyze the effect of the averaged 
EoS parameter during reheating (with appropriate initial conditions) on
the spectrum of GWs.
We shall briefly discuss the derivation of the spectrum of GWs and arrive 
at the shape of the spectrum in the domains $k < \kre$ and $k > \kre$.
In the next section, we shall compare the results with those that arise 
in the case of perturbative reheating. 

It should be clear from the expression~\eqref{PGW_spectrum_main2} that
we shall require the quantities~$\mathcal{E}_k$ and $\mathcal{F}_k$ to 
arrive at the spectrum of GWs.
Using Eqs.~\eqref{eq:Ek-Fk}, we can express~$\mathcal{E}_k$ 
and $\mathcal{F}_k$ as
\begin{subequations}\label{eq:cal-Ek-Fk}
\begin{eqnarray}
\mathcal{E}_k 
&=&\l(1-\f{i\,k}{\kre}\r)\,\chi_k^{_\mathrm{RH}}(\Are)
+ \Are\,\f{\d\chi_k^{_\mathrm{RH}}(\Are)}{\d A},\\
\mathcal{F}_k 
&=& \l(1+\f{i\,k}{\kre}\r)\,\chi_k^{_\mathrm{RH}}(\Are)
+ \Are\,\f{\d\chi_k^{_\mathrm{RH}}(\Are)}{\d A}.
\end{eqnarray}
\end{subequations}
Also, recall that the transfer function at the end of the epoch 
of reheating~$\chi_k^{_\mathrm{RH}}(\Are)$ is given by 
Eq.~\eqref{transfer solution reheating case1}, with the coefficients 
$C_k$ and $D_k$ being described by Eqs.~\eqref{eq:Ck-Dk}.
During radiation domination, we have $H^2\,A^4=\Hre^2\,\Are^4$. 
Since $\kf=\af\,\HI$ and $\kre=\are\,\Hre$, we find that we can 
write $\Are=(\kf/\kre)^{1/\gamma}$.
As a result, the Bessel functions in the 
expression~\eqref{transfer solution reheating case1}
for~$\chi_k^{_\mathrm{RH}}(\Are)$ depend on the 
ratio $(k/\kre)$.
Note that, in contrast, the coefficients $C_k$ and $D_k$ depend
only on the ratio~$k/\kf$.
As we mentioned earlier, we have been interested in arriving 
at the spectrum over wave numbers such that $k< 10^{-2}\,\kf$.
For small $z$, the Bessel function $J_\alpha(z)$ behaves 
as (see, for instance, Ref.~\cite{gradshteyn2007})
\begin{equation}
\lim_{z \ll 1} J_\alpha(z) 
\simeq \f{1}{\Gamma(1+\alpha)}\,\l(\f{z}{2}\r)^\alpha,
\label{eq:J-sz}
\end{equation}
where $\Gamma(z)$ denotes the Gamma function.
Clearly, in such a limit, the Bessel functions involving the largest 
negative value for the index~$\alpha$ can be expected to dominate.
Since $0 \leq w_\phi \leq 1$, the quantities $\nu$ and 
$\gamma$ are always positive [cf. Eq.~\eqref{eq:n-g}]. 
Hence, in the limit $k \ll \kf$, we find that it is the term involving 
$D_k$ in Eq.~\eqref{transfer solution reheating case1} that will dominate.
For the above reasons, the quantity~$\chi_k^{_\mathrm{RH}}(\Are)$ 
can be approximated as follows:
\begin{equation}
\chi_k^{_\mathrm{RH}}(\Are)
\simeq \Are^{-\nu}\, D_k\, J_{\nu/\gamma}\l(\f{k}{\gamma\,\kre}\r),
\label{eq:chi-RH-Are-af}
\end{equation}
with the coefficient $D_k$ being given by
\begin{equation}
D_k \simeq -\f{\pi}{\Gamma(-\nu/\gamma)}\,
\csc\l(\frac{\pi\,\nu}{\gamma}\r)\,
\l(\f{k}{2\,\gamma\, \kf}\r)^{-\nu/\gamma}.\label{eq:Dk-sk}
\end{equation}

Let us first arrive at the shape of the spectrum in the domain $k \ll \kre$.
In such a domain, we can use the form~\eqref{eq:J-sz} for the 
Bessel function $J_{\nu/\gamma}[k/(\gamma\,\kre)]$ that
appears in the expression~\eqref{eq:chi-RH-Are-af} above
for $\chi_k^{_\mathrm{RH}}(\Are)$.
On doing so and utilizing the identity $\Gamma(z)\,\Gamma(1-z)
=\pi/\mathrm{sin}\,(\pi\,z)$~\cite{gradshteyn2007}, we find 
that, in the domain $k \ll \kre$, the quantity 
$\chi_k^{_\mathrm{RH}}(\Are)$ reduces to unity.
Under the same conditions, we find that the quantity 
$\d \chi_k^{_\mathrm{RH}}(\Are)/\d A$ vanishes.
Therefore, it should be evident from the expressions~\eqref{eq:cal-Ek-Fk}
that, in the limit $k \ll \kre$, $\mathcal{E}_k\simeq \mathcal {F}_k
\simeq 1$.
In other words, the spectrum of GWs today is scale invariant over this domain 
and its present day amplitude is given by
\begin{equation}
\ogw(k)\,h^2
\simeq  \l(\f{g_{r,0}}{g_{r,\mathrm{eq}}}\r)^{1/3}\,
\Omega_{_{\mathrm{R}}}\, h^2\; \f{\pt(k)}{24}
\simeq \Omega_{_{\mathrm{R}}}\, h^2\; 
\f{\HI^2}{12\,\pi^2\Mpl^2}.\label{eq:ogw-sk}
\end{equation}
In arriving at the final expression, we have assumed that 
$g_{r,0}\simeq g_{r,\mathrm{eq}}$ and have made use of the tensor 
power spectrum~\eqref{eq:pt-i} arising in de Sitter inflation.
We should mention that this result is the same as in the case of 
instantaneous reheating.
This result should come as a surprise since these large scale modes 
are on super-Hubble scales during the epoch of reheating and hence
are not influenced by it.

Let us now turn to the domain $k\gg \kre$.
Since the limit $k\ll \kf$ continues to be valid, the term involving
$D_k$ in Eq.~\eqref{transfer solution reheating case1} remains the 
dominant term.
Therefore, $\chi_k^{_\mathrm{RH}}(\Are)$ is again described by
Eq.~\eqref{eq:chi-RH-Are-af}, with $D_k$ given by Eq.~\eqref{eq:Dk-sk}.
However, the argument of the Bessel function in Eq.~\eqref{eq:chi-RH-Are-af} 
is now large.
For large $z$, the Bessel function $J_\alpha(z)$ behaves as (see, 
for instance, Ref.~\cite{gradshteyn2007})
\begin{equation}
\lim_{z \gg 1} J_\alpha(z) 
\simeq \sqrt{\f{2}{\pi\,z}}\,
\mathrm{cos}\l[z-\pi\,\alpha-(\pi/4)\r].\label{eq:J-lz}
\end{equation}
Therefore, in the domain $k\gg \kre$, we find that the quantity
$\chi_k^{_\mathrm{RH}}(\Are)$ and its derivative 
$\d\chi_k^{_\mathrm{RH}}(\Are)/\d A$ behave as
\begin{subequations}
\begin{eqnarray}
\chi_k^{_\mathrm{RH}}(\Are)
&\simeq& -\f{1}{\sqrt{\pi}}\;\Gamma\l(1+\f{\nu}{\gamma}\r)\,
\l(\f{k}{2\,\gamma\,\kre}\r)^{-(\nu/\gamma)-(1/2)}\,
\mathrm{cos} \l(\frac{k}{2\,\gamma\,\kre}-\f{\pi\,\nu}{\gamma}-\f{\pi}{4}\r),\\
\Are\,\f{\d\chi_k^{_\mathrm{RH}}(\Are)}{\d A}
&\simeq& -\f{2\,\gamma}{\sqrt{\pi}}\;\Gamma\l(1+\f{\nu}{\gamma}\r)\,
\l(\f{k}{2\,\gamma\,\kre}\r)^{-(\nu/\gamma)+(1/2)}\,
\mathrm{sin} \l(\frac{k}{2\,\gamma\,\kre}-\f{\pi\,\nu}{\gamma}-\f{\pi}{4}\r),
\end{eqnarray}
\end{subequations}
with $\nu$ and $\gamma$ being given by Eq.~\eqref{eq:n-g}.
On substituting these expressions in Eqs.~\eqref{eq:cal-Ek-Fk}, we 
obtain the corresponding~$\mathcal{E}_k$ and~$\mathcal{F}_k$ to be
\begin{equation}
\mathcal{E}_k 
\simeq \mathcal{F}_k^\ast 
\simeq -\f{2\,i\,\gamma}{\sqrt{\pi}}\;\Gamma\l(1+\f{\nu}{\gamma}\r)\,
\l(\f{k}{2\,\gamma\,\kre}\r)^{-(\nu/\gamma)+(1/2)}\,
\mathrm{exp}\; i\,\l(\frac{k}{2\,\gamma\,\kre}-\f{\pi\,\nu}{\gamma}
-\f{\pi}{4}\r)
\end{equation}
so that 
\begin{equation}
\vert \mathcal{E}_k\vert^2 = \vert \mathcal{F}_k\vert^2 
= \f{4\,\gamma^2}{\pi}\, \Gamma^2\l(1+\f{\nu}{\gamma}\r)\,
\l(\f{k}{2\,\gamma\,\kre}\r)^{\ngw},
\end{equation}
where we have defined $\ngw$ to be
\begin{equation}
\ngw=1-\f{2\,\nu}{\gamma}=-\f{2\,(1-3\,w_\phi)}{1+3\,w_\phi}.
\end{equation}
If we substitute these results in the expression~\eqref{presentrelic},
we obtain the spectrum of GWs today in the domain $k\gg \kre$ to be
\begin{equation}
\ogw(k)\,h^2
\simeq  \l(\f{g_{r,0}}{g_{r,\mathrm{eq}}}\r)^{1/3}\,
\Omega_{_{\mathrm{R}}}\, h^2\; \f{\pt(k)}{24}\,\vert \mathcal{E}_k\vert^2
\simeq \Omega_{_{\mathrm{R}}}\, h^2\; 
\f{\HI^2}{12\,\pi^2\Mpl^2}\,
\f{4\,\gamma^2}{\pi}\, \Gamma^2\l(1+\f{\nu}{\gamma}\r)\,
\l(\f{k}{2\,\gamma\,\kre}\r)^{\ngw}.\label{GW spectrum- second domain}
\end{equation} 
In other words, for wave numbers such that $k\gg \kre$, the spectrum of 
GWs today has the index~$\ngw$.
Notably, the index vanishes when $w_\phi=1/3$.
Also, while the spectrum is blue for $w_\phi>1/3$, it is red for~$w_\phi< 1/3$.
Moreover, in the extreme cases wherein $w_\phi$ vanishes or is unity,
we have $\ngw=-2$ and $\ngw=1$, respectively.

\begin{figure}[t!]
\centering
\includegraphics[width=12.50cm]{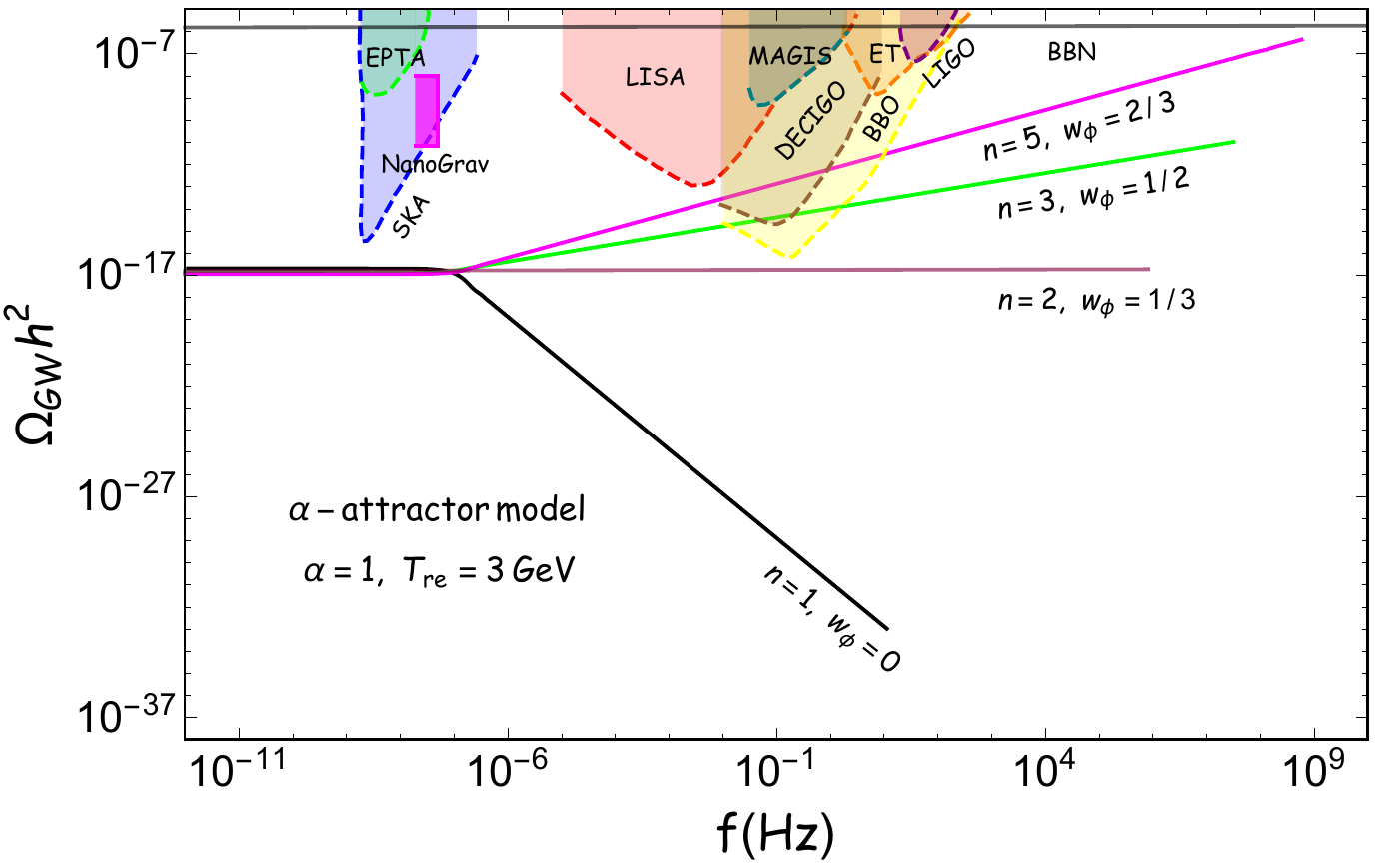}
\caption{The behavior of the dimensionless energy density of primordial GWs 
observed today, viz. $\ogw(f)$, has been plotted over a wide range of frequencies.
The spectrum has been obtained analytically and it corresponds to the case wherein 
the post-inflationary phase is described by the EoS parameter~$w_\phi$
and reheating is expected to occur instantaneously at a given time.
We have considered the scenario wherein the inflationary potential is described
by the $\alpha$-attractor model~\eqref{eq:aap}.
We have illustrated the spectra for the cases wherein $n=(1,2,3,5)$ (in black, 
brown, green and magenta), which correspond to $w_\phi=(0,1/3,1/2,2/3)$.
We should mention that we have chosen the parameters such that $\Tre = 3\, 
\mathrm{GeV}$ in all the cases.
In the figure, we have also included the sensitivity curves of the different 
ongoing and forthcoming GW observatories (in varied colors, on top).
Note that, as expected, the spectrum is strictly scale invariant for frequencies
such that $f < f_\mathrm{re}=\kre/(2\,\pi)$.
However, for larger frequencies such that $f>f_\mathrm{re}$, while the spectrum 
has a red tilt for $w_\phi < 1/3$, it has a blue tilt for $w_\phi> 1/3$.
Interestingly, we find that, for a suitably large value of $w_\phi$, 
the spectrum of GWs already intersect the sensitivity curves of some of 
the observatories over a certain range of frequencies.
Moreover, we find that, for a high value of $w_\phi$, the spectra cross the 
BBN bound of $\ogw\,h^2<10^{-6}$ at suitably large 
frequencies.}\label{plot_analytic1}
\end{figure}

On utilizing the expression~\eqref{transfer solution reheating case1} for 
the transfer function during reheating and the expressions~\eqref{eq:Ek-Fk} 
to determine the quantities~$\mathcal{E}_k$ and~$\mathcal{F}_k$, we can 
arrive at the complete spectrum of GWs by substituting the expressions in
Eq.~\eqref{PGW_spectrum_main2}. 
In Fig.~\ref{plot_analytic1}, we have plotted the spectrum of GWs today that 
arise in the case of the $\alpha$-attractor model~\eqref{eq:aap} for a set of 
values of the EoS parameter $w_\phi$.
In plotting the spectra, we have chosen the other parameters in such a fashion 
that the reheating temperature is $\Tre = 3\, \mathrm{GeV}$ in all the cases.
The figure clearly illustrates the qualitative features we discussed above:
(i)~the spectrum is strictly scale invariant for $k<\kre$, and (ii)~the spectrum
has the index $\ngw$ for $k> \kre$.
We have plotted the spectra for $w_{\phi} = (0,1/3,1/2,2/3)$, which correspond 
to the values $n=(1,2,3,5)$ for the index in the potential~\eqref{eq:aap}.
We should mention that these cases lead to the indices $\ngw = (-2,0,2/5,2/3)$, 
as expected.
In the figure, we have also included the sensitivity curves of the some of the 
current and forthcoming GW observatories (for a discussion on the sensitivity 
curves, see Ref.~\cite{Moore:2014lga} and the associated web-page).  
Interestingly, we find that, for a set of inflationary and reheating parameters,
the spectra already intersect the sensitivity curves. 
Moreover, we find that the BBN bound, viz. $\ogw(\kf)\,h^2 
\leq 10^{-6}$ (in this context, see, for instance, Ref.~\cite{Pagano:2015hma}
and the reviews~\cite{Caprini:2018mtu,Guzzetti:2016mkm}),
can be violated for $w_\phi>1/3$, which leads to constraints
on the EoS parameter~$w_\phi$ for a given~$\kf$ and vice-versa. 
As we have emphasized, Fig.~\ref{plot_analytic1} depicts the interesting 
dependence of the value of~$\kf$ on the inflationary model parameter~$n$ 
due to different initial conditions at the beginning of reheating. 
This interdependence of $\kf$ and the EoS parameter~$w_\phi$ can be translated 
into the constraints on the reheating temperature~$\Tre$ and scalar spectral 
index~$\ns$ through the aforementioned BBN bound.
For instance, in the figure, the spectrum corresponding to $w_\phi=2/3$
clearly crosses the BBN bound at large frequencies.
These clearly suggest that observations of the spectrum of GWs today can lead 
to interesting constraints on the primordial physics. 


\section{Spectrum of GWs in the case of perturbative reheating}\label{sec:actual}

In the perturbative reheating scenario, the inflaton continuously transfers 
its energy to radiation after the end of the inflationary epoch.
As a result, the effective EoS parameter during the reheating era, say, 
$w_\mathrm{eff}$, becomes time dependent.
It can be expressed as
\begin{eqnarray}
w_\mathrm{eff} = \f{3\,w_{\phi}\,\rho_{\phi} 
+ \rhoR}{3\,\l(\rho_{\phi} + \rhoR\r)},\label{effective EoS}
\end{eqnarray}
where, recall that, the evolution of the energy densities of the inflaton and 
radiation, viz. $\rho_{\phi}$ and $\rhoR$, are governed by the Boltzmann 
equations~\eqref{eq:be}, while $w_{\phi}$ is the EoS parameter describing 
the inflaton.
Such a time dependence of the effective EoS parameter has been explicitly 
demonstrated earlier (see, for example, Refs.~\cite{Maity:2018dgy,Haque:2020zco}). 
It has been illustrated that, while immediately after the termination of 
inflation, $w_\mathrm{eff}$ is approximately equal to $w_{\phi}$, after 
a certain time, the effective EoS parameter smoothly transits from 
$w_{\phi}$ to $1/3$, which indicates the onset of the epoch of radiation 
domination.
We should emphasize again here that such a reheating scenario is different 
from the case considered in the previous section where the inflaton energy 
density is assumed to be converted instantaneously into radiation after a 
certain period of time. 
Specifically, in the previous reheating scenario, $w_\mathrm{eff}$ 
remains equal to $w_{\phi}$ during the whole of reheating era and, 
at a particular time, $w_\mathrm{eff}$ sharply changes to $1/3$.
These differences in the dynamics of the reheating scenarios should be
reflected in spectrum of GWs today.
The corresponding features in the GW spectrum can, in principle, help us 
probe the microscopic mechanisms operating during the era of reheating.
 
\begin{figure}[!t]
\includegraphics[height=5.25cm,width=8.50cm]{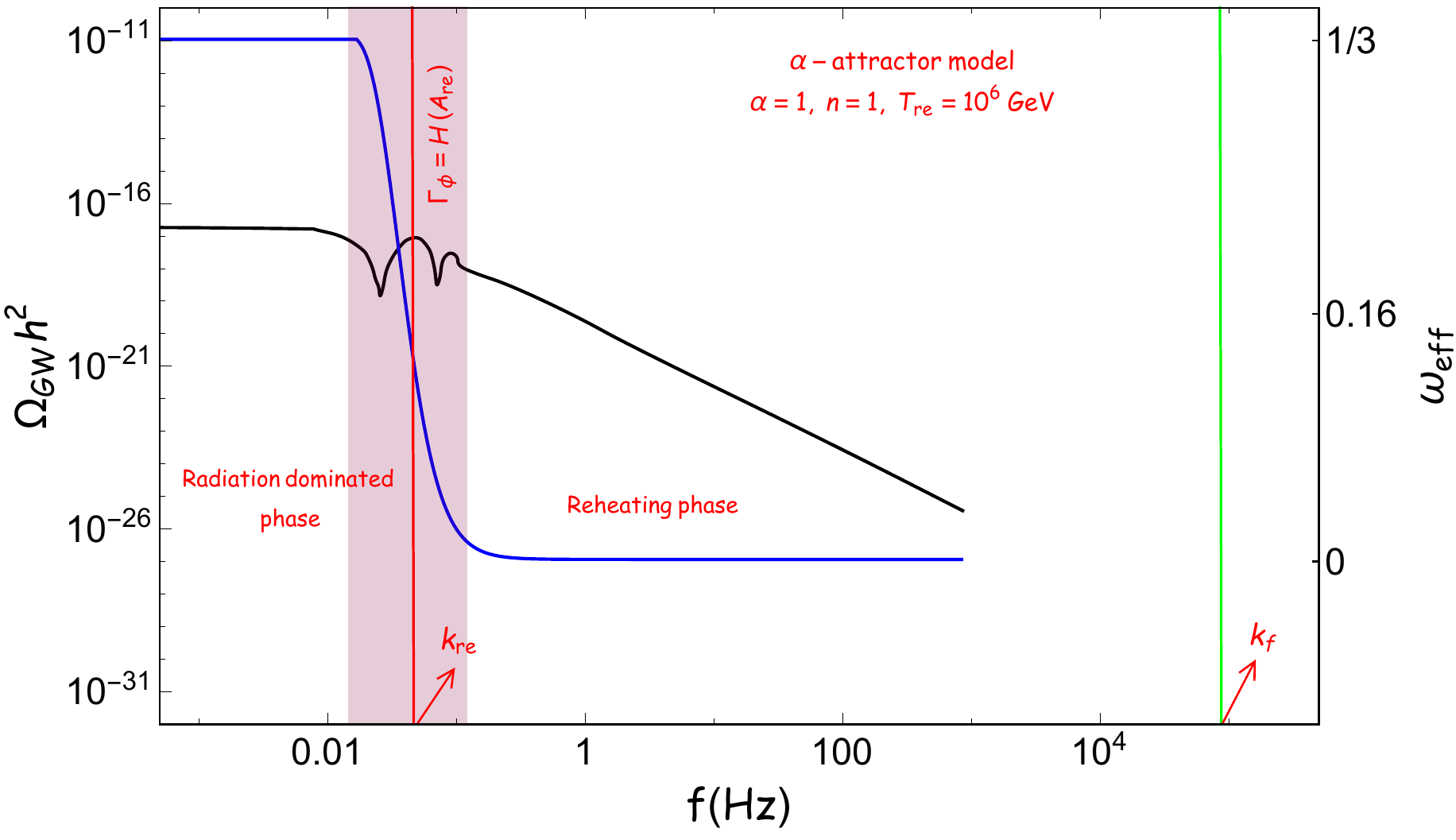}
\hskip 5pt
\includegraphics[height=5.25cm,width=8.40cm]{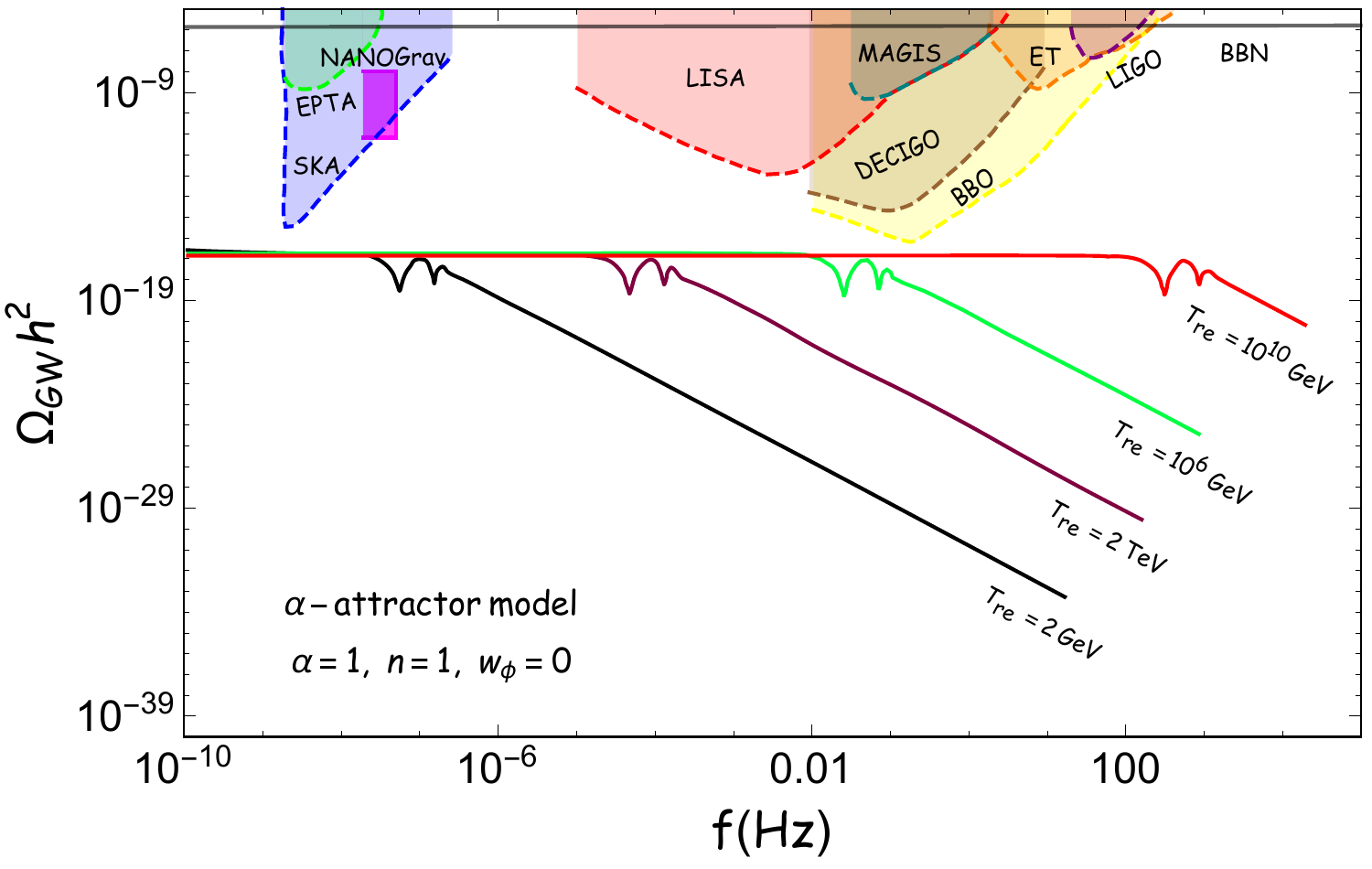}
\caption{The behavior of the dimensionless energy density of primordial GWs
today, viz. $\ogw(f)$, has been plotted over a small (in black, on the left) 
as well as a wide range of frequencies (in red, green, brown and black, on 
the right).
We have considered the scenario wherein the inflationary potential is described
by the $\alpha$-attractor model~\eqref{eq:aap} with $n=1$, 
which corresponds to $w_\phi = 0$.
We have plotted the spectrum of GWs for the following values of the reheating
temperature~$\Tre$:~$10^{10}\, \mathrm{GeV}$ (in red, on the right), $10^{6}\, 
\mathrm{GeV}$ (in black on the left and green on the right), $2\, \mathrm{TeV}$ 
and $2\, \mathrm{GeV}$ (in brown and black on the right).
Note that, we have also illustrated the behavior of the effective EoS 
parameter~$w_\mathrm{eff}$ (in blue, in the figure on the left) 
as a function of the frequency~$f$, which has been determined using 
the relation $f = a\,H/(2\,\pi)$.
In other words, $w_\mathrm{eff}(f)$ (marked on the $y$-axis on the right 
hand side of the figure on the left) represents the effective EoS parameter at 
the instant when the mode with the frequency~$f$ reenters the Hubble radius. 
We have also indicated the frequencies associated with the wave numbers~$\kre$
and $\kf$ (as vertical red and green lines, on the left).
Moreover, we have demarcated the regime (in pink) wherein the transition from 
$w_\mathrm{eff}=0$ to $w_\mathrm{eff}=1/3$ occurs.
We should point out that the spectrum of GWs exhibit oscillations in the region
of the transition.
Further, we have included the sensitivity curves of different ongoing and 
forthcoming GW observatories (in the figure on the right).}\label{plot_perturbative1}
\end{figure}

\begin{figure}[t!]
\includegraphics[height=5.25cm,width=8.50cm]{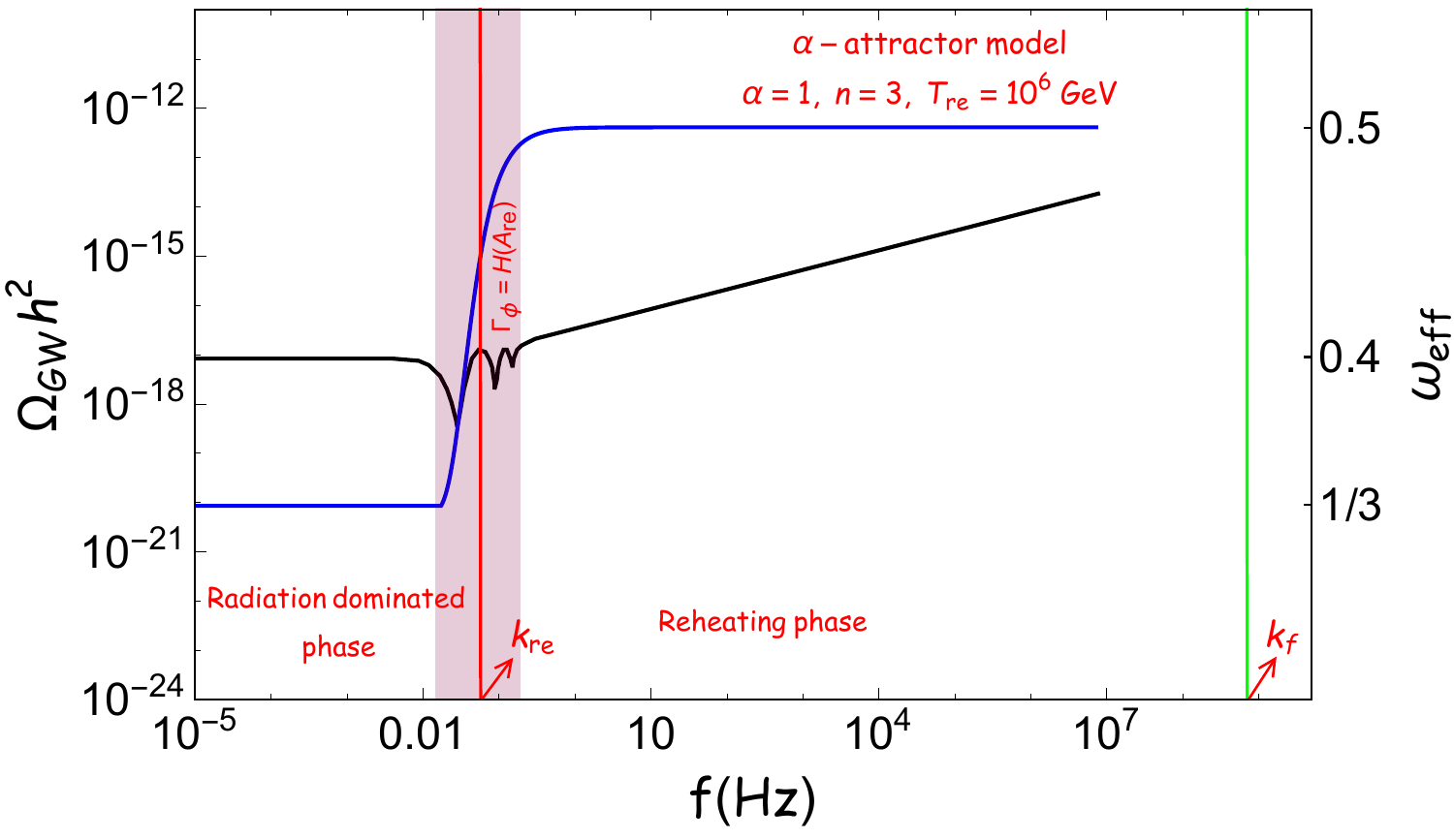}
\hskip 5pt
\includegraphics[height=5.25cm,width=8.50cm]{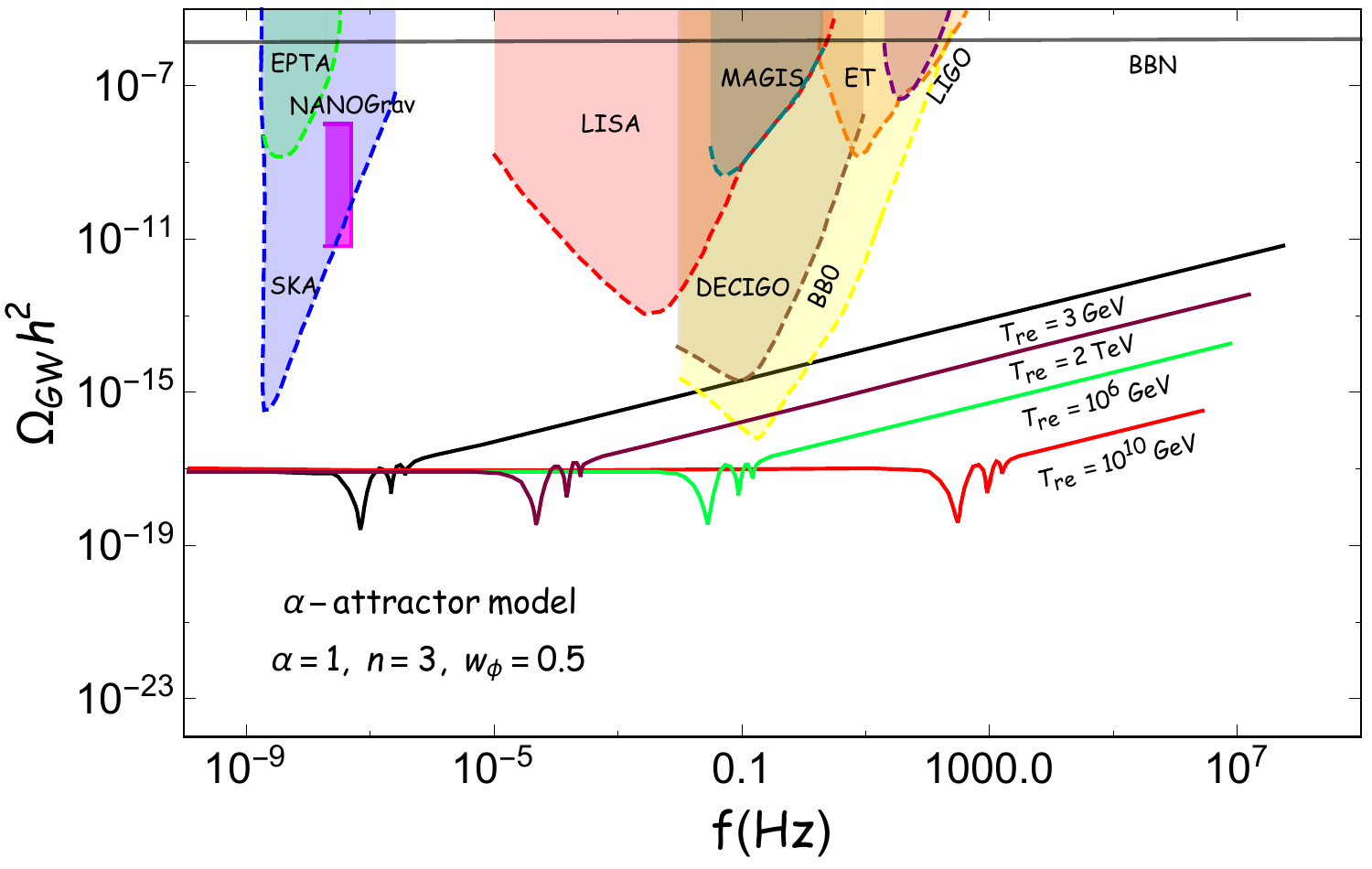}
\caption{The spectrum of GWs today have been illustrated in the same manner as
in the last figure.
But, in contrast to the previous figure wherein we had considered the case 
$w_\phi=0$ [or, equivalently, $n=1$ in the potential~\eqref{eq:aap}], we 
have set $w_\phi=0.5$ (i.e. $n=3$) in arriving at the plots above.
Note that, as in the case of $w_\phi=0$, the spectrum is scale invariant 
over frequencies corresponding to $k < \kre$.
However, the spectrum spectrum has a strong blue tilt at higher frequencies.
Importantly, for some values of the reheating temperature, the spectra intersect 
the sensitivity curves of the various GW observatories which immediately 
translate to constraints on the parameters $w_\phi$ and $\Tre$ that 
characterize the epoch of reheating.
In the figure, we have also included the BBN constraint (as the horizontal black 
line, on the right), which corresponds to~$\ogw\,h^2< 10^{-6}$.}
\label{plot_perturbative2}
\end{figure}

We shall now proceed to compute the spectrum of GWs at the present time, 
i.e. $\ogw(k)$ or, equivalently, $\ogw(f)$, in the case of the perturbative 
reheating scenario. 
As we had discussed, we shall analyze this case numerically.
With the solution to the Hubble parameter~$H(A)$ at hand, we proceed to 
solve for the transfer function~$\chi_k(A)$ during the epoch of reheating, 
as we had outlined in section~\ref{sec:egw-drh}. 
The numerical solutions are determined using the initial 
conditions~\eqref{boundary condition inflation1}
and~\eqref{boundary condition inflation2}. 
With the solutions at hand, we arrive at the spectrum of GWs at the 
current epoch for different sets of reheating temperature and the EoS 
parameter $w_\phi$ describing the inflaton. 
The results we have obtained are illustrated in Figs.~\ref{plot_perturbative1} 
and~\ref{plot_perturbative2} for the cases of $w_{\phi} < 1/3$ 
and~$w_{\phi} > 1/3$, respectively.
Let us now highlight a few points concerning the results plotted in the two
figures.

Let us first broadly understand the spectra in Fig.~\ref{plot_perturbative1} 
wherein we have plotted the results for~$w_{\phi} = 0$.
In the figure, we have illustrated the dimensionless energy density of GWs today
as a function of the frequency~$f$.
We have considered the case wherein the inflationary potential is described by 
the $\alpha$-attractor model~\eqref{eq:aap},
and have plotted the results for~$n=1$ (which corresponds to $w_\phi=0$) 
and a set of values of~$\Tre$.
We have also included behavior of the effective EoS parameter $w_\mathrm{eff}$,
which we have plotted as a function of frequency using the relation $f=a\,H/(2\pi)$.
The plot indicates the evolution of the parameter $w_\mathrm{eff}$ as the modes
with different frequencies~$f$ reenter the Hubble radius (in this context, also
see the earlier efforts~\cite{Maity:2018dgy,Haque:2020zco}).
Note that larger wave numbers or, equivalently, larger frequencies reenter the 
Hubble radius earlier than the smaller ones.
The plot clearly highlights the transition from the inflaton dominated universe to 
the epoch of radiation domination, achieved through the mechanism of perturbative
reheating.
The transition is clearly reflected in the behavior of the effective EoS parameter 
which changes smoothly from $w_\mathrm{eff}=0$ at early times (i.e. at large 
frequencies) to $w_\mathrm{eff}=1/3$ at late times (i.e. at small frequencies).
In the figure, we have indicated the frequencies associated with the wave numbers 
$\kf$ and $\kre$ and have also marked the domain of the transition to highlight 
these points.

Let us now point out a few more aspects of the results presented in 
Fig.~\ref{plot_perturbative1}.
In the figure, we have also plotted the spectrum of GWs for a few different values of
the reheating temperature.
Further, we have included the sensitivity curves of different current and forthcoming
GW observatories.
Note that the plots suggest that the spectra of GWs remain scale invariant over 
wave numbers~$k < \kre$ which reenter the Hubble radius during the radiation
dominated epoch.
This result should not come as surprise.
As we have pointed our earlier, these modes are on super-Hubble scales during the 
period of reheating and hence are unaffected by the process. 
Therefore, they carry the scale invariant nature of the spectrum of GWs generated
during inflation.
However, modes with wave numbers $\kre < k < \kf$ reenter the Hubble radius 
during the epoch of reheating and hence they carry the signatures of the 
mechanism of reheating.
For the value of $w_\phi=0$ we have worked with in Fig.~\ref{plot_perturbative1}, 
we find that the spectrum exhibits a strong red tilt for $k > \kre$.
In fact, we find that the red tilt occurs over this range of wave numbers whenever
$w_\phi <1/3$.
Moreover, for lower values of the reheating temperature, the red tilt begins to 
occur at smaller wave numbers.
Such a behavior can be attributed to the fact that, when the reheating temperature
is lower, the epoch of reheating lasts longer.
Since reheating is delayed, the mode with wave number $\kre$ reenters the Hubble 
radius at a later time or, equivalently, leaves the Hubble radius during inflation
at an earlier time thereby suggesting that it will have a smaller wave number.
We should mention here that these features in the spectrum of GWs are similar to
the behavior in the simpler reheating scenario we had discussed in the last section.

Interestingly, we find that the perturbative reheating scenario leaves tell tale
imprints on the spectrum of GWs which can possibly help us decipher finer details
of the mechanism of reheating.
We find that the spectrum exhibits a burst of oscillations near~$\kre$.
It should be clear from Fig.~\ref{plot_perturbative1} that the oscillations occur 
over modes which leave the Hubble radius during the period of the transition when
$w_\mathrm{eff}$ changes from its initial value of $w_\phi$ to the final 
value of~$1/3$.
Recall that, in the perturbative reheating scenario, the reheating temperature 
is identified as the temperature associated with the energy density of radiation
at the instance when $H(\Are) = \Gamma_{\phi}$. 
Consequently, it is at this point of time that the change in the effective EoS 
parameter with respect to the scale factor is the maximum.
This aspect is reflected in the peak that arises in the spectrum of GWs exactly 
at the wave number $\kre$ which re-enters the Hubble radius when $H(\Are) = 
\Gamma_{\phi}$. 
We should also mention that these features in the spectrum of GWs spectrum are 
not limited only to the case of $w_{\phi} = 0$ and $\Tre = 10^{6}\, \mathrm{GeV}$,
but also arise for all possible sets of values of~$(w_{\phi} < 1/3, \Tre)$. 
In order to highlight this point, in Fig.~\ref{plot_perturbative1}, we have 
illustrated the spectrum $\ogw(f)$ for different values of reheating temperature 
$\Tre$. 
However, note that, the frequency around which the spectrum begins to exhibit 
a red tilt increases as the reheating temperature increases. 
This is expected for the reason we discussed above, viz. that the period of 
reheating is shorter for a higher reheating temperature as a result of which
the wave number~$\kre$ of the mode which re-enters at the end of reheating 
turns out to be larger.
Lastly, we should mention that, for $w_{\phi} < 1/3$, the spectrum of GWs 
is indeed compatible with the BBN constraints.

To illustrate the dependence of the spectrum of GWs on $w_\phi$, in 
Fig.~\ref{plot_perturbative2}, we have plotted the spectrum for~$w_{\phi} 
= 1/2$ [i.e. for $n=3$ in the potential~\eqref{eq:aap}] 
and a set of values of the reheating temperature, just as in
Fig.~\ref{plot_perturbative1}.
Clearly, the spectrum of GWs is scale invariant over frequencies corresponding 
to $k< \kre$.
But, in contrast to the $w_{\phi} = 0$ case, the spectrum has a strong blue
tilt at higher frequencies.
Also, as expected, the higher the reheating temperature, the larger is $\kre$,
for reasons we have discussed earlier.
Moreover, we find that the spectrum exhibits a burst of oscillation around $\kre$,
exactly as in the $w_{\phi} = 0$ case.
Further, the maximum of the oscillation occurs at the instance when $\kre$ reenters 
the Hubble radius and the width of the oscillation coincides with the period of 
transition from  $w_\mathrm{eff} = w_\phi=0.5$ to $w_\mathrm{eff}=1/3$.
Finally, we should mention that the spectra exhibit a blue tilt for $k > \kre$
whenever $w_\phi>1/3$.

The above arguments clearly indicate that the details of epoch of reheating 
significantly affects the spectrum of GWs. 
Therefore, the characteristic features of $\ogw(f)$ can considerably aid us 
in garnering adequate amount of information regarding the reheating phase. 
Upon comparing Figs.~\ref{plot_analytic1} and~\ref{plot_perturbative1} 
(or~\ref{plot_perturbative2}), it is clear that the perturbative reheating 
mechanism, wherein the transfer of energy from the inflaton to radiation 
occurs smoothly, leads to oscillations in the spectrum of GWs in contrast
to the simpler model wherein the transition to radiation domination 
occurs instantaneously.
We believe that such quantitative differences can provide us with stronger
constraints on the mechanism of reheating.
Specifically,
\begin{itemize}
\item 
The presence of the oscillating feature in the spectrum of GWs, in particular,
the width of the oscillation can provide us information concerning the time 
scale over which $w_\mathrm{eff}$ makes the transition from $w_{\phi}$ 
to $1/3$.
\item 
As we have discussed above, the peak of the oscillation occurs at $k = \kre$.
Thus, identifying the location of the peak of the oscillation in the spectrum 
can help us determine the Hubble scale at the end of reheating or, equivalently, 
the decay rate $\Gamma_\phi$ of the inflaton to radiation.
In other words, the observation of the peak can indicate the strength of the
coupling between the inflaton and radiation in given a decay channel.
\end{itemize}


\section{Spectrum of GWs near the end of the  inflation}\label{sec:actualinflation}

Until now, while discussing the tensor power spectrum generated during inflation, 
for simplicity, we had assumed that inflation was of the de Sitter form.
This had led to a scale invariant power spectrum for scales such that $k \ll \kf$ 
[cf. Eq.~\eqref{eq:pt-i}], where $\kf$ denotes the wave number that leaves the 
Hubble radius at the end of inflation.
However, potentials such as the $\alpha$-attractor model~\eqref{eq:aap} of our 
interest actually lead to slow roll inflation and, as we had mentioned earlier, 
in such cases, there will arise a small tensor spectral tilt.
Moreover, even the slow roll approximation will cease to be valid towards the 
end of inflation.
Therefore, to understand the nature of the inflationary tensor power spectrum 
close to the wave number~$\kf$, the easiest method seems to evaluate the 
spectrum numerically.

\begin{figure}[t!]
\includegraphics[width=8.50cm]{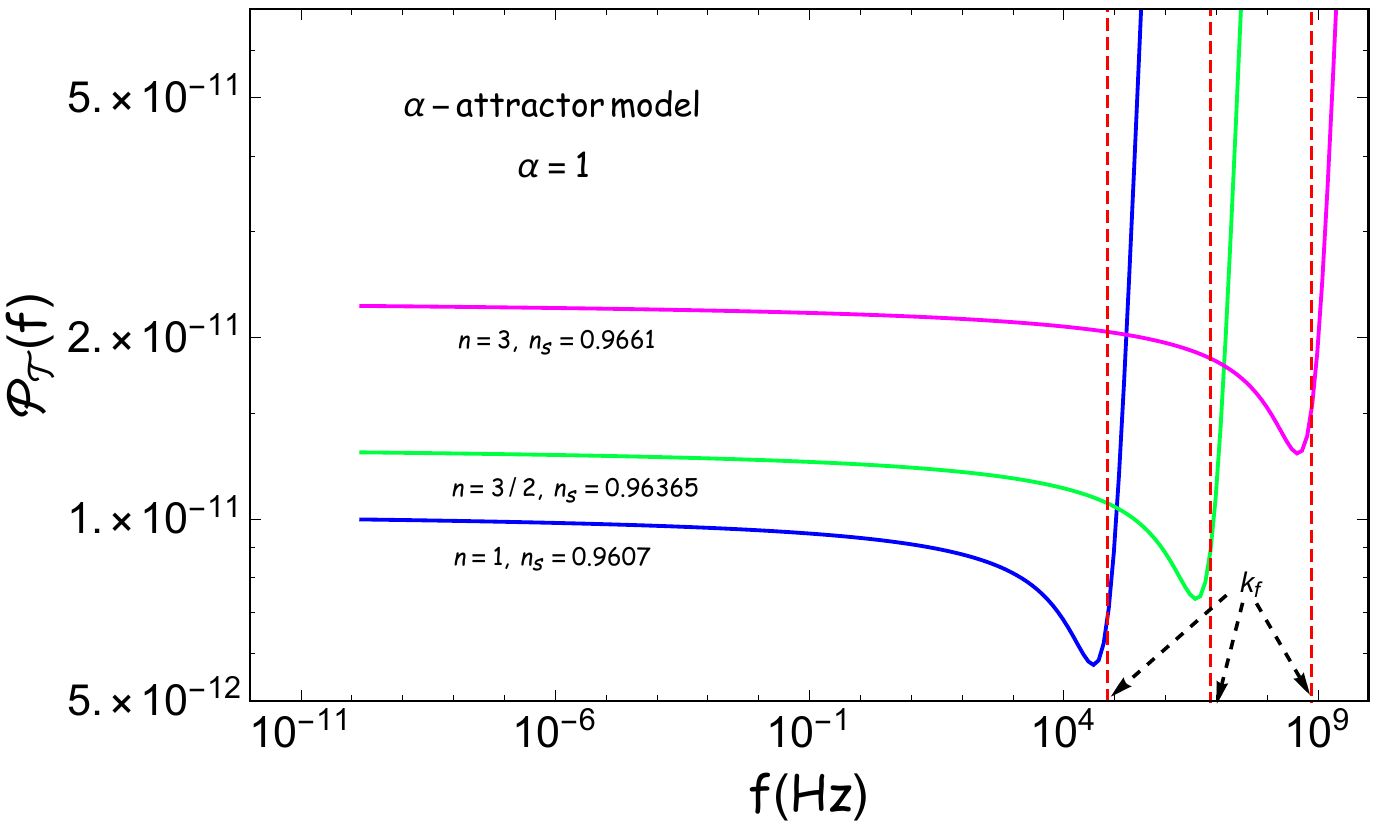}
\hskip 5pt
\includegraphics[width=8.10cm]{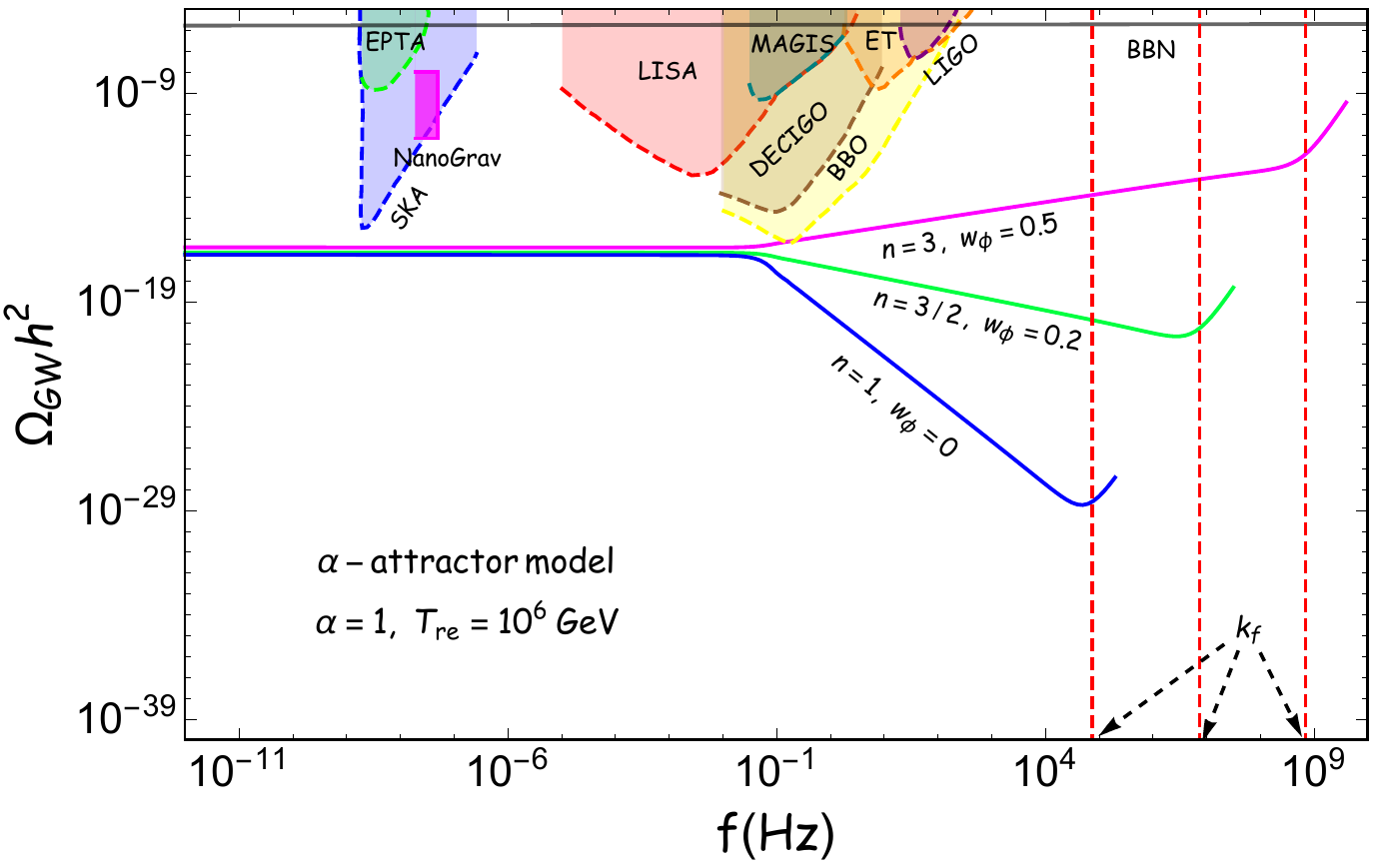}
\caption{The inflationary tensor power spectrum arising in the 
$\alpha$-attractor model of interest (on the left) as well as 
the corresponding spectrum of GWs today (on the right) have 
been plotted for frequencies close to the wave number~$\kf$.
The inflationary spectra have been computed numerically and
we have made use of the analytical solutions for the tensor
transfer function $\chi_k$ during the post-inflationary epochs
to arrive at the spectrum of GWs today.
We have plotted the spectrum of GWs today, viz. $\ogw(f)$, for a 
few different values of the reheating EoS parameter~$w_\phi$
and a specific reheating temperature.
We find that, while the inflationary power spectrum begins to 
behave as $k^2$ near~$\kf$, the corresponding spectra of GWs 
today behave as~$k^4$.}
\label{numericalsolution}
\end{figure}

There exists a standard procedure to evaluate the spectrum of perturbations generated
during inflation (in this context, see, for instance, Ref.~\cite{Hazra:2012yn}).
The modes are typically evolved from the Bunch-Davies initial conditions when they 
are well inside the Hubble radius and the spectra are evaluated in the super-Hubble 
domain when the amplitude of the perturbations have frozen.
Such an approach works well for the large scale modes. 
But, since we are interested in the tensor power spectrum over small scales, in 
particular, with wave numbers close to $\kf$, these modes would not be able to 
spend adequate amount of time in the super-Hubble regime.
Hence, in these situations, the best approach would be evaluate the spectrum at 
the end of inflation.
In Fig.~\ref{numericalsolution}, we have plotted the inflationary tensor power 
spectrum computed numerically in the $\alpha$-attractor model of our interest.
Actually, in the figure, we have also plotted the spectra of GWs today~$\ogw(f)$ 
for a few sets of values of the EoS parameter $w_\phi$ and a specific value
of the reheating temperature.
Having computed the inflationary spectra numerically, we have used the analytical
forms for the tensor transfer function post-inflation to arrive at the~$\ogw(f)$.
Note that the inflationary spectral shape begins to change for wave numbers close 
to $\kf$.
In fact, we find that, the inflationary tensor power spectrum $\pt(k)$ behaves 
as~$k^2$ for wave numbers close to and beyond~$\kf$.
This is not surprising and occurs due to the fact that these modes have either 
hardly left or remain inside the Hubble radius at the end of inflation.
Therefore, the modes are essentially of the Minkowskian form leading to 
the~$k^2$ behavior of the power spectrum.
From the structure of energy density of GWs [cf. Eq.~\eqref{eq:rho-gw-at}], 
it is easy to establish that the corresponding $\ogw(k)$ would behave as~$k^4$
over this domain of wave numbers.
It is easy to see from Fig.~\ref{numericalsolution} that $\ogw(f)$ indeed 
behaves as expected around and beyond $\kf$.
Further, from the figure, we can see that $\kf$ is crucially dependent on the 
structure of the inflationary potential. 
For reheating dynamics described by an effective EoS parameter~$w_\phi$, we can 
write $\kf$ in terms of the potential parameter~$n$, the reheating 
parameters~$(\Tre\,\Nre$) and the inflationary parameter~$N_\ast$ as follows:
\begin{equation}
\kf=\af\, H_\mathrm{f}
=k_\ast\,\f{H_\mathrm{f}}{\HI}\,\mathrm{e}^{N_\ast}
=k_\ast\,\l(\f{V_\mathrm{f}}{2\,\HI^2\,\Mpl^2}\r)^{1/2}\,\mathrm{e}^{N_\ast} 
=k_\ast\,\l(\f{\pi^2\,g_{r, \mathrm{re}}}{90}\r)^{1/2}\,
\f{\Tre^2}{\HI\,\Mpl}\,\mathrm{e}^{N_\ast+[3\,n/(n+1)]\,\Nre},
\end{equation}
where we have been careful to distinguish between the value of $H_{k_\ast}\simeq 
\HI$ and $H_\mathrm{f}$, i.e. the Hubble parameters evaluated at the moment 
when the pivot scale $k_\ast$ crosses the Hubble radius and at the end of 
the inflation, respectively.


\section{CMB, spectrum of GWs, and the microscopic reheating 
parameters}\label{sec:probe}

As is well known, the observations of the anisotropies in the CMB by 
missions such as Planck can be explained in a simple and successful 
manner by invoking an early phase of 
inflation~\cite{Mukhanov:1990me,Martin:2003bt,Martin:2004um,
Bassett:2005xm,Sriramkumar:2009kg,Baumann:2008bn,Baumann:2009ds,
Sriramkumar:2012mik,Linde:2014nna,Martin:2015dha}). 
Nevertheless, the characterization of the inflaton is far from complete 
because of the lack of adequate observational constraints, particularly
over scales smaller than the CMB scales.
The spectrum of GWs is possibly the only probe which can provide us direct 
access to the physics operating during the epochs of inflation and reheating.
In this section, we shall discuss the manner in which we can extract the 
properties of the inflaton by the combining the observations of the CMB 
and GWs.

As we have already mentioned, the spectrum of GWs carries signatures which 
reflect some details of the mechanism of reheating. 
Recall that, the primary aspect of the reheating  phase is the time evolution 
of the EoS parameter from the initial value of $w_{\phi}$ associated with 
the inflaton to the final value of~$1/3$ corresponding to radiation. 
The phase can be generically divided into three stages based on the underlying 
physical processes that operate.
In what follows, we shall discuss these stages and the corresponding 
imprints on~$\ogw(f)$. 

\begin{figure}[t!]
\includegraphics[height=5.25cm,width=8.50cm]{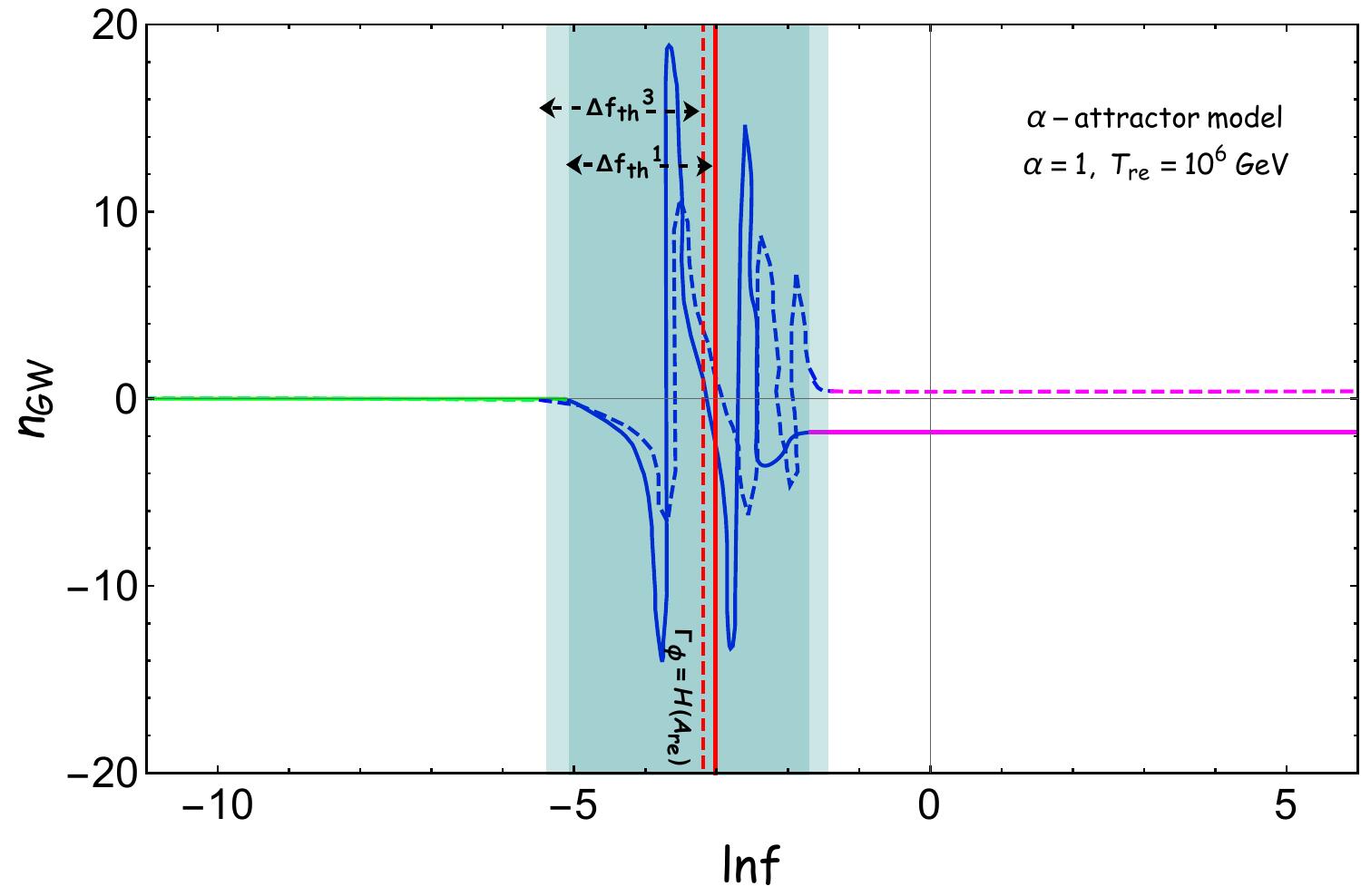}
\hskip 10pt
\includegraphics[height=5.25cm,width=8.50cm]{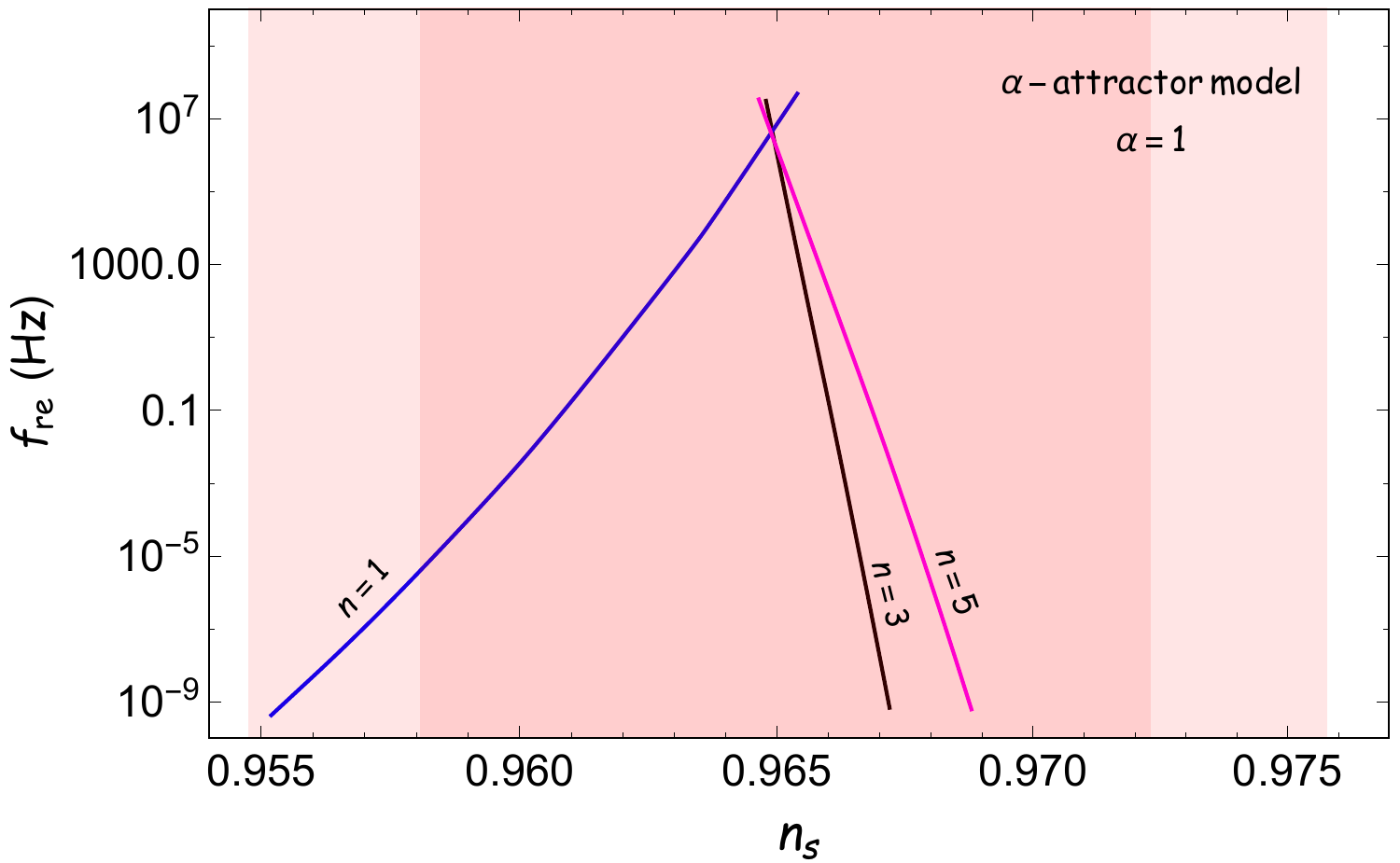}
\caption{\textit{Left:}\/~The variation of the index $\ngw$ associated the 
spectrum of primordial GWs observed today in the case of the perturbative
reheating scenario has been plotted as a function of frequency~$f$ for two 
different values of the inflaton EoS parameter $w_\phi=(0,1/2)$ (as 
solid and dashed lines) for the $\alpha$-attractor potential~\eqref{eq:aap}.
We have fixed value of the reheating temperature to be $\Tre=10^6\, 
\mathrm{GeV}$ in arriving at the plot.
In the figure (on the left), we have also explicitly highlighted the 
behavior of $\ngw$ for modes that re-enter the Hubble radius during 
the following regimes: (i)~reheating phase dominated by the inflaton
(as solid and dashed lines, in magenta), (ii)~the period of rapid 
transition of the EoS parameter from $w_\phi$ to $1/3$ (as solid 
and dashed lines in blue), and (iii)~the radiation dominated epoch 
(in green).
We have also demarcated the domain in frequency associated with the 
thermalization time scales in the figure (as shaded regions in dark 
and light blue).
These quantities have been denoted as $\Delta f_\mathrm{th}^1$ and 
$\Delta f_\mathrm{th}^3$ for $w_\phi=0$ and $1/2$, 
respectively.
\textit{Right:}\/~We have illustrated the variation of the frequency 
$f_\mathrm{re}=\kre/(2\,\pi)$ as a function of the scalar spectral 
index $\ns$ for three different values of $n=(1,3,5)$ (in blue, 
black and pink) for the $\alpha$-attractor model. 
We have also indicated the $1$-$\sigma$ and $2$-$\sigma$ confidence 
regions (as light and dark bands in red) associated with the constraint 
on scalar spectral index~$\ns$ from Planck~\cite{Akrami:2018odb}.}
\label{plotnomega}
\end{figure}

To facilitate the discussion, let us introduce the spectral index $\ngw 
=\d\,\mathrm{ln}\,\ogw/\d\, \mathrm{ln}\, k$ associated with the spectrum
of GWs.
Interestingly, we find that all the three stages leave distinct imprints on 
the spectral index~$\ngw$, and we have illustrated the behavior of $\ngw(f)$ 
for the $\alpha$-attractor model~\eqref{eq:aap} in Fig.~\ref{plotnomega}. 
Note that the first and longest stage is when $H \ll \Gamma_{\phi}$, i.e. when 
the inflaton is decaying very slowly and hence the background dynamics is 
dominated by the EoS parameter~$w_{\phi}$ governing the inflaton.
The spectral index $\ngw$ associated with modes which reenter the Hubble 
radius during the stage is given by $\ngw = 2\, (3\, w_{\phi} -1)/(3\,
w_{\phi}+ 1)$.
In Fig.~\ref{plotnomega}, we have indicated the $\ngw(f)$ associated with 
$w_\phi=0$ and $1/2$. 
In the subsequent stage, as the Hubble parameter approaches $\Gamma_{\phi}$,
the decay of the inflaton becomes increasingly efficient, and the effective
EoS parameter begins to change rapidly.
The corresponding effects are reflected in the variation of $\ngw$ over modes 
which reenter the Hubble radius during the transition, as highlighted in
Fig.~\ref{plotnomega}.
However, this intermediate stage is the shortest among the three stages and 
it ends when $\Hre = \Gamma_{\phi}$, i.e. when the rate of decay of the 
inflaton to radiation is at its maximum. 
The most important of the three stages is the final stage of thermalization 
which is characterized by the time scale~$\Delta t_\mathrm{th}$. 
It is the time scale over which the decay products of the inflaton thermalize 
among themselves.
In fact, it is this mechanism that determines the actual initial temperature 
of the radiation dominated phase contrary to the conventional definition of 
reheating temperature~$\Tre$ defined when $H = \Gamma_{\phi}$. 
The thermalization process and the associated time 
scale~$\Delta t_\mathrm{th}$ would crucially depend upon nature of all the 
decay products of the inflaton as well as the detailed dynamics of the decay 
process.
These details will determine the manner in which the EoS parameter changes during 
this stage and its variation will be imprinted in the behavior of $\ngw(f)$ as we
have illustrated in Fig.~\ref{plotnomega}.
Note that there arises a frequency range, say, $\Delta f_\mathrm{th}$, associated
with the time scale $\Delta t_\mathrm{th}$, and the variation of the spectral 
index $\ngw$ over this domain can provide us with clues to the physics operating
during the stage.

Therefore, from the CMB observations and the spectrum of GWs, we can, in 
principle, extract the following essential information regarding the nature 
of the inflaton and the underlying physical process taking place during 
reheating.  

\vskip 4pt\noindent
(i) {\it Effective inflaton EoS parameter~$w_{\phi}$:}\/~The nature of inflaton 
potential near its minimum or, equivalently, the effective EoS parameter $w_\phi$ 
associated with the decay of the inflaton can be determined from the spectral index 
of GWs through the relation
\begin{equation}
w_{\phi} = \f{1}{3}\, \l(\f{2 + \ngw}{2 - \ngw}\r).
\end{equation}

\vskip 4pt\noindent
(ii)~{\it Inflaton decay width $\Gamma_{\phi}$:}~Once we have arrived at the EoS 
parameter describing the inflaton, the effective inflaton decay constant~$\Gamma_\phi$ 
can be determined from the CMB and the spectrum of GWs in the following fashion. 
As we have already mentioned, for modes with wave numbers 
$k < \kre$, the amplitude of the tensor perturbations will remain approximately 
constant from the time the modes leave the Hubble radius during inflation till 
they renter the Hubble radius during the epoch of radiation domination.
Due to this reason, over these range of modes, the spectrum of GWs at late times
retains the same shape as the spectrum of tensor perturbations generated during
inflation.
Therefore, using Eq.~\eqref{eq:ogw-sk}, the scale invariant amplitude of the 
spectrum of GWs can be utilized to estimate the approximate energy scale near 
the end of inflation.
In the limit of high reheating temperature, we have
\begin{equation} 
\ogw\, h^2 \simeq \Omega_{_{\mathrm{R}}}\,h^2\,\frac{\HI^2}{6\,\pi^2\,\Mpl^2} 
\label{infscale}
\end{equation}
and, as a result, 
\begin{equation}
\HI \simeq \l(\f{6\,\pi^2\,\Mpl^2\,\ogw\, h^2}{\Omega_{_{\mathrm{R}}}\,
h^2}\r)^{1/2}
\end{equation}
which, in turn, allows us to express the energy density at the end of inflation as 
follows:
\begin{equation}
\rho_\mathrm{f} 
\simeq \f{18\,\pi^2\, \Mpl^4\, \ogw\, h^2}{\Omega_{_{\mathrm{R}}}\, h^2}.
\end{equation}
From the observed spectrum, we can, in principle, determine wave numbers $\kre$
and $\kf$, which are the wave numbers that reenters at the end of the epoch of 
reheating and leaves the Hubble radius at the end of inflation, respectively. 
In Fig.~\ref{plotnomega}, we have illustrated the dependence of $\kre$ on 
the scalar spectral~$\ns$ for different values of the parameter~$n$ of the 
$\alpha$-attractor model. 
In a manner similar to the existence of a maximum possible reheating temperature, 
we notice that there arises a maximum possible value for the frequency associated 
with the wave number~$\kre$.
We find that, generically, $f_\mathrm{re}^\mathrm{max} \simeq 10^7\, \mathrm{Hz}$,  
irrespective of the values of the other parameters involved. 
It would be interesting to study further implications of this point. 
Nonetheless, the aforementioned wave numbers satisfy the following relations
\begin{equation}
\kf = a_\mathrm{f}\, H_\mathrm{f} \simeq a_\mathrm{f}\, \HI,\quad
\kre=a_\mathrm{re}\, \Hre = \Are\, a_\mathrm{f}\, \Gamma_\phi.
\end{equation}
Considering perturbative reheating, one can obtain an approximate analytical 
expression for the normalized scale factor~$\Are$ at the end of the reheating 
to be (in this context, see Ref.~\cite{Haque:2020zco})
\begin{equation}
\Are=\l(\f{4\, \rho_\mathrm{f}\,(1+w_\phi)^2}{\mathcal{G}^4\,\beta\,
(5-3\,w_\phi)^2}\r)^{-1/(1-3\,w_\phi)},\quad
\beta=\f{\pi^2\,g_{r, \mathrm{re}}}{30},\quad
\mathcal{G}=\l(\f{43}{11\,g_\mathrm{re}}\r)^{1/3}\,
\l(\f{a_0\,HI}{k_\ast}\r)\,T_0\, \mathrm{e}^{-N_\ast}.
\end{equation}
The primary assumption in arriving at the above expressions is that the energy 
scale does not change significantly throughout the entire period of inflation.
With all the above expressions at hand, we find that the inflaton decay 
constant $\Gamma_\phi$ can be written in terms of the observable 
quantities as 
\begin{equation}
\Gamma_\phi 
= \f{\kre\, \HI}{\Are\, \kf}
=\frac{\kre}{\kf}\, 
\l(\f{6\,\pi^2\,\Mpl^2\,\ogw\, h^2}{\Omega_{_{\mathrm{R}}}\, h^2}\r)^{1/2} 
\l(\f{72\, \pi^2\,\Mpl^4\, \ogw\, h^2\,(1+w_\phi)^2}{\Omega_{_{\mathrm{R}}}\,
h^2\,\mathcal{G}^4\,\beta\,(5-3\,w_\phi)^2}\r)^{1/(1-3\omega_{\phi})}. 
\label{gammaphi}
\end{equation}
One can then immediately obtain the following analytic expression for the reheating temperature:
\begin{equation}
\Tre = \f{\mathcal{G}}{\Are}
=  \l(\frac{43}{11\,g_\mathrm{re}}\r)^{1/3}\, 
\frac{\kf\,\HI}{k_\ast\,H_0}\,
\l(\f{72\,\pi^2\, \Mpl^4\, \ogw\, h^2\, (1+w_\phi)^2}{\Omega_{_{\mathrm{R}}}\,
h^2 \mathcal{G}^4\,\beta\,(5-3\,w_\phi)^2}\r)^{1/(1-3\,w_{\phi})}\, T_0.
\end{equation}
So far, we have expressed the inflation decay constant and the reheating 
temperature in terms of the observables associated with the CMB and the
spectrum of GWs today. 
To understand the exact nature of the coupling, the subsequent thermalization 
processes right after reheating (i.e. when $\Gamma_{\phi} = \Hre$) becomes
important. 
Therefore, let us now compute the thermalization time scale. 

\vskip 4pt\noindent
(iii)~{\it Thermalization time scale $\Delta t_\mathrm{th}$:}\/~Thermalization
is an important non-equilibrium phenomenon that is ubiquitous in nature. 
At the end of the phase of reheating, when the rate at which the inflaton 
decays into radiation has attained a maximum, the subsequent thermalization 
phase leads to the epoch of radiation domination. 
In this process, thermalization time scale $\Delta t_\mathrm{th}$ is an
important observable which crucially depends on the nature of the initial 
state as well as the interactions among the internal degrees of freedom. 
Also, it is the initial state which encodes the information about the
coupling between the inflaton and the other fields to which the energy is
being transferred. 
Hence, if we can arrive at $\Delta t_\mathrm{th}$ from the spectrum of
GWs, valuable information regarding the fundamental nature of the coupling
parameters between the inflation and other fields at very high energies can, 
in principle, be extracted.
The thermalization time scale is defined as
\begin{equation}
\label{ther}
\Delta t_\mathrm{th}=t_\mathrm{th}-t_\mathrm{re},
\end{equation}
where $t_\mathrm{re}$ is the time corresponding to the end of reheating 
and $t_\mathrm{th}$ denotes the time at the end of the thermalization 
process, which leads to the beginning of the actual radiation dominated 
epoch. 
In order to obtain an approximate analytic expression, during this regime, 
we shall assume that the scale factor behaves as $a\propto t^{2/3\,(1+w)}$. 
This can be justified since we can express the effective EoS parameter
during the thermalization phase using the following perturbative 
expansion:
\begin{equation}
w= \f{1}{t_\mathrm{th} - t_\mathrm{re}}\, 
\int_{t_\mathrm{re}}^{t_\mathrm{th}} \d t\, 
\l[w_{_\mathrm{R}} + \l(w_{\phi} - w_{_{\mathrm{R}}}\r)\, 
\f{\rho_{\phi}}{\rho_{_\mathrm{R}}} + \cdots\r] 
\simeq \f{1}{3} + \l(w_{\phi} -\frac{1}{3}\r)\,x 
+ {\cal O}(x^2),\quad x= \f{1}{t_\mathrm{th} - t_\mathrm{re}}\, 
\int_{t_\mathrm{re}}^{t_\mathrm{th}}\d t\,\frac{\rho_{\phi}}{\rho_{_\mathrm{R}}},
\end{equation}
where $w_{_\mathrm{R}} = 1/3$ is the EoS parameter describing radiation.
Note that, during the thermalization phase $\rho_{\phi} \ll \rho_{_\mathrm{R}}$ 
and, hence, $x \ll 1$.
This particular fact enables us to obtain the leading order expression for the
thermalization time scale in terms of the observable quantities that we discussed.
On utilizing the above form for the EoS parameter, we find that the leading order
behavior of the scale factor at $\Gamma_{\phi} = \Hre$ can be written as
\begin{equation}\label{are}
a_\mathrm{re}
\simeq \l(\f{t_\mathrm{re}}{t_1}\r)^{2/[3\,(1+\omega_{_\mathrm{R}})]}\, 
\l[1 - \frac{2\,x\,(w_{\phi} - w{_\mathrm{R}})}{3\, (1+w_{_\mathrm{R}})^2}\; 
\mathrm{ln}\, \left(\frac{t_\mathrm{re}}{t_1}\right) + \cdots \r],
\end{equation}
where $t_1$ is a constant we have introduced for purposes of normalization. 
Upon using the relations $\kre = a_\mathrm{re}\, \Hre$, $k_\mathrm{th}
= a_\mathrm{th}\, H_\mathrm{th}$ and $\Hre = \Gamma_{\phi}$, we can 
express the reheating time $t_\mathrm{re}$ and the thermalization time 
$t_\mathrm{th}$ in terms of the wave numbers $\kre$ and $k_\mathrm{th}$ as
\begin{equation}
t_\mathrm{re}
= \l(\f{k_\mathrm{th}}{p_{_\mathrm{R}}}\r)^{1/(p_{_\mathrm{R}}-1)}\, t_1^{p_{_\mathrm{R}}/(p_\mathrm{R}-1)} 
+ \mathcal{O}(x),\quad
t_\mathrm{th} = \l(\f{k_\mathrm{re}}{p_{_\mathrm{R}}}\r)^{1/(p_{_\mathrm{R}}-1)}  
t_1^{p_{_\mathrm{R}}/(p_{_\mathrm{R}}-1)} + \mathcal{O} (x).  
\end{equation}
We can also express $\Gamma_{\phi}$ in terms of normalized time $t_1$ as
\begin{equation}
\label{gamma}
\Gamma_\phi \sim \l(\f{p_{_\mathrm{R}}}{t_1}\r)^{p_{_\mathrm{R}}/(p_{_\mathrm{R}}-1)}\;
\kre^{-1/(p_{_\mathrm{R}}-1)}+ {\cal O}(x),
\end{equation}
where we have introduced the quantity $p_{_\mathrm{R}}
=2/[3\,(1+\omega_{_\mathrm{R}})]$. 
Utilizing all the above equations, we finally obtain the final expression 
for the thermalization time scale to the leading order in $x$ to be
\begin{equation} 
\renewcommand*{\arraystretch}{2.5}
\Delta t_\mathrm{th}
\sim \l\{\begin{array}{ll}
\l[\l(\f{k_\mathrm{th}}{p_{_\mathrm{R}}}\r)^{1/(p_{_\mathrm{R}}-1)}
- \l(\f{\kre}{p_{_\mathrm{R}}}\r)^{1/(p_{_\mathrm{R}}-1)}\r]\, p_{_\mathrm{R}}^{p_{_\mathrm{R}}/(p_{_\mathrm{R}}-1)}\;
\l(\f{\kre^{-1/(p_{_\mathrm{R}}-1)}}{\Gamma_\phi}\r)
+ \mathcal{O}(x), &\mbox{in terms of $\Gamma_{\phi}$},\\
\l[\l(\f{k_\mathrm{th}}{p_{_\mathrm{R}}}\r)^{1/(p_{_\mathrm{R}}-1)}
- \l(\f{\kre}{p_{_\mathrm{R}}}\r)^{1/(p_{_\mathrm{R}}-1)}\r]\,
p_{_\mathrm{R}}^{p_{_\mathrm{R}}/(p_{_\mathrm{R}}-1)}\;
\l(\frac{G\, \kre^{-p_{_\mathrm{R}}/(p_{_\mathrm{R}}-1)}}{\HI\, \Tre}\r)
+ {\cal O}(x), &\mbox{in terms of $\Tre$}.
\end{array}\r.\label{thermalization}
\end{equation}
In the above expressions, an important point one should remember is that the 
value of the wave numbers $\kre$ and  $k_\mathrm{th}$ are, in general, 
dependent on the decay width of the inflaton. 
Therefore, the overall thermalization time scale is a non-trivial function 
of $\Gamma_{\phi}$.

The thermalization time scale crucially depends on the initial number density
of the decayed particles compared with the thermalized ones. 
The initial number density at the instant when $\Gamma_{\phi} = \Hre$ can be 
approximately estimated to be $n_\mathrm{i} \simeq \Mpl^2\, \Hre^2/m_{\phi} 
= \Mpl^2\, \Gamma^2_{\phi}/m_{\phi}$, and, in arriving at this expression, it 
has been assumed that the momentum of the decay products is as large as the 
mass $m_\phi$ of the inflaton~\cite{Kurkela:2011ti,Harigaya:2013vwa}. 
If we consider the particles to have thermalized at the reheating temperature
$\Tre$, then the number density can again be approximately determined to be 
$n_\mathrm{th} \simeq \Tre^3 \simeq \Gamma_{\phi}^{3/2} \Mpl^{3/2}$. 
Hence, the ratio of the particle number densities $n_\mathrm{th}$ and 
$n_\mathrm{i}$ turns out to be
\begin{equation}
\f{n_\mathrm{th}}{n_\mathrm{i}} 
\simeq \f{m_{\phi}}{\sqrt{\Gamma_\phi\,\Mpl}} 
= \l(\f{m_{\phi}^2\, \Are\, \kf}{\Mpl\,\kre\,\HI}\r)^{1/2}.\label{ninth}
\end{equation}
This is one of the crucial parameters in the context of the thermalizing 
plasma which dictates the kind of physical processes that occur during 
thermalization~\cite{Ellis:1987rw,McDonald:1999hd,Allahverdi:2000ss,
Kurkela:2011ti,Harigaya:2013vwa}. 
At this stage, we are unable to extract the inflaton mass $m_{\phi}$ from
the observations in a model independent manner. 
However, given the mass of the inflaton, along with the CMB observations,
Eq.~\eqref{ninth} will have two generic possibilities, viz.
\begin{itemize}
\item 
${n_\mathrm{i} < n_\mathrm{th}}$:\/~The particle number density is smaller 
than the thermalized ones, which means that, at the end of reheating, the 
universe is under occupied. 
For example if one considers marginal inflaton-scalar ($\xi$) coupling such 
$\beta\, \phi\, \xi^3$, inflaton-Fermion ($\psi$) Yukawa coupling $\beta\, 
\phi\, \bar{\psi}\, \psi$, the inflaton decay width behaves as $\Gamma_{\phi} 
\sim \beta^2\, m_{\phi}$ which implies that 
$n_\mathrm{i}/n_\mathrm{th} \sim \sqrt{m_\phi/\Mpl} < 1$.
\item 
$n_\mathrm{i} > n_\mathrm{th}$:\/~The particle number density is larger than 
the thermalized ones, i.e. at the end of reheating, the universe is over 
occupied. 
For example, if one considers any Planck suppressed operator containing a 
coupling between the inflaton and the reheating field, the decay width behaves 
as $\Gamma_{\phi} \sim  m_{\phi}^3/\Mpl^2$, which implies that $n_\mathrm{i}/
n_\mathrm{th} \sim \sqrt{\Mpl}/m_{\phi} > 1$. 
\end{itemize}

\vskip 4pt\noindent 
(iv) {\it Determination of microscopic interactions:}\/~From our discussion 
above for the two cases, it is clear that if we can determine the value of
$n_\mathrm{th}/n_\mathrm{i}$ from the combined observations of the CMB and 
GWs, the fundamental nature of the inflaton-reheating field coupling such 
as `$\beta$' can be extracted. 
Furthermore, interestingly, it has been pointed out that, depending on the 
aforementioned two conditions for the initial, non-thermal states generated by 
the end of reheating, the behavior of the thermalization time scale in terms 
of the microscopic variables will be very different, and will behave as (in 
this context, see Ref.~\cite{Kurkela:2011ti})
\begin{equation}
\renewcommand*{\arraystretch}{1.5}
\Delta t_\mathrm{th} 
\sim \l\{\begin{array}{ll}
\alpha^{-2}\, m_{\phi}^{1/2}\, T_\mathrm{th}^{-3/2}, 
& \mbox{for under-occupied initial states such
that $n_\mathrm{i} < n_\mathrm{th}$},\\
\alpha^{-2}\, T_\mathrm{th}^{-1}, 
& \mbox{for the over-occupied initial state scuh that 
$n_\mathrm{i} > n_\mathrm{th}$},\\
\end{array}\r.\label{alpha}
\end{equation}
where~$\alpha$ denotes the gauge interaction strength among the decayed particles,
and $T_\mathrm{th}$ is the final thermalization temperature. 
Hence, it is extremely important to recognize that, once we know the inflaton 
mass $m_{\phi}$ and the final thermalization temperature $T_\mathrm{th}$ 
[the latter can be computed once the reheating dynamics is fixed, by using Eqs.~\eqref{ninth} and~\eqref{alpha}], the gauge interaction strength can, 
in principle, be computed in terms of the observable quantities through the
relations
\begin{equation}
\renewcommand*{\arraystretch}{2.5}
\alpha \sim 
\l\{\begin{array}{ll}
T_\mathrm{th}^{-1/2 }\, 
\l[\l(\f{k_\mathrm{th}}{p_{_\mathrm{R}}}\r)^{1/(p_{_\mathrm{R}}-1)}
- \l(\f{\kre}{p_{_\mathrm{R}}}\r)^{1/(p_{_\mathrm{R}}-1)}\r]^{-1/2}\;
p_{_\mathrm{R}}^{-p_{_\mathrm{R}}/[2\,(p_{_\mathrm{R}}-1)]}\;
\l(\f{\kre^{-1/(p_{_\mathrm{R}}-1)}}{\Gamma_\phi}\r)^{-1/2}, 
& \mbox{for $H_I > \f{m_{\phi}^2\, \Are\, \kf}{\Mpl\, \kre}$},\\
\l(\frac{m_{\phi}}{T_\mathrm{th}^3}\r)^{1/2}\; 
\l[\l(\f{k_{th}}{p_{_\mathrm{R}}}\r)^{1/(p_{_\mathrm{R}}-1)}
- \l(\f{k_{re}}{p_{_\mathrm{R}}}\r)^{1/(p_{_\mathrm{R}}-1)}\r]^{-1/2}\; p_{_\mathrm{R}}^{p_{_\mathrm{R}}/[-2\,(p_{_\mathrm{R}}-1)]}\;
\l(\f{\kre^{-1/(p_{_\mathrm{R}}-1)}}{\Gamma_\phi}\r)^{-1/2},
& \mbox{for $H_I < \frac{m_{\phi}^2\, \Are\, \kf}{\Mpl\, \kre}$}.
\end{array}\r. \label{alpha2}
\end{equation}
We expect to carry out a detailed study on these important issues in a 
future publication. 
Having examined the effects of the epoch of reheating on the spectrum of 
GWs today, let us now turn to discuss the effects that arise due to a 
secondary phase of reheating.
As we shall see, such a phase can have an important implication for 
the recent observational results reported by NANOGrav~\cite{Arzoumanian:2020vkk,Pol:2020igl}. 
 

\section{Spectrum of GWs with late time entropy production and implications 
for the recent NANOGrav observations}\label{sec:NANOGrav}

In Secs.~\ref{sec:averaged} and~\ref{sec:actual}, while arriving at the 
spectrum of GWs $\ogw(f)$ today, we had assumed that the perturbations were 
generated during inflation and had evolved through the epochs of reheating 
and radiation domination. 
In such a scenario, the entropy of the universe is conserved from the end 
of reheating until today.
In fact, we had earlier utilized the conservation of entropy to relate the 
temperature~$\Tre$ at the end of reheating to the temperature~$T_0$ today. 
Recall that, for simplicity, we had assumed that the spectrum of tensor
perturbations generated during inflation was strictly scale invariant 
[cf. Eq.~\eqref{eq:pt-i}].
We had also found that, for wave numbers $k < \kre$, the evolution of the 
tensor perturbations through the standard epochs of reheating and radiation 
domination does not alter the shape of the spectrum of GWs observed today, 
i.e. $\ogw(f)$ remains scale invariant for $f < \fre=\kre/(2\,\pi)$.
 
Over the last decade or so, there has been an interest in examining 
scenarios wherein there arises a short, secondary phase of reheating 
some time after the original phase of reheating which immediately 
follows the inflationary epoch (see, for example, 
Ref.~\cite{Kawasaki:2004rx}).
It has been shown that such a modified scenario can also be consistent 
with the the various observations~\cite{Nakayama:2009ce,Kuroyanagi:2013ns,
Hattori:2015xla}. 
A secondary phase of entropy production can occur due to the decay of an
additional scalar field (which we shall denote as~$\sigma$) that can be 
present, such as the non-canonical scalar fields often considered in 
high energy physics or the moduli fields encountered in string 
theory\footnote{It is for this reason that the secondary phase is 
sometimes referred to as the moduli dominated epoch.
The scalar field could have emerged from an extra-dimensional modulus field 
or due to some higher curvature effects \cite{Banerjee:2017lxi,Elizalde:2018rmz}}.
In this section, we shall discuss the effects of such a secondary phase of
reheating (which occurs apart from the primary phase of reheating considered 
earlier) on the spectrum of GWs observed today.
As we shall see, the secondary phase of entropy production leads to unique 
imprints on the spectrum of GWs which has interesting implications for the 
recent observations by NANOGrav~\cite{Arzoumanian:2020vkk,Pol:2020igl}.

Let us first calculate the reheating temperature associated with the secondary
phase of reheating. 
We can expect the entropy to be conserved during the radiation dominated epoch 
sandwiched between the two phases of reheating. 
On following the chronology of evolution mentioned above and, upon demanding 
the conservation of entropy, we can arrive at the relation between the 
temperature $\Tre$ at the end of the first phase of reheating and the temperature
at the beginning of the second phase of reheating, say, $T_{\sigma R}$.
We find that they can be related as follows:
\begin{equation}
g_{s,\mathrm{re}}\, \are^3\, \Tre^3 
= g_{s,\sigma R}\, a_{\sigma R}^3\, T_{\sigma R}^3,\label{entropy conservation1}
\end{equation}
where $(g_{s,\mathrm{re}},g_{s,\sigma R})$ and  $(\are,a_{\sigma R})$ denote 
the relativistic degrees of contributing to the entropy and the scale factor 
at the end of primary reheating phase and at the start of the second phase of 
reheating (or, equivalently, at the end of the first epoch of radiation 
domination), respectively. 
Using the above relation, we can express the original reheating 
temperature~$\Tre$ in terms of the temperature~$T_{\sigma R}$ 
at the beginning of the secondary phase of reheating as
\begin{equation}
\Tre= \l(\f{g_{s,\sigma R}}{g_{s,\mathrm{re}}}\r)^{1/3}\, 
\mathrm{e}^{N_{_\mathrm{RD}}^{(1)}}\;T_{\sigma R},\label{temperature1}
\end{equation}
where $N_{_\mathrm{RD}}^{(1)} = \mathrm{ln}\, (a_{\sigma R}/\are)$ denotes 
the duration of  first epoch of radiation domination in terms of the number 
of e-folds. 
Similarly, we can relate the temperature at the end of the secondary phase 
of reheating, say, $T_\sigma$, to the temperature $T_0$ today by demanding 
the conservation of entropy after the onset of the second epoch of radiation 
domination.
On doing so, we obtain that  
\begin{equation}
T_{\sigma} = \l(\f{43}{11\, g_{s,\sigma}}\r)^{1/3}\,
\l(\f{a_0}{a_{\sigma}}\r)\,T_0,\label{temperature2}
\end{equation}
where $g_{s,\sigma}$ and $a_{\sigma}$ represent the degrees of freedom 
contributing to the entropy and the scale factor at the end of the 
secondary phase of reheating. 
If $a_\mathrm{eq}$ denotes the scale factor at the epoch of matter-radiation
equality, then the above expression for $T_{\sigma}$ can be written as
\begin{equation}
T_{\sigma} = \l(\f{43}{11\, g_{s,\sigma}}\r)^{1/3}\,
\l(\f{a_0}{a_\mathrm{eq}}\r)\,
\mathrm{e}^{N_{_\mathrm{RD}}^{(2)}}\,T_0.\label{modified temperature2}
\end{equation}
The factor $a_0/a_\mathrm{eq}$ can be expressed in terms of the quantity
$a_0/a_k$ through the relation
\begin{equation}
\f{a_0}{a_\mathrm{eq}} 
=  \l(\frac{a_0}{a_k}\r)\,\mathrm{e}^{-\l[N_k+\Nre+N_{_\mathrm{RD}}^{(1)}\, 
+N_{\mathrm{sre}}+N_{_\mathrm{RD}}^{(2)}\r]},\label{connection}
\end{equation}
where $a_k$ denotes the scale factor when the mode with the wave 
number~$k$ crosses the Hubble radius during inflation, while~$N_k$ 
represents the number of e-folds from the time corresponding 
to $a_k$ to the end of inflation.
Moreover, recall that, $\Nre$ denotes the duration of the first phase 
of reheating.
It should be evident that the quantities $N_\mathrm{sre}$ and 
$N_{_\mathrm{RD}}^{(2)}$ represent the duration (in terms of 
e-folds) of the secondary phase of (say, moduli dominated) reheating 
and the second epoch of radiation domination, respectively. 
With $k$ set to be the pivot scale $k_\ast$, on substituting the above 
expression for $a_0/a_\mathrm{eq}$ in Eq.~\eqref{modified temperature2},
we obtain that
\begin{equation}
T_{\sigma} = \l(\f{43}{11\, g_{s,\sigma}}\r)^{1/3}\,
\l(\f{a_0\,\HI}{k_\ast}\r)\,
\mathrm{e}^{-\l[N_\ast+\Nre+N_{_\mathrm{RD}}^{(1)}
+N_{\mathrm{sre}}\r]}\;T_0,\label{modified temperature3}
\end{equation}
which is the temperature at the end of the secondary phase of reheating.

Note that the above expression for $T_\sigma$ can be inverted to 
write $N_{_\mathrm{RD}}^{(1)}$ as
\begin{equation}
\mathrm{e}^{N_{_{\mathrm{RD}}}^{(1)}} 
= \l(\frac{43}{11\, g_{s,\sigma}}\r)^{1/3}\,
\l(\f{a_0\,\HI}{k_ast}\r)\,
\mathrm{e}^{-\l[N_\ast+\Nre+N_{\mathrm{sre}}\r]}\,\l(\f{T_0}{T_{\sigma}}\r).
\end{equation}
This relation, along with Eq.~(\ref{temperature1}), immediately leads to 
the following expression for the original reheating temperature~$\Tre$ in 
terms of the parameters associated with the late time entropy production:
\begin{equation}
\Tre = \l(\f{43}{11\, g_{s,re}}\r)^{1/3}\,
\l(\f{a_0\,\HI}{k_\ast}\r)\, F^{-1/3}\,
\mathrm{e}^{-(N_\ast+\Nre)}\,T_0.\label{final reheating temperature late entropy}
\end{equation}
In this relation, the factor $F$ represents the ratio of the entropy at the 
end and at the beginning of the secondary phase of reheating, and it is given by
\begin{equation}
F = \f{s(T_{\sigma})\,a_{\sigma}^3}{s(T_{\sigma R})~a_{\sigma R}^3},\label{F1}
\end{equation}
where $s(T)$ denotes the entropy at the temperature~$T$. 
If we now assume that the secondary phase of reheating is described 
by the EoS parameter $w_\sigma$, then we can arrive at the 
following useful relations between the Hubble parameter and the 
temperature at the end and at the beginning of the secondary phase 
of reheating:
\begin{equation}
H_\sigma=\l(\f{\gamma_1\,T_\sigma}{\gamma_2\,F^{1/3}\,
T_{\sigma R}}\r)^{3\,(1+w_\sigma)/2}\,H_{\sigma R},\quad
T_\sigma=\l(\f{\gamma_ 1}{\gamma_2\,F^{1/3}}\r)^{3\,(1
+w_\sigma)/(1-3\,w_\sigma)}\, 
\l(\f{g_{r,\sigma R}}{g_{r,\sigma}}\r)^{1/(1-3\,w_\sigma)}\,
T_{\sigma R},
\end{equation}
where the quantities $\gamma_1$ and $\gamma_2$ are defined as
\begin{equation}
\gamma_1=\l(\f{g_{r,\mathrm{re}}}{g_{r,\sigma R}}\r)^{1/4},\quad
\gamma_2=\l(\f{g_{s,\mathrm{re}}}{g_{s,\sigma}}\r)^{1/3}.
\end{equation}
Clearly, the factor~$F$ controls the extent of entropy produced at late times. 
And, in the absence of such entropy production, the factor~$F$ reduces to unity.
Also, in such a case, the expression~\eqref{final reheating temperature late entropy} 
for the reheating temperature~$\Tre$ reduces to the earlier expression~\eqref{eq:Tre}, 
as required.

\begin{figure}[t!]
\centering
\includegraphics[width=12.50cm]{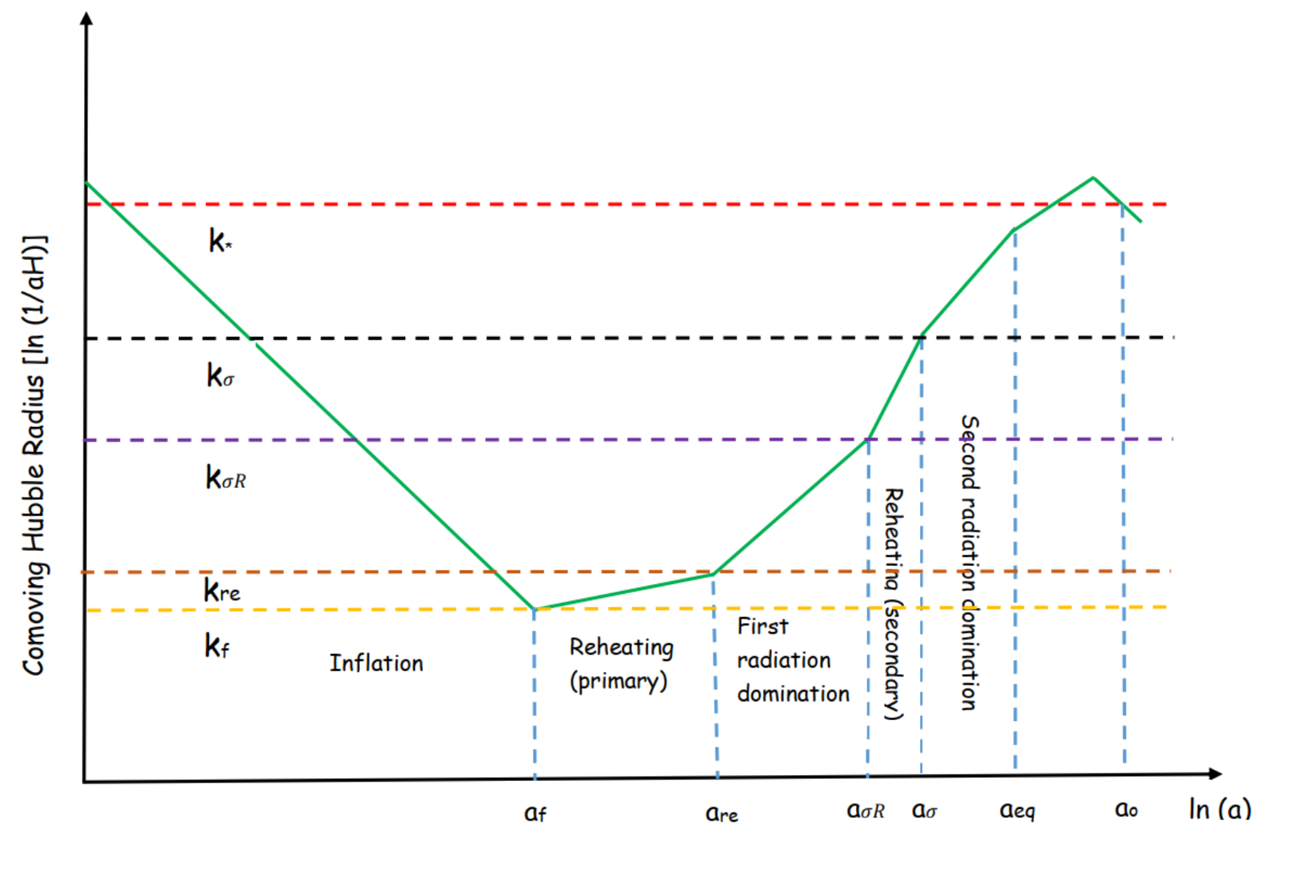}
\caption{A schematic diagram illustrating the evolution of the 
comoving Hubble radius $(a\,H)^{-1}$ plotted (in green) against 
the number of e-folds~$N=\mathrm{ln}\,a$. 
In the diagram, we have also delineated the various epochs that 
are relevant for our discussion.
In the above plot, we have assumed that the EoS 
parameter~$w_\phi$ describing the primary 
reheating phase is less than~$1/3$. 
However, we have assumed that the EoS parameter, say, 
$w_\sigma$, associated with a string modulus or 
a non-canonical scalar field driving the secondary 
phase of reheating is greater than~$1/3$.
In the figure, apart from the wave numbers $k_\ast$, $\kre$ 
and~$\kf$ we had encountered earlier (indicated as red, 
brown and yellow, dashed lines), we have indicated the 
scales $k_{\sigma R}$ and $k_{\sigma}$ (as dashed lines 
in purple and black) which correspond to wave numbers 
that reenter the Hubble radius at the beginning and at 
the end of the second phase of reheating, respectively.}
\label{diagram-scales}
\end{figure}

Let us now turn to discuss the spectrum of GWs that arises in such a 
modified scenario.
As we mentioned above, for simplicity, we shall assume that the secondary 
phase of reheating is described by the EoS parameter~$w_\sigma$. 
In order to arrive at $\ogw(k)$ in the new scenario, we shall follow the 
calculations described in Sec.~\ref{sec:averaged} wherein we have evaluated 
the spectrum analytically. 
To highlight all the relevant scales involved and also to aid our 
discussion below, in Fig.~\ref{diagram-scales}, we have illustrated 
the evolution of the comoving Hubble radius in the modified scenario.
Specifically, in the figure, we have indicated the new scales $k_{\sigma R}$ 
and $k_\sigma$ which are the wave numbers that reenter the Hubble radius at 
the start and at the end of the secondary phase of reheating. 

Before we go on to illustrate the results, let us try to understand the 
shape of $\ogw(k)$ that we can expect in the modified scenario involving 
late time production of entropy.
\begin{itemize}
\item 
{\it For wave numbers $k < k_{\sigma}$}:\/~As we mentioned above, $k_{\sigma}$ 
represents the wave number of the mode that reenters the Hubble radius at 
the onset of the second epoch of radiation domination or, equivalently, at
the end of the secondary phase of reheating. 
Therefore, the range of wave numbers $k < k_{\sigma}$ (but with wave numbers
larger than those corresponding to the CMB scales) reenter the the Hubble 
radius during the second epoch of radiation domination. 
Since they are on super-Hubble scales prior to their reentry, they are not 
influenced by the background dynamics during the earlier epochs. 
Hence, the spectrum of GWs for these range of modes can be expected to be 
scale invariant, which implies that the corresponding spectral index~$\ngw$ 
vanishes identically.
\item 
{\it For wave numbers $k_{\sigma} < k < k_{\sigma R}$}:\/~Recall that, 
$k_{\sigma R}$ denotes the wave number that reenters the Hubble radius 
at the beginning of the secondary phase of reheating. 
As in the case of modes that reenter the primary phase of reheating, we 
can expect the spectrum of GWs over this range of wave numbers to exhibit 
a spectral tilt which depends on the EoS parameter~$w_\sigma$.
We find that, over this range of wave numbers, $\ogw(k)$ behaves as
\begin{equation}
\ogw(k) \sim k^{2\,(3\,w_\sigma - 1)/(3\,w_\sigma + 1)}\label{2}.
\end{equation}
Consequently, the spectral index of the primordial GWs over this range of 
wave numbers turns out to be $\ngw = 2\,(3\,w_\sigma - 1)/(3\,w_\sigma + 1)$. 
In others words, over the domain $k_{\sigma} < k < k_{\sigma R}$, the spectrum 
has a blue tilt for $w_\sigma > 1/3$ and a red tilt for $w_\sigma < 1/3$.
\item 
{\it For wave numbers $k_{\sigma R} < k < \kre$}:\/~These range of wave numbers 
reenter the Hubble radius during the first epoch of radiation domination.
Hence, we can expect the spectrum of GWs to be scale invariant over this range. 
It is important to recognize that the amplitude of the spectrum $\ogw(k)$ over
this range will be greater or lesser than the amplitude over $k < k_\sigma$ (i.e.
over wave numbers which reenter the Hubble radius during the second epoch of
radiation domination) depending on whether the EoS parameter~$w_\sigma$ 
(characterizing the second phase of reheating) is greater than or less than~$1/3$.
\item 
{\it For wave numbers $\kre < k < \kf$}:\/~These correspond to wave numbers 
that reenter the Hubble radius during the first phase of reheating and, as 
we have discussed before, the spectrum of GWs over this range of wave numbers 
is expected to behave as
\begin{equation}
\ogw(k) \sim k^{2\,(3\,w_\phi - 1)/(3\,w_\phi + 1)}.\label{4}
\end{equation}
In other words, the corresponding spectral index is given by $\ngw = 2\,(3\,w_\phi 
- 1)/(3\,w_\phi + 1)$, which has a blue or red tilt depending on whether $w_\phi$
is greater than or less than $1/3$.
\end{itemize}

\begin{figure}[t!]
\includegraphics[height=5.25cm,width=8.50cm]{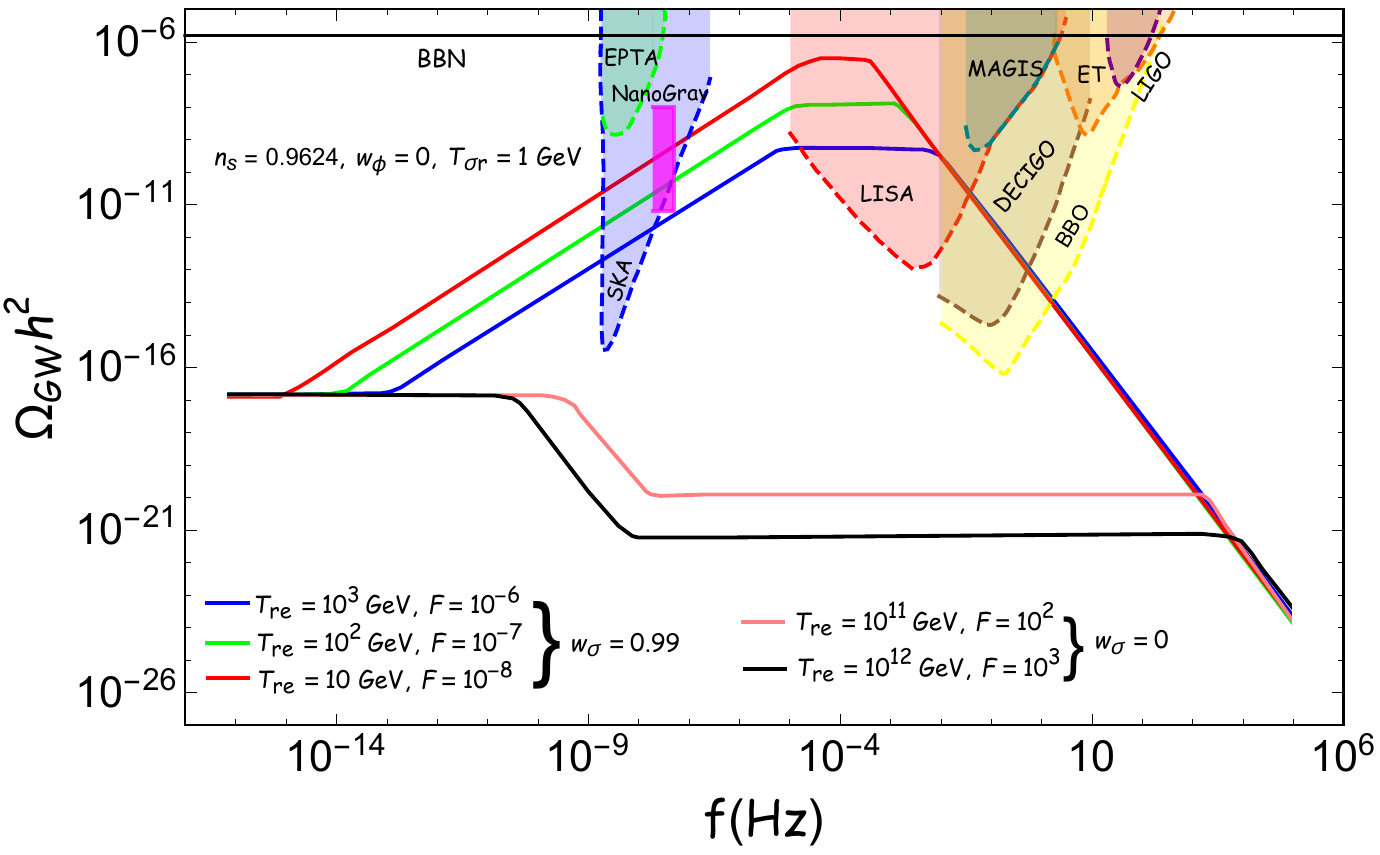}
\includegraphics[height=5.25cm,width=8.50cm]{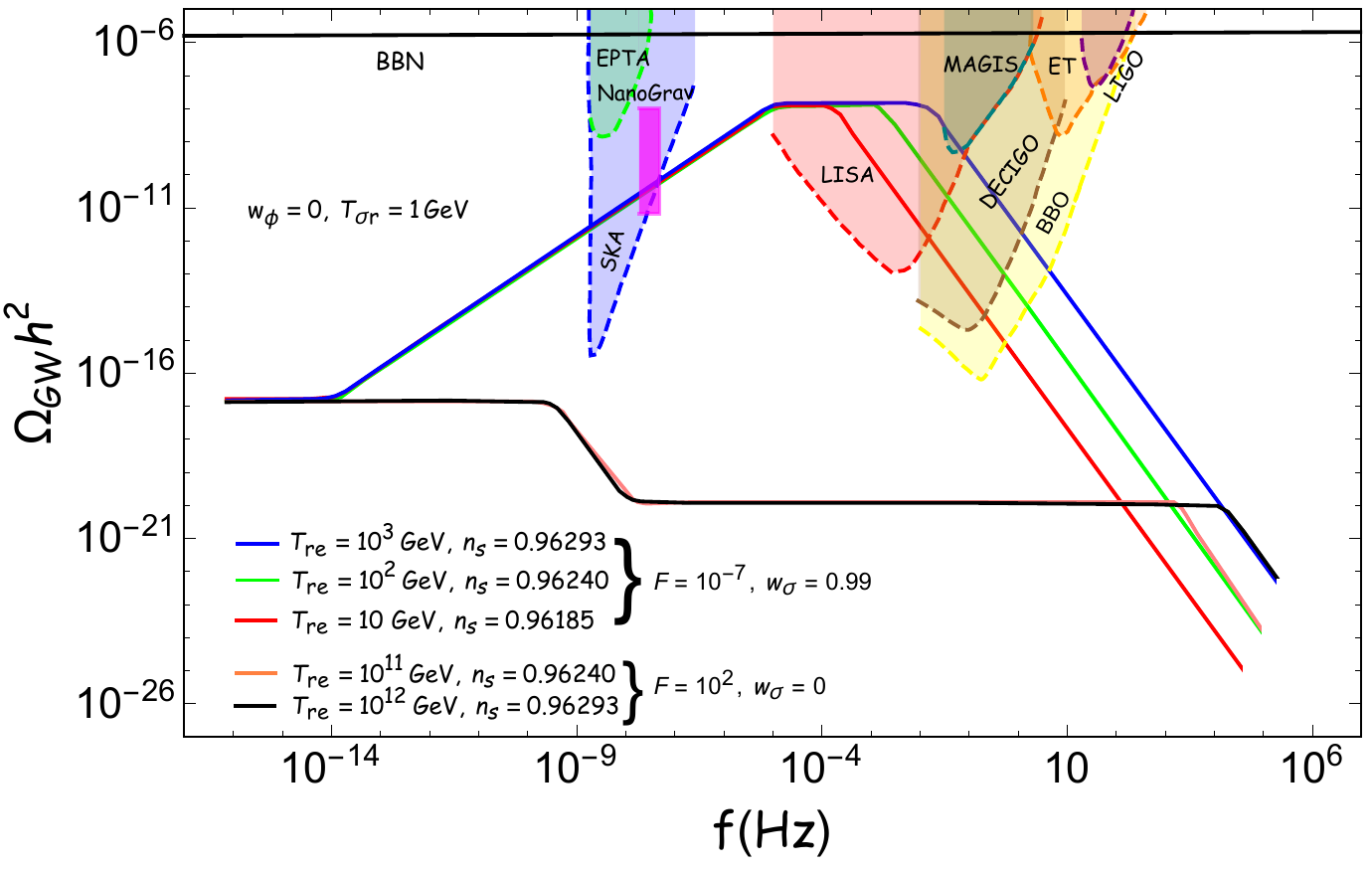}
\caption{The spectrum of GWs observed today $\ogw(f)$ has been plotted in 
the modified scenario with late time production of entropy. 
We have illustrated the results for the cases wherein $\ns$ is fixed and the 
quantity $F$ is varied (on the left) as well as for the cases wherein $F$ is
fixed and $\ns$ is varied (on the right).
We have set $w_\phi = 0$ and $T_{\sigma R} = 1\,\mathrm{GeV}$ in arriving
at the above plots.
Also, we have considered the extreme values for $w_\sigma$ to demonstrate
the maximum levels of impact that the generation of entropy at late times 
can have on the spectrum of GWs.
Interestingly, we find that for a set of values of the parameters associated 
the secondary phase of reheating, the spectrum $\ogw$ can have amplitudes as 
suggested by the recent observations 
by NANOGrav~\cite{Arzoumanian:2020vkk,Pol:2020igl}.}\label{plot_latetimeentropy}
\end{figure}

The behavior we have highlighted above can be clearly seen in 
Fig.~\ref{plot_latetimeentropy} wherein we have plotted the 
quantity $\ogw(f)$ in scenarios involving the second phase of 
reheating.
In arriving at the plots in the figure, we have set the inflaton the 
EoS parameter to be $w_{\phi}=0$ and have assumed that $T_{\sigma R}
= 1\,\mathrm{GeV}$.
Note that, for a given inflaton EoS parameter~$w_{\phi}$, we have two
parameters, viz. $\ns$ and $F$, which control the global shape of the 
spectrum of GWs.
We have plotted the spectrum for different values of~$F$ with a fixed 
value of~$\ns$ as well as for different values of $\ns$ with a fixed~$F$.
In order to highlight the effects due to the additional generation of 
entropy, we have plotted the results for the limiting values of zero 
and unity for the EoS parameter~$w_\sigma$ governing the second phase 
of reheating.
We should point that, if a canonical scalar field also dominates the 
secondary phase of reheating, then for $V(\sigma)\propto \sigma^{2\,n}$, 
we have $w_\sigma = (n-1)/(n+1)$ so that for the above mentioned
limiting values can be achieved for $n=1$ and $n \to \infty$, 
respectively.
One can also consider a more exotic, non-canonical, scalar field with a
Lagrangian density of the form ${\mathcal L} \sim (\partial \sigma)^\mu 
- \sigma^{2\,n}$, where $\mu$ is a rational number, to drive the second 
phase of reheating.
In such a case, it can be shown that the EoS parameter is given by (in 
this context, see, for example, Ref.~\cite{Unnikrishnan:2012zu})
\begin{equation}
w_{\sigma} = \f{n-\mu}{n\,(2\,\mu -1) + \mu},
\label{noncanEoS}
\end{equation}
with the expression reducing to the canonical result for $\mu=1$, 
as required.
Such a model can lead to the extreme values of $w_\sigma = 0$ (for 
$n = \mu$) and $w_\sigma \simeq 1$ for [$n \simeq \mu/
(1-\mu)$] that we have considered, without unnaturally large values 
for a dimensionless number, as it occurs in the canonical case. 
It seems worthwhile to explore such models in some detail as they could 
have interesting phenomenological implications.

We find that the effects on $\ogw(f)$ over the range $f < \fre$ due to
the late time creation of entropy have important implications for the 
recent observations by the NANOGrav mission. 
Recall that, recent observations by the NANOGrav mission suggest a stochastic 
GW background with an amplitude of $\ogw\, h^2\simeq 10^{-11}$ around the 
frequency of~$10^{-8}\,\mathrm{Hz}$~\cite{Arzoumanian:2020vkk,Pol:2020igl}.
Clearly, the frequency lies in the domain $f < \fre$. 
In the absence of a second phase of reheating, evidently, the amplitude 
of $\ogw$ in the nano-Hertz range of frequencies is rather small, much 
below the sensitivity of the NANOGrav mission, as we had seen in  
Secs.~\ref{sec:averaged} and~\ref{sec:actual}. 
However, as we have discussed, the late time decay of an additional scalar 
field such as the moduli field leads to a spectrum with a blue tilt for 
$w_\sigma > 1/3$ over the frequency range $f_\sigma < f< f_{\sigma R}$,
where $f_\sigma$ and $f_{\sigma R}$ are the frequencies associated with 
the wave numbers $k_\sigma$ and $k_{\sigma R}$.
Therefore, in such a modified scenario, it is possible to construct situations 
that result in $\ogw$ of the strength indicated by NANOGrav, albeit with rather
large values for $\omega_\sigma$ and relatively low values of the reheating
temperature of $10 < \Tre < 10^3\,\mathrm{GeV}$, as illustrated in 
Fig.~\ref{plot_latetimeentropy}. 
To motivate high values for the EoS parameter, as we mentioned, it seems 
interesting to consider a non-canonical model of a scalar field that leads 
to an EoS parameter as in Eq.~\eqref{noncanEoS}.
We should clarify that, to avoid pathological behavior, in the model,
we can consider parameters lying within the domains $n>0$ and $\mu> 0$.
In such a case, one can obtain that $w_{\sigma} \sim 1$ for $\mu$ in the 
range $0 < \mu < 1$. 
Moreover, note that, in the modified scenario with late time entropy production, 
to be compatible with the NANOGrav results, the reheating temperature $\Tre$ has 
to be less than $10^{3}\,\mathrm{GeV}$, which implies a low decay width for the 
inflaton.
Further, from Fig.~\ref{plot_latetimeentropy}, it can be easily seen that 
the modified scenario can be strongly constrained by many of the forthcoming 
GW observatories such as SKA, BBO, LISA and DECIGO.

\begin{figure}[t!]
\centering
\includegraphics[width=8.00cm]{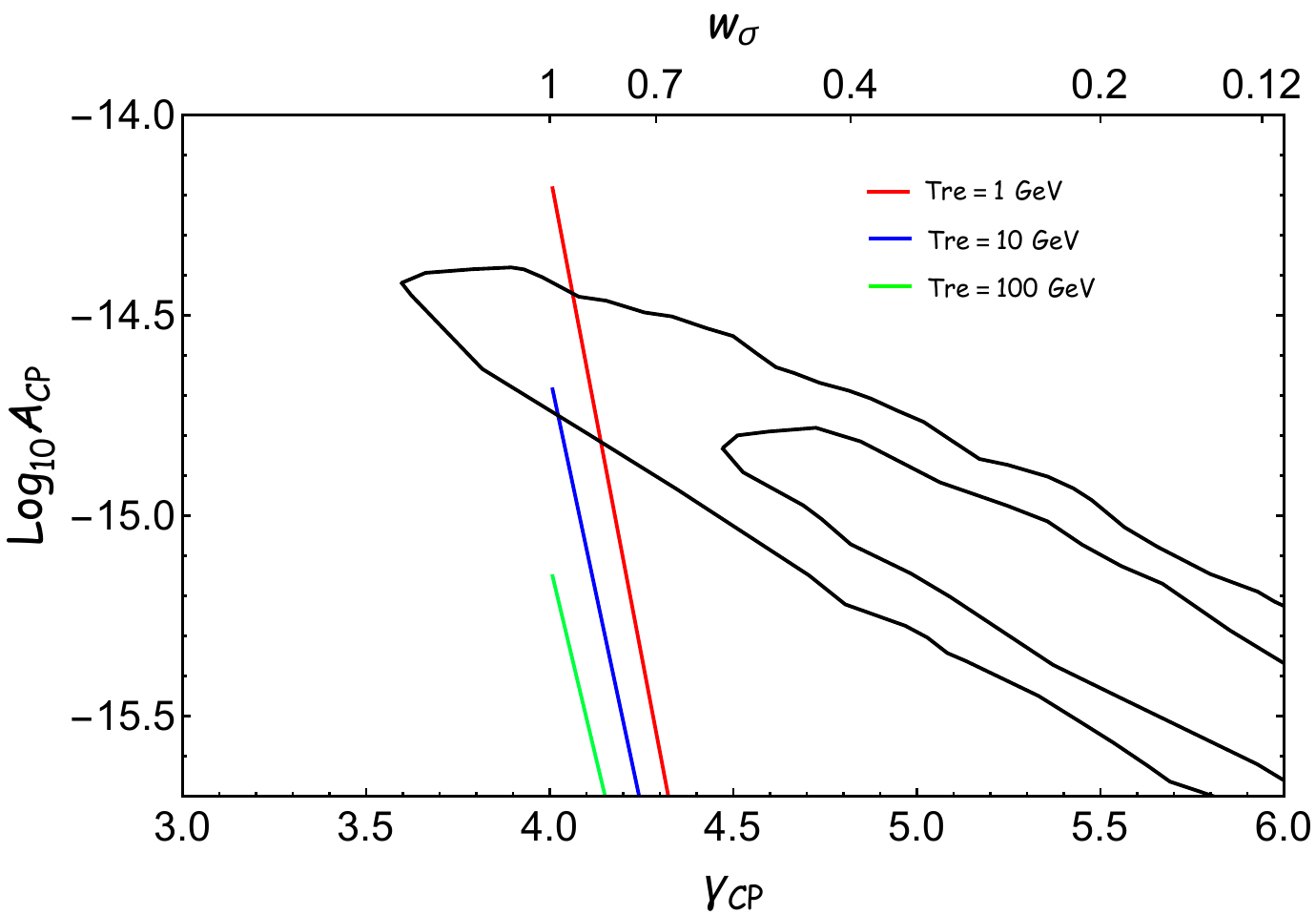}
\hskip 5pt
\includegraphics[width=8.75cm]{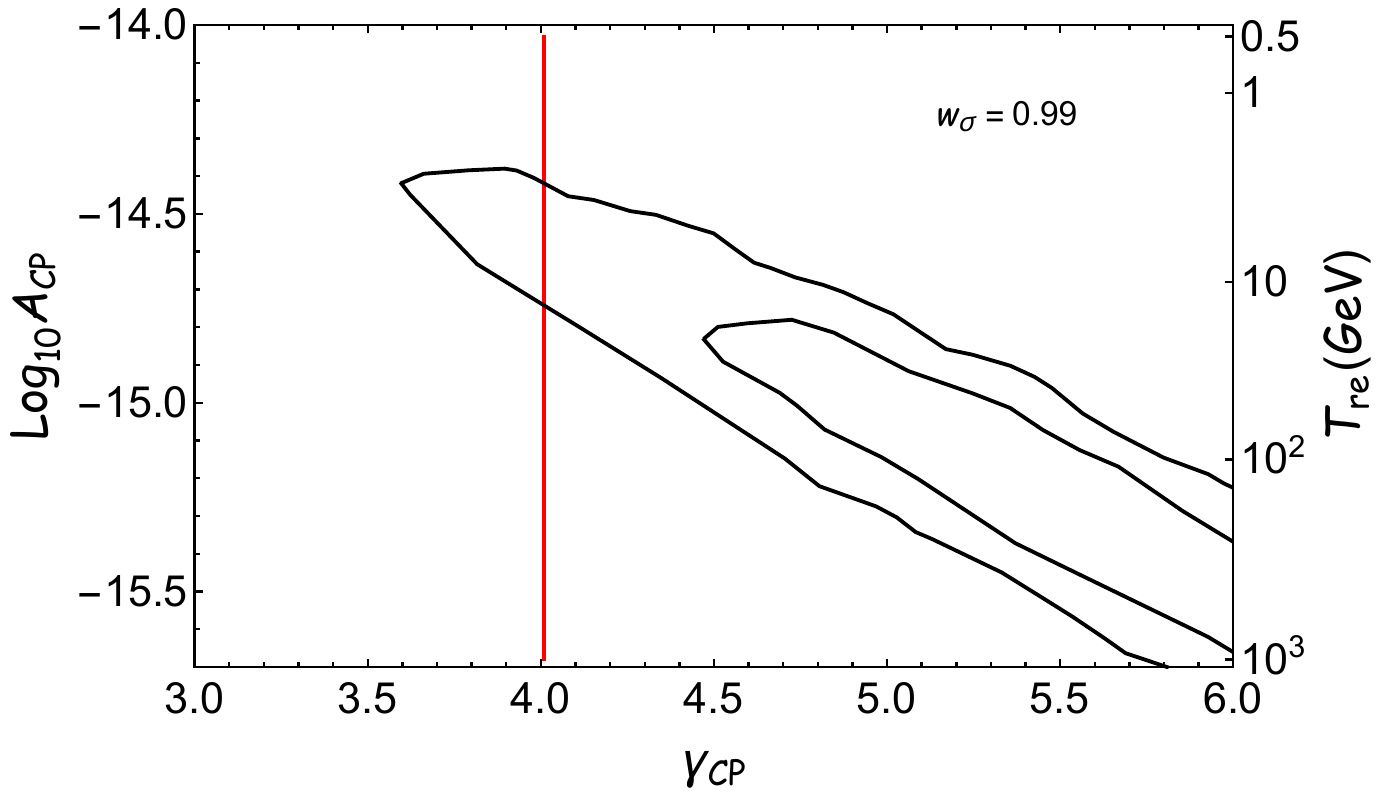}
\caption{The constraints from NANOGrav on the parameters $A_{_\mathrm{CP}}$ 
and $\gamma_{\mathrm{CP}}$ have been utilized to illustrate the corresponding
constraints on the EoS parameter~$w_\sigma$ describing the secondary phase 
of reheating and the reheating temperature~$\Tre$ associated with the primary
phase.
In the figures, we have included the 1-$\sigma$ and 2-$\sigma$ contours (in
black) arrived at by the NANOGrav analysis based on the five-frequency 
power-law fit~\cite{Arzoumanian:2020vkk}. 
We have projected the results into the $\gamma_{_\mathrm{CP}}$-$A_{_\mathrm{CP}}$  
plane for the following two possibilities: (i)~a scenario wherein~$\Tre$ is 
fixed and the parameter~$w_\sigma$ is varied (on the left), and (ii)~a scenario 
wherein~$w_\sigma$ is fixed and~$\Tre$ is varied (on the right). 
Note that we have assumed that $w_\phi = 0$ and $T_{\sigma R} = 0.1\,\mathrm{GeV}$ 
in arriving at the above plots.}\label{fig:gammacp}
\end{figure}

In the pulsar-timing data considered by NANOGrav, the spectrum of the 
characteristic strain $h_\mathrm{c}(f)$ induced by the GWs is assumed 
to be a power law of the form~\cite{Arzoumanian:2020vkk,Pol:2020igl}
\begin{equation}
h_\mathrm{c}(f) = A_{_\mathrm{CP}}\,
\l(\f{f}{f_\mathrm{yr}}\r)^{(3-\gamma_{_\mathrm{CP}})/2},
\label{eq:cs}
\end{equation}
where $A_{_\mathrm{CP}}$ refers to the amplitude at the reference frequency 
$f_\mathrm{yr}=1\,\mathrm{yr}^{-1} =3.17\times10^{-8}\, \mathrm{Hz}$, and 
$\gamma_{_\mathrm{CP}}$ is the timing-residual cross-power spectral density.
The dimensionless energy density of GWs today~$\ogw(f)$ is related to 
characteristic strain~$h_\mathrm{c}(f)$ induced by the GWs through the 
relation (in this context, see the recent review~\cite{Yokoyama:2021hsa})
\begin{equation} \label{omegahc}
\ogw(f)=\f{2\,\pi^2\,f^2}{3\,H_0^2}\,h_\mathrm{c}^2(f).
\end{equation}
Upon utilizing the form~\eqref{eq:cs} for the characteristic strain,
the energy density of GWs today can be expressed  in terms of the 
amplitude $A_{_\mathrm{CP}}$ and the index $\gamma_{_{\mathrm{CP}}}$ 
as follows:
\begin{equation}
\ogw(f)=\frac{2\,\pi^2\,f_\mathrm{yr}^2}{3\,H_0^2}\, 
A_{_\mathrm{CP}}^2\,\l(\f{f}{f_\mathrm{yr}}\r)^{5-\gamma_{_\mathrm{CP}}}.
\label{eq:omegaACP}
\end{equation}
As we have discussed above, in the scenario with late time entropy production, 
it is the power associated with the modes that reenter the Hubble radius during 
the secondary phase of reheating that are consistent with the NANOGrav results
[cf. Fig.~\ref{plot_latetimeentropy}].
Over this domain of wave numbers, since the index of the spectrum of GWs is 
given by $\ngw=2\,(3\,w_\sigma - 1)/(3\,w_\sigma + 1)$, where $w_\sigma$ 
is the EoS parameter describing the secondary phase of reheating, clearly, 
we can set $\gamma_{_\mathrm{CP}}=5-\ngw$.
We can utilize the constraints from the NANOGRav results on the parameters 
$A_{_\mathrm{CP}}$ and $\gamma_{_{\mathrm{CP}}}$ to arrive at the corresponding 
constraints on, say, the EoS parameter $w_\sigma$ and the reheating 
temperature~$\Tre$ associated with the primary phase.
We have illustrated these constraints in Fig.~\ref{fig:gammacp}.
We have illustrated the constraints for the two possibilities, viz. with the 
reheating temperature~$\Tre$ associated with the primary phase fixed and the
EoS parameter~$w_\sigma$ describing the secondary phase of reheating 
varied as well as with~$w_\sigma$ fixed and~$\Tre$ varied.
These constraints suggest that only low reheating temperatures (say, 
$\Tre< 10\, \mathrm{GeV}$) and very high values of the EoS 
parameter~$w_\sigma$ (with $w_\sigma \simeq 1$) are compatible with 
the NANOGrav data.


\section{Conclusions}\label{sec:c}

In this work, we have attempted to understand the effects of reheating on 
the spectrum of primordial GWs observed today. 
As a specific example, we had considered the so-called $\alpha$-attractor 
model of inflation and had evolved the tensor perturbations from the end 
of inflation through the epochs of reheating and radiation domination to
eventually arrive at the spectrum of GWs today. 
Moreover, we had considered two different scenarios to achieve reheating.
In the first scenario, the epoch is described by a constant EoS 
parameter~$w_\phi$ and the transition to radiation domination
is expected to occur instantaneously~\cite{Dai:2014jja}.
Such a simpler modeling of the reheating mechanism had allowed us to study
the evolution of the GWs analytically.
In the second and more realistic scenario of perturbative reheating wherein
the evolution of the energy densities are governed by the Boltzmann equations,
the inflaton decays gradually and the transition to the epoch of radiation
domination occurs smoothly.
The effective EoS parameter in such a scenario changes continuously from its initial
value of $w_\mathrm{eff} = w_\phi$ to the final value of 
$w_\mathrm{eff} = 1/3$.
As it proves to be difficult to obtain analytical solutions in the perturbative
reheating scenario, we had examined the problems at hand numerically. 

Note that we are interested in the spectrum of GWs today over scales that are 
considerably smaller than the CMB scales.
These scales reenter the Hubble radius either during the phase of reheating or 
during the epoch of radiation domination.
In both the models of reheating we have considered, the spectrum of GWs today
$\ogw(f)$ is scale invariant over wave numbers that reenter the Hubble radius 
during the epoch of radiation domination (i.e. for $k <\kre$). 
The scale invariant amplitude over this domain can help us extract the 
inflationary energy scale in terms of the present radiation abundance 
since $\HI^2/\Mpl^2 \sim 6\,\pi^2\,\ogw/ \Omega_{_\mathrm{R}}$ 
[cf. Eq.~\eqref{infscale}].
The spectrum of GWs today has a tilt over wave numbers $\kre <k <\kf$ which 
reenter the Hubble radius during the phase of reheating.
The spectral tilt $\ngw = 2\,(3\,w_{\phi} - 1)/(3\,w_\phi + 1)$ is red 
or blue depending on whether $w_\phi < 1/3$ or $w_\phi > 1/3$. 
Clearly, the observation of the tilt will allow us to determine the reheating 
parameters such as the EoS parameter~$w_\phi$ and the reheating temperature~$\Tre$.
These will allow us to determine the behavior of the inflaton near the minimum 
of the potential. 
Moreover, the constraint on the reheating parameters, in turn, can help us 
constrain the inflationary parameters such as the scalar spectral index~$\ns$. 
Further, in the realistic perturbative reheating scenario, there arises an important 
feature in the spectrum of GWs around wave numbers that reenter the Hubble radius 
towards the end of the phase of reheating. 
The spectrum exhibits distinct oscillations near the frequency $\fre$ and we find 
that the width of the oscillation reflects the time scale over which the EoS parameter
changes from the inflaton dominated value of $w_\phi$ to that of $1/3$ in the 
radiation dominated epoch (cf. Figs.~\ref{plot_perturbative1} and 
\ref{plot_perturbative2}). 
In fact, the peak of the oscillation occurs at $\fre$, which can be associated with 
the end of perturbative reheating (i.e. when $H =\Gamma_{\phi}$). 
We find that, at this instant, the rate of change of the effective
EoS parameter exhibits a maximum.
This can help us further in determining the inflaton decay rate $\Gamma_{\phi}$ 
in terms of the observed quantities, as expressed in Eq.~\eqref{gammaphi}.
The end of reheating is to be followed by the most important stage of 
thermalization. 
From the width of the oscillation in the spectrum, one can extract information 
about thermalization time scale $\Delta t_\mathrm{th}$ 
[cf. Eq.~\eqref{thermalization}] as well as the nature of the thermalization 
process depending upon over-occupied or under occupied initial state set by 
the end of reheating [cf. Eq.~\eqref{alpha}].
It turns out that the ratio of the thermalized particle density to the initial decaying 
particle density (i.e. $n_\mathrm{th}/n_\mathrm{i}$) depends on the inflaton mass, 
inflationary energy scale and the wave numbers $\kre$ and $\kf$.
Therefore, given the inflaton mass, the spectrum of primordial GWs at the present 
epoch can be utilized to determine the ratio $n_\mathrm{th}/n_\mathrm{i}$, which 
is one of the important parameters describing the thermalizing plasma.
Once the value of $n_\mathrm{th}/n_\mathrm{i}$ is extracted, the strength of the 
gauge interaction, say, $\alpha$, among the reheating decay products can be obtained
[cf. Eq.~\eqref{alpha2}].
We should mention that the oscillatory feature in the spectrum of GWs is encountered 
for all possible values of $w_{\phi}$ and $\Tre$. 
We find that the range of frequencies around which the oscillations arise shifts 
towards higher frequencies as the reheating temperature increases. 
This can be attributed to the fact that, for a higher value of $\Tre$, the duration of 
reheating proves to be shorter and, as a result, the wave number~$\kre$ which reenters
at the end of reheating becomes larger.

Apart from the effects due to the standard phase of reheating, we had also 
considered the signatures on the spectrum of GWs due to a secondary phase 
of reheating.
Such a secondary phase can arise due to the decay of additional scalar fields 
such as the moduli fields, which can lead to the production of entropy at late 
times.
The moduli dominated phase, which can be described by a constant EoS 
parameter~$w_\sigma$ (as the primary phase of reheating), leads 
to a tilt in the spectrum over the range of wave numbers that reenter 
the Hubble radius during the epoch. 
Remarkably, we have shown that, for $w_\sigma > 1/3$ and certain values of 
the reheating temperature, the production of entropy at late times can lead to 
$\ogw(k)$ that correspond to the strengths of the stochastic GW background 
suggested by the recent NANOGrav observations~\cite{Arzoumanian:2020vkk,Pol:2020igl}.
We should clarify that such a possibility cannot arise in the conventional 
reheating scenario wherein the entropy remains conserved from the end of
reheating until the present epoch.
In fact, the assumption of the secondary phase of reheating and the NANOGrav 
observations indicate a low reheating temperature of $\Tre \lesssim 10^{3}\,
\mathrm{GeV}$. We are currently examining different models to understand the wider implications
of such interesting possibilities.
 

\acknowledgments

M.R.H would like to thank the Ministry of Human Resource Development, 
Government of India (GoI), for financial assistance.
D.M and L.S wish to acknowledge support from the Science and Engineering 
Research Board~(SERB), Department of Science and Technology~(DST), GoI,
through the Core Research Grant CRG/2020/003664.
L.S also wishes to acknowledge support from SERB, DST, GoI, through the 
Core Research Grant No CRG/2018/002200. 


\bibliographystyle{apsrev4-1}
\bibliography{references} 

\begin{thebibliography}{136}%
\makeatletter
\providecommand \@ifxundefined [1]{%
 \@ifx{#1\undefined}
}%
\providecommand \@ifnum [1]{%
 \ifnum #1\expandafter \@firstoftwo
 \else \expandafter \@secondoftwo
 \fi
}%
\providecommand \@ifx [1]{%
 \ifx #1\expandafter \@firstoftwo
 \else \expandafter \@secondoftwo
 \fi
}%
\providecommand \natexlab [1]{#1}%
\providecommand \enquote  [1]{``#1''}%
\providecommand \bibnamefont  [1]{#1}%
\providecommand \bibfnamefont [1]{#1}%
\providecommand \citenamefont [1]{#1}%
\providecommand \href@noop [0]{\@secondoftwo}%
\providecommand \href [0]{\begingroup \@sanitize@url \@href}%
\providecommand \@href[1]{\@@startlink{#1}\@@href}%
\providecommand \@@href[1]{\endgroup#1\@@endlink}%
\providecommand \@sanitize@url [0]{\catcode `\\12\catcode `\$12\catcode
  `\&12\catcode `\#12\catcode `\^12\catcode `\_12\catcode `\%12\relax}%
\providecommand \@@startlink[1]{}%
\providecommand \@@endlink[0]{}%
\providecommand \url  [0]{\begingroup\@sanitize@url \@url }%
\providecommand \@url [1]{\endgroup\@href {#1}{\urlprefix }}%
\providecommand \urlprefix  [0]{URL }%
\providecommand \Eprint [0]{\href }%
\providecommand \doibase [0]{http://dx.doi.org/}%
\providecommand \selectlanguage [0]{\@gobble}%
\providecommand \bibinfo  [0]{\@secondoftwo}%
\providecommand \bibfield  [0]{\@secondoftwo}%
\providecommand \translation [1]{[#1]}%
\providecommand \BibitemOpen [0]{}%
\providecommand \bibitemStop [0]{}%
\providecommand \bibitemNoStop [0]{.\EOS\space}%
\providecommand \EOS [0]{\spacefactor3000\relax}%
\providecommand \BibitemShut  [1]{\csname bibitem#1\endcsname}%
\let\auto@bib@innerbib\@empty
\bibitem [{\citenamefont {Hawking}(1982)}]{Hawking:1982cz}%
  \BibitemOpen
  \bibfield  {author} {\bibinfo {author} {\bibfnamefont {S.}~\bibnamefont
  {Hawking}},\ }\href {\doibase 10.1016/0370-2693(82)90373-2} {\bibfield
  {journal} {\bibinfo  {journal} {Phys.\ Lett.\ B}\ }\textbf {\bibinfo {volume}
  {115}},\ \bibinfo {pages} {295} (\bibinfo {year} {1982})}\BibitemShut
  {NoStop}%
\bibitem [{\citenamefont {Guth}\ and\ \citenamefont {Pi}(1982)}]{Guth:1982ec}%
  \BibitemOpen
  \bibfield  {author} {\bibinfo {author} {\bibfnamefont {A.~H.}\ \bibnamefont
  {Guth}}\ and\ \bibinfo {author} {\bibfnamefont {S.}~\bibnamefont {Pi}},\
  }\href {\doibase 10.1103/PhysRevLett.49.1110} {\bibfield  {journal} {\bibinfo
   {journal} {Phys.\ Rev.\ Lett.}\ }\textbf {\bibinfo {volume} {49}},\ \bibinfo
  {pages} {1110} (\bibinfo {year} {1982})}\BibitemShut {NoStop}%
\bibitem [{\citenamefont {Starobinsky}(1982)}]{Starobinsky:1982ee}%
  \BibitemOpen
  \bibfield  {author} {\bibinfo {author} {\bibfnamefont {A.~A.}\ \bibnamefont
  {Starobinsky}},\ }\href {\doibase 10.1016/0370-2693(82)90541-X} {\bibfield
  {journal} {\bibinfo  {journal} {Phys.\ Lett.\ B}\ }\textbf {\bibinfo {volume}
  {117}},\ \bibinfo {pages} {175} (\bibinfo {year} {1982})}\BibitemShut
  {NoStop}%
\bibitem [{\citenamefont {Bardeen}\ \emph {et~al.}(1983)\citenamefont
  {Bardeen}, \citenamefont {Steinhardt},\ and\ \citenamefont
  {Turner}}]{Bardeen:1983qw}%
  \BibitemOpen
  \bibfield  {author} {\bibinfo {author} {\bibfnamefont {J.~M.}\ \bibnamefont
  {Bardeen}}, \bibinfo {author} {\bibfnamefont {P.~J.}\ \bibnamefont
  {Steinhardt}}, \ and\ \bibinfo {author} {\bibfnamefont {M.~S.}\ \bibnamefont
  {Turner}},\ }\href {\doibase 10.1103/PhysRevD.28.679} {\bibfield  {journal}
  {\bibinfo  {journal} {Phys.\ Rev.\ D}\ }\textbf {\bibinfo {volume} {28}},\
  \bibinfo {pages} {679} (\bibinfo {year} {1983})}\BibitemShut {NoStop}%
\bibitem [{\citenamefont {Mukhanov}\ \emph {et~al.}(1992)\citenamefont
  {Mukhanov}, \citenamefont {Feldman},\ and\ \citenamefont
  {Brandenberger}}]{Mukhanov:1990me}%
  \BibitemOpen
  \bibfield  {author} {\bibinfo {author} {\bibfnamefont {V.~F.}\ \bibnamefont
  {Mukhanov}}, \bibinfo {author} {\bibfnamefont {H.~A.}\ \bibnamefont
  {Feldman}}, \ and\ \bibinfo {author} {\bibfnamefont {R.~H.}\ \bibnamefont
  {Brandenberger}},\ }\href {\doibase 10.1016/0370-1573(92)90044-Z} {\bibfield
  {journal} {\bibinfo  {journal} {Phys. Rept.}\ }\textbf {\bibinfo {volume}
  {215}},\ \bibinfo {pages} {203} (\bibinfo {year} {1992})}\BibitemShut
  {NoStop}%
\bibitem [{\citenamefont {Martin}(2004)}]{Martin:2003bt}%
  \BibitemOpen
  \bibfield  {author} {\bibinfo {author} {\bibfnamefont {J.}~\bibnamefont
  {Martin}},\ }\href {\doibase 10.1590/S0103-97332004000700005} {\bibfield
  {journal} {\bibinfo  {journal} {Braz. J. Phys.}\ }\textbf {\bibinfo {volume}
  {34}},\ \bibinfo {pages} {1307} (\bibinfo {year} {2004})},\ \Eprint
  {http://arxiv.org/abs/astro-ph/0312492} {arXiv:astro-ph/0312492} \BibitemShut
  {NoStop}%
\bibitem [{\citenamefont {Martin}(2005)}]{Martin:2004um}%
  \BibitemOpen
  \bibfield  {author} {\bibinfo {author} {\bibfnamefont {J.}~\bibnamefont
  {Martin}},\ }\href {\doibase 10.1007/11377306_7} {\bibfield  {journal}
  {\bibinfo  {journal} {Lect. Notes Phys.}\ }\textbf {\bibinfo {volume}
  {669}},\ \bibinfo {pages} {199} (\bibinfo {year} {2005})},\ \Eprint
  {http://arxiv.org/abs/hep-th/0406011} {arXiv:hep-th/0406011} \BibitemShut
  {NoStop}%
\bibitem [{\citenamefont {Bassett}\ \emph {et~al.}(2006)\citenamefont
  {Bassett}, \citenamefont {Tsujikawa},\ and\ \citenamefont
  {Wands}}]{Bassett:2005xm}%
  \BibitemOpen
  \bibfield  {author} {\bibinfo {author} {\bibfnamefont {B.~A.}\ \bibnamefont
  {Bassett}}, \bibinfo {author} {\bibfnamefont {S.}~\bibnamefont {Tsujikawa}},
  \ and\ \bibinfo {author} {\bibfnamefont {D.}~\bibnamefont {Wands}},\ }\href
  {\doibase 10.1103/RevModPhys.78.537} {\bibfield  {journal} {\bibinfo
  {journal} {Rev. Mod. Phys.}\ }\textbf {\bibinfo {volume} {78}},\ \bibinfo
  {pages} {537} (\bibinfo {year} {2006})}\BibitemShut {NoStop}%
\bibitem [{\citenamefont {Sriramkumar}(2009)}]{Sriramkumar:2009kg}%
  \BibitemOpen
  \bibfield  {author} {\bibinfo {author} {\bibfnamefont {L.}~\bibnamefont
  {Sriramkumar}},\ }\href@noop {} {\  (\bibinfo {year} {2009})},\ \Eprint
  {http://arxiv.org/abs/0904.4584} {arXiv:0904.4584 [astro-ph.CO]} \BibitemShut
  {NoStop}%
\bibitem [{\citenamefont {Baumann}\ and\ \citenamefont
  {Peiris}(2009)}]{Baumann:2008bn}%
  \BibitemOpen
  \bibfield  {author} {\bibinfo {author} {\bibfnamefont {D.}~\bibnamefont
  {Baumann}}\ and\ \bibinfo {author} {\bibfnamefont {H.~V.}\ \bibnamefont
  {Peiris}},\ }\href {\doibase 10.1166/asl.2009.1019} {\bibfield  {journal}
  {\bibinfo  {journal} {Adv. Sci. Lett.}\ }\textbf {\bibinfo {volume} {2}},\
  \bibinfo {pages} {105} (\bibinfo {year} {2009})},\ \Eprint
  {http://arxiv.org/abs/0810.3022} {arXiv:0810.3022 [astro-ph]} \BibitemShut
  {NoStop}%
\bibitem [{\citenamefont {Baumann}(2009)}]{Baumann:2009ds}%
  \BibitemOpen
  \bibfield  {author} {\bibinfo {author} {\bibfnamefont {D.}~\bibnamefont
  {Baumann}},\ }in\ \href {\doibase 10.1142/9789814327183_0010} {\emph
  {\bibinfo {booktitle} {{Theoretical Advanced Study Institute in Elementary
  Particle Physics}: {Physics of the Large and the Small}}}}\ (\bibinfo {year}
  {2009})\ \Eprint {http://arxiv.org/abs/0907.5424} {arXiv:0907.5424 [hep-th]}
  \BibitemShut {NoStop}%
\bibitem [{\citenamefont {Sriramkumar}(2012)}]{Sriramkumar:2012mik}%
  \BibitemOpen
  \bibfield  {author} {\bibinfo {author} {\bibfnamefont {L.}~\bibnamefont
  {Sriramkumar}},\ }\enquote {\bibinfo {title} {{On the generation and
  evolution of perturbations during inflation and reheating}},}\ in\ \href
  {\doibase 10.1142/9789814322072_0008} {\emph {\bibinfo {booktitle}
  {{Vignettes in Gravitation and Cosmology}}}},\ \bibinfo {editor} {edited by\
  \bibinfo {editor} {\bibfnamefont {L.}~\bibnamefont {Sriramkumar}}\ and\
  \bibinfo {editor} {\bibfnamefont {T.~R.}\ \bibnamefont {Seshadri}}}\
  (\bibinfo {year} {2012})\BibitemShut {NoStop}%
\bibitem [{\citenamefont {Linde}(2014)}]{Linde:2014nna}%
  \BibitemOpen
  \bibfield  {author} {\bibinfo {author} {\bibfnamefont {A.}~\bibnamefont
  {Linde}},\ }in\ \href {\doibase 10.1093/acprof:oso/9780198728856.003.0006}
  {\emph {\bibinfo {booktitle} {{100e Ecole d'Ete de Physique: Post-Planck
  Cosmology}}}}\ (\bibinfo {year} {2014})\ \Eprint
  {http://arxiv.org/abs/1402.0526} {arXiv:1402.0526 [hep-th]} \BibitemShut
  {NoStop}%
\bibitem [{\citenamefont {Martin}(2016)}]{Martin:2015dha}%
  \BibitemOpen
  \bibfield  {author} {\bibinfo {author} {\bibfnamefont {J.}~\bibnamefont
  {Martin}},\ }\href {\doibase 10.1007/978-3-319-44769-8_2} {\bibfield
  {journal} {\bibinfo  {journal} {Astrophys. Space Sci. Proc.}\ }\textbf
  {\bibinfo {volume} {45}},\ \bibinfo {pages} {41} (\bibinfo {year} {2016})},\
  \Eprint {http://arxiv.org/abs/1502.05733} {arXiv:1502.05733 [astro-ph.CO]}
  \BibitemShut {NoStop}%
\bibitem [{\citenamefont {Grishchuk}(1974)}]{Grishchuk:1974ny}%
  \BibitemOpen
  \bibfield  {author} {\bibinfo {author} {\bibfnamefont {L.~P.}\ \bibnamefont
  {Grishchuk}},\ }\href@noop {} {\bibfield  {journal} {\bibinfo  {journal} {Zh.
  Eksp. Teor. Fiz.}\ }\textbf {\bibinfo {volume} {67}},\ \bibinfo {pages} {825}
  (\bibinfo {year} {1974})}\BibitemShut {NoStop}%
\bibitem [{\citenamefont {Starobinsky}(1979)}]{Starobinsky:1979ty}%
  \BibitemOpen
  \bibfield  {author} {\bibinfo {author} {\bibfnamefont {A.~A.}\ \bibnamefont
  {Starobinsky}},\ }\href@noop {} {\bibfield  {journal} {\bibinfo  {journal}
  {JETP Lett.}\ }\textbf {\bibinfo {volume} {30}},\ \bibinfo {pages} {682}
  (\bibinfo {year} {1979})}\BibitemShut {NoStop}%
\bibitem [{\citenamefont {Guzzetti}\ \emph {et~al.}(2016)\citenamefont
  {Guzzetti}, \citenamefont {Bartolo}, \citenamefont {Liguori},\ and\
  \citenamefont {Matarrese}}]{Guzzetti:2016mkm}%
  \BibitemOpen
  \bibfield  {author} {\bibinfo {author} {\bibfnamefont {M.~C.}\ \bibnamefont
  {Guzzetti}}, \bibinfo {author} {\bibfnamefont {N.}~\bibnamefont {Bartolo}},
  \bibinfo {author} {\bibfnamefont {M.}~\bibnamefont {Liguori}}, \ and\
  \bibinfo {author} {\bibfnamefont {S.}~\bibnamefont {Matarrese}},\ }\href
  {\doibase 10.1393/ncr/i2016-10127-1} {\bibfield  {journal} {\bibinfo
  {journal} {Riv. Nuovo Cim.}\ }\textbf {\bibinfo {volume} {39}},\ \bibinfo
  {pages} {399} (\bibinfo {year} {2016})},\ \Eprint
  {http://arxiv.org/abs/1605.01615} {arXiv:1605.01615 [astro-ph.CO]}
  \BibitemShut {NoStop}%
\bibitem [{\citenamefont {Caprini}\ and\ \citenamefont
  {Figueroa}(2018)}]{Caprini:2018mtu}%
  \BibitemOpen
  \bibfield  {author} {\bibinfo {author} {\bibfnamefont {C.}~\bibnamefont
  {Caprini}}\ and\ \bibinfo {author} {\bibfnamefont {D.~G.}\ \bibnamefont
  {Figueroa}},\ }\href {\doibase 10.1088/1361-6382/aac608} {\bibfield
  {journal} {\bibinfo  {journal} {Class. Quant. Grav.}\ }\textbf {\bibinfo
  {volume} {35}},\ \bibinfo {pages} {163001} (\bibinfo {year} {2018})},\
  \Eprint {http://arxiv.org/abs/1801.04268} {arXiv:1801.04268 [astro-ph.CO]}
  \BibitemShut {NoStop}%
\bibitem [{\citenamefont {Abbott}\ \emph
  {et~al.}(2016{\natexlab{a}})\citenamefont {Abbott} \emph
  {et~al.}}]{TheLIGOScientific:2016agk}%
  \BibitemOpen
  \bibfield  {author} {\bibinfo {author} {\bibfnamefont {B.}~\bibnamefont
  {Abbott}} \emph {et~al.} (\bibinfo {collaboration} {LIGO Scientific,
  Virgo}),\ }\href {\doibase 10.1103/PhysRevLett.116.131103} {\bibfield
  {journal} {\bibinfo  {journal} {Phys. Rev. Lett.}\ }\textbf {\bibinfo
  {volume} {116}},\ \bibinfo {pages} {131103} (\bibinfo {year}
  {2016}{\natexlab{a}})},\ \Eprint {http://arxiv.org/abs/1602.03838}
  {arXiv:1602.03838 [gr-qc]} \BibitemShut {NoStop}%
\bibitem [{\citenamefont {Abbott}\ \emph
  {et~al.}(2016{\natexlab{b}})\citenamefont {Abbott} \emph
  {et~al.}}]{TheLIGOScientific:2016qqj}%
  \BibitemOpen
  \bibfield  {author} {\bibinfo {author} {\bibfnamefont {B.}~\bibnamefont
  {Abbott}} \emph {et~al.} (\bibinfo {collaboration} {LIGO Scientific,
  Virgo}),\ }\href {\doibase 10.1103/PhysRevD.93.122003} {\bibfield  {journal}
  {\bibinfo  {journal} {Phys. Rev. D}\ }\textbf {\bibinfo {volume} {93}},\
  \bibinfo {pages} {122003} (\bibinfo {year} {2016}{\natexlab{b}})},\ \Eprint
  {http://arxiv.org/abs/1602.03839} {arXiv:1602.03839 [gr-qc]} \BibitemShut
  {NoStop}%
\bibitem [{\citenamefont {Abbott}\ \emph
  {et~al.}(2016{\natexlab{c}})\citenamefont {Abbott} \emph
  {et~al.}}]{TheLIGOScientific:2016wfe}%
  \BibitemOpen
  \bibfield  {author} {\bibinfo {author} {\bibfnamefont {B.}~\bibnamefont
  {Abbott}} \emph {et~al.} (\bibinfo {collaboration} {LIGO Scientific,
  Virgo}),\ }\href {\doibase 10.1103/PhysRevLett.116.241102} {\bibfield
  {journal} {\bibinfo  {journal} {Phys. Rev. Lett.}\ }\textbf {\bibinfo
  {volume} {116}},\ \bibinfo {pages} {241102} (\bibinfo {year}
  {2016}{\natexlab{c}})},\ \Eprint {http://arxiv.org/abs/1602.03840}
  {arXiv:1602.03840 [gr-qc]} \BibitemShut {NoStop}%
\bibitem [{\citenamefont {Abbott}\ \emph
  {et~al.}(2016{\natexlab{d}})\citenamefont {Abbott} \emph
  {et~al.}}]{Abbott:2016blz}%
  \BibitemOpen
  \bibfield  {author} {\bibinfo {author} {\bibfnamefont {B.~P.}\ \bibnamefont
  {Abbott}} \emph {et~al.} (\bibinfo {collaboration} {LIGO Scientific,
  Virgo}),\ }\href {\doibase 10.1103/PhysRevLett.116.061102} {\bibfield
  {journal} {\bibinfo  {journal} {Phys. Rev. Lett.}\ }\textbf {\bibinfo
  {volume} {116}},\ \bibinfo {pages} {061102} (\bibinfo {year}
  {2016}{\natexlab{d}})},\ \Eprint {http://arxiv.org/abs/1602.03837}
  {arXiv:1602.03837 [gr-qc]} \BibitemShut {NoStop}%
\bibitem [{\citenamefont {Abbott}\ \emph
  {et~al.}(2016{\natexlab{e}})\citenamefont {Abbott} \emph
  {et~al.}}]{Abbott:2016nmj}%
  \BibitemOpen
  \bibfield  {author} {\bibinfo {author} {\bibfnamefont {B.~P.}\ \bibnamefont
  {Abbott}} \emph {et~al.} (\bibinfo {collaboration} {LIGO Scientific,
  Virgo}),\ }\href {\doibase 10.1103/PhysRevLett.116.241103} {\bibfield
  {journal} {\bibinfo  {journal} {Phys. Rev. Lett.}\ }\textbf {\bibinfo
  {volume} {116}},\ \bibinfo {pages} {241103} (\bibinfo {year}
  {2016}{\natexlab{e}})},\ \Eprint {http://arxiv.org/abs/1606.04855}
  {arXiv:1606.04855 [gr-qc]} \BibitemShut {NoStop}%
\bibitem [{\citenamefont {Abbott}\ \emph
  {et~al.}(2017{\natexlab{a}})\citenamefont {Abbott} \emph
  {et~al.}}]{Abbott:2017vtc}%
  \BibitemOpen
  \bibfield  {author} {\bibinfo {author} {\bibfnamefont {B.~P.}\ \bibnamefont
  {Abbott}} \emph {et~al.} (\bibinfo {collaboration} {LIGO Scientific,
  VIRGO}),\ }\href {\doibase 10.1103/PhysRevLett.118.221101,
  10.1103/PhysRevLett.121.129901} {\bibfield  {journal} {\bibinfo  {journal}
  {Phys. Rev. Lett.}\ }\textbf {\bibinfo {volume} {118}},\ \bibinfo {pages}
  {221101} (\bibinfo {year} {2017}{\natexlab{a}})},\ \bibinfo {note} {[Erratum:
  Phys. Rev. Lett.121,no.12,129901(2018)]},\ \Eprint
  {http://arxiv.org/abs/1706.01812} {arXiv:1706.01812 [gr-qc]} \BibitemShut
  {NoStop}%
\bibitem [{\citenamefont {Abbott}\ \emph
  {et~al.}(2017{\natexlab{b}})\citenamefont {Abbott} \emph
  {et~al.}}]{Abbott:2017gyy}%
  \BibitemOpen
  \bibfield  {author} {\bibinfo {author} {\bibfnamefont {B.~P.}\ \bibnamefont
  {Abbott}} \emph {et~al.} (\bibinfo {collaboration} {LIGO Scientific,
  Virgo}),\ }\href {\doibase 10.3847/2041-8213/aa9f0c} {\bibfield  {journal}
  {\bibinfo  {journal} {Astrophys. J. Lett.}\ }\textbf {\bibinfo {volume}
  {851}},\ \bibinfo {pages} {L35} (\bibinfo {year} {2017}{\natexlab{b}})},\
  \Eprint {http://arxiv.org/abs/1711.05578} {arXiv:1711.05578 [astro-ph.HE]}
  \BibitemShut {NoStop}%
\bibitem [{\citenamefont {Abbott}\ \emph
  {et~al.}(2017{\natexlab{c}})\citenamefont {Abbott} \emph
  {et~al.}}]{Abbott:2017oio}%
  \BibitemOpen
  \bibfield  {author} {\bibinfo {author} {\bibfnamefont {B.~P.}\ \bibnamefont
  {Abbott}} \emph {et~al.} (\bibinfo {collaboration} {LIGO Scientific,
  Virgo}),\ }\href {\doibase 10.1103/PhysRevLett.119.141101} {\bibfield
  {journal} {\bibinfo  {journal} {Phys. Rev. Lett.}\ }\textbf {\bibinfo
  {volume} {119}},\ \bibinfo {pages} {141101} (\bibinfo {year}
  {2017}{\natexlab{c}})},\ \Eprint {http://arxiv.org/abs/1709.09660}
  {arXiv:1709.09660 [gr-qc]} \BibitemShut {NoStop}%
\bibitem [{\citenamefont {Abbott}\ \emph
  {et~al.}(2017{\natexlab{d}})\citenamefont {Abbott} \emph
  {et~al.}}]{TheLIGOScientific:2017qsa}%
  \BibitemOpen
  \bibfield  {author} {\bibinfo {author} {\bibfnamefont {B.~P.}\ \bibnamefont
  {Abbott}} \emph {et~al.} (\bibinfo {collaboration} {LIGO Scientific,
  Virgo}),\ }\href {\doibase 10.1103/PhysRevLett.119.161101} {\bibfield
  {journal} {\bibinfo  {journal} {Phys. Rev. Lett.}\ }\textbf {\bibinfo
  {volume} {119}},\ \bibinfo {pages} {161101} (\bibinfo {year}
  {2017}{\natexlab{d}})},\ \Eprint {http://arxiv.org/abs/1710.05832}
  {arXiv:1710.05832 [gr-qc]} \BibitemShut {NoStop}%
\bibitem [{\citenamefont {Abbott}\ \emph
  {et~al.}(2020{\natexlab{a}})\citenamefont {Abbott} \emph
  {et~al.}}]{LIGOScientific:2020stg}%
  \BibitemOpen
  \bibfield  {author} {\bibinfo {author} {\bibfnamefont {R.}~\bibnamefont
  {Abbott}} \emph {et~al.} (\bibinfo {collaboration} {LIGO Scientific,
  Virgo}),\ }\href {\doibase 10.1103/PhysRevD.102.043015} {\bibfield  {journal}
  {\bibinfo  {journal} {Phys. Rev. D}\ }\textbf {\bibinfo {volume} {102}},\
  \bibinfo {pages} {043015} (\bibinfo {year} {2020}{\natexlab{a}})},\ \Eprint
  {http://arxiv.org/abs/2004.08342} {arXiv:2004.08342 [astro-ph.HE]}
  \BibitemShut {NoStop}%
\bibitem [{\citenamefont {Abbott}\ \emph
  {et~al.}(2020{\natexlab{b}})\citenamefont {Abbott} \emph
  {et~al.}}]{Abbott:2020uma}%
  \BibitemOpen
  \bibfield  {author} {\bibinfo {author} {\bibfnamefont {B.~P.}\ \bibnamefont
  {Abbott}} \emph {et~al.} (\bibinfo {collaboration} {LIGO Scientific,
  Virgo}),\ }\href {\doibase 10.3847/2041-8213/ab75f5} {\bibfield  {journal}
  {\bibinfo  {journal} {Astrophys. J. Lett.}\ }\textbf {\bibinfo {volume}
  {892}},\ \bibinfo {pages} {L3} (\bibinfo {year} {2020}{\natexlab{b}})},\
  \Eprint {http://arxiv.org/abs/2001.01761} {arXiv:2001.01761 [astro-ph.HE]}
  \BibitemShut {NoStop}%
\bibitem [{\citenamefont {Abbott}\ \emph
  {et~al.}(2020{\natexlab{c}})\citenamefont {Abbott} \emph
  {et~al.}}]{Abbott:2020khf}%
  \BibitemOpen
  \bibfield  {author} {\bibinfo {author} {\bibfnamefont {R.}~\bibnamefont
  {Abbott}} \emph {et~al.} (\bibinfo {collaboration} {LIGO Scientific,
  Virgo}),\ }\href {\doibase 10.3847/2041-8213/ab960f} {\bibfield  {journal}
  {\bibinfo  {journal} {Astrophys. J.}\ }\textbf {\bibinfo {volume} {896}},\
  \bibinfo {pages} {L44} (\bibinfo {year} {2020}{\natexlab{c}})},\ \Eprint
  {http://arxiv.org/abs/2006.12611} {arXiv:2006.12611 [astro-ph.HE]}
  \BibitemShut {NoStop}%
\bibitem [{\citenamefont {Abbott}\ \emph
  {et~al.}(2017{\natexlab{e}})\citenamefont {Abbott} \emph
  {et~al.}}]{TheLIGOScientific:2016dpb}%
  \BibitemOpen
  \bibfield  {author} {\bibinfo {author} {\bibfnamefont {B.~P.}\ \bibnamefont
  {Abbott}} \emph {et~al.} (\bibinfo {collaboration} {LIGO Scientific,
  Virgo}),\ }\href {\doibase 10.1103/PhysRevLett.118.121101} {\bibfield
  {journal} {\bibinfo  {journal} {Phys. Rev. Lett.}\ }\textbf {\bibinfo
  {volume} {118}},\ \bibinfo {pages} {121101} (\bibinfo {year}
  {2017}{\natexlab{e}})},\ \bibinfo {note} {[Erratum: Phys.Rev.Lett. 119,
  029901 (2017)]},\ \Eprint {http://arxiv.org/abs/1612.02029} {arXiv:1612.02029
  [gr-qc]} \BibitemShut {NoStop}%
\bibitem [{\citenamefont {Punturo}\ \emph {et~al.}(2010)\citenamefont {Punturo}
  \emph {et~al.}}]{Punturo:2010zz}%
  \BibitemOpen
  \bibfield  {author} {\bibinfo {author} {\bibfnamefont {M.}~\bibnamefont
  {Punturo}} \emph {et~al.},\ }\href {\doibase 10.1088/0264-9381/27/19/194002}
  {\bibfield  {journal} {\bibinfo  {journal} {Class. Quant. Grav.}\ }\textbf
  {\bibinfo {volume} {27}},\ \bibinfo {pages} {194002} (\bibinfo {year}
  {2010})}\BibitemShut {NoStop}%
\bibitem [{\citenamefont {Sathyaprakash}\ \emph {et~al.}(2012)\citenamefont
  {Sathyaprakash} \emph {et~al.}}]{Sathyaprakash:2012jk}%
  \BibitemOpen
  \bibfield  {author} {\bibinfo {author} {\bibfnamefont {B.}~\bibnamefont
  {Sathyaprakash}} \emph {et~al.},\ }\href {\doibase
  10.1088/0264-9381/29/12/124013} {\bibfield  {journal} {\bibinfo  {journal}
  {Class. Quant. Grav.}\ }\textbf {\bibinfo {volume} {29}},\ \bibinfo {pages}
  {124013} (\bibinfo {year} {2012})},\ \bibinfo {note} {[Erratum:
  Class.Quant.Grav. 30, 079501 (2013)]},\ \Eprint
  {http://arxiv.org/abs/1206.0331} {arXiv:1206.0331 [gr-qc]} \BibitemShut
  {NoStop}%
\bibitem [{\citenamefont {Crowder}\ and\ \citenamefont
  {Cornish}(2005)}]{Crowder:2005nr}%
  \BibitemOpen
  \bibfield  {author} {\bibinfo {author} {\bibfnamefont {J.}~\bibnamefont
  {Crowder}}\ and\ \bibinfo {author} {\bibfnamefont {N.~J.}\ \bibnamefont
  {Cornish}},\ }\href {\doibase 10.1103/PhysRevD.72.083005} {\bibfield
  {journal} {\bibinfo  {journal} {Phys. Rev.}\ }\textbf {\bibinfo {volume}
  {D72}},\ \bibinfo {pages} {083005} (\bibinfo {year} {2005})},\ \Eprint
  {http://arxiv.org/abs/gr-qc/0506015} {arXiv:gr-qc/0506015 [gr-qc]}
  \BibitemShut {NoStop}%
\bibitem [{\citenamefont {Corbin}\ and\ \citenamefont
  {Cornish}(2006)}]{Corbin:2005ny}%
  \BibitemOpen
  \bibfield  {author} {\bibinfo {author} {\bibfnamefont {V.}~\bibnamefont
  {Corbin}}\ and\ \bibinfo {author} {\bibfnamefont {N.~J.}\ \bibnamefont
  {Cornish}},\ }\href {\doibase 10.1088/0264-9381/23/7/014} {\bibfield
  {journal} {\bibinfo  {journal} {Class. Quant. Grav.}\ }\textbf {\bibinfo
  {volume} {23}},\ \bibinfo {pages} {2435} (\bibinfo {year} {2006})},\ \Eprint
  {http://arxiv.org/abs/gr-qc/0512039} {arXiv:gr-qc/0512039 [gr-qc]}
  \BibitemShut {NoStop}%
\bibitem [{\citenamefont {Baker}\ \emph {et~al.}(2019)\citenamefont {Baker}
  \emph {et~al.}}]{Baker:2019pnp}%
  \BibitemOpen
  \bibfield  {author} {\bibinfo {author} {\bibfnamefont {J.}~\bibnamefont
  {Baker}} \emph {et~al.},\ }\href@noop {} {\  (\bibinfo {year} {2019})},\
  \Eprint {http://arxiv.org/abs/1907.11305} {arXiv:1907.11305 [astro-ph.IM]}
  \BibitemShut {NoStop}%
\bibitem [{\citenamefont {Seto}\ \emph {et~al.}(2001)\citenamefont {Seto},
  \citenamefont {Kawamura},\ and\ \citenamefont {Nakamura}}]{Seto:2001qf}%
  \BibitemOpen
  \bibfield  {author} {\bibinfo {author} {\bibfnamefont {N.}~\bibnamefont
  {Seto}}, \bibinfo {author} {\bibfnamefont {S.}~\bibnamefont {Kawamura}}, \
  and\ \bibinfo {author} {\bibfnamefont {T.}~\bibnamefont {Nakamura}},\ }\href
  {\doibase 10.1103/PhysRevLett.87.221103} {\bibfield  {journal} {\bibinfo
  {journal} {Phys. Rev. Lett.}\ }\textbf {\bibinfo {volume} {87}},\ \bibinfo
  {pages} {221103} (\bibinfo {year} {2001})},\ \Eprint
  {http://arxiv.org/abs/astro-ph/0108011} {arXiv:astro-ph/0108011} \BibitemShut
  {NoStop}%
\bibitem [{\citenamefont {Kawamura}\ \emph {et~al.}(2011)\citenamefont
  {Kawamura} \emph {et~al.}}]{Kawamura:2011zz}%
  \BibitemOpen
  \bibfield  {author} {\bibinfo {author} {\bibfnamefont {S.}~\bibnamefont
  {Kawamura}} \emph {et~al.},\ }\href {\doibase 10.1088/0264-9381/28/9/094011}
  {\bibfield  {journal} {\bibinfo  {journal} {Class. Quant. Grav.}\ }\textbf
  {\bibinfo {volume} {28}},\ \bibinfo {pages} {094011} (\bibinfo {year}
  {2011})}\BibitemShut {NoStop}%
\bibitem [{\citenamefont {Sato}\ \emph {et~al.}(2017)\citenamefont {Sato} \emph
  {et~al.}}]{Sato:2017dkf}%
  \BibitemOpen
  \bibfield  {author} {\bibinfo {author} {\bibfnamefont {S.}~\bibnamefont
  {Sato}} \emph {et~al.},\ }\href {\doibase 10.1088/1742-6596/840/1/012010}
  {\bibfield  {journal} {\bibinfo  {journal} {J. Phys. Conf. Ser.}\ }\textbf
  {\bibinfo {volume} {840}},\ \bibinfo {pages} {012010} (\bibinfo {year}
  {2017})}\BibitemShut {NoStop}%
\bibitem [{\citenamefont {Kawamura}(2019)}]{Kawamura:2019jqt}%
  \BibitemOpen
  \bibfield  {author} {\bibinfo {author} {\bibfnamefont {S.}~\bibnamefont
  {Kawamura}} (\bibinfo {collaboration} {DECIGO working group}),\ }\href
  {\doibase 10.22323/1.356.0019} {\bibfield  {journal} {\bibinfo  {journal}
  {PoS}\ }\textbf {\bibinfo {volume} {KMI2019}},\ \bibinfo {pages} {019}
  (\bibinfo {year} {2019})}\BibitemShut {NoStop}%
\bibitem [{\citenamefont {Amaro-Seoane}\ \emph {et~al.}(2013)\citenamefont
  {Amaro-Seoane} \emph {et~al.}}]{AmaroSeoane:2012km}%
  \BibitemOpen
  \bibfield  {author} {\bibinfo {author} {\bibfnamefont {P.}~\bibnamefont
  {Amaro-Seoane}} \emph {et~al.},\ }\href@noop {} {\bibfield  {journal}
  {\bibinfo  {journal} {GW Notes}\ }\textbf {\bibinfo {volume} {6}},\ \bibinfo
  {pages} {4} (\bibinfo {year} {2013})},\ \Eprint
  {http://arxiv.org/abs/1201.3621} {arXiv:1201.3621 [astro-ph.CO]} \BibitemShut
  {NoStop}%
\bibitem [{\citenamefont {Amaro-Seoane}\ \emph {et~al.}(2017)\citenamefont
  {Amaro-Seoane} \emph {et~al.}}]{Audley:2017drz}%
  \BibitemOpen
  \bibfield  {author} {\bibinfo {author} {\bibfnamefont {P.}~\bibnamefont
  {Amaro-Seoane}} \emph {et~al.} (\bibinfo {collaboration} {LISA}),\
  }\href@noop {} {\  (\bibinfo {year} {2017})},\ \Eprint
  {http://arxiv.org/abs/1702.00786} {arXiv:1702.00786 [astro-ph.IM]}
  \BibitemShut {NoStop}%
\bibitem [{\citenamefont {Barausse}\ \emph {et~al.}(2020)\citenamefont
  {Barausse} \emph {et~al.}}]{Barausse:2020rsu}%
  \BibitemOpen
  \bibfield  {author} {\bibinfo {author} {\bibfnamefont {E.}~\bibnamefont
  {Barausse}} \emph {et~al.},\ }\href {\doibase 10.1007/s10714-020-02691-1}
  {\bibfield  {journal} {\bibinfo  {journal} {Gen. Rel. Grav.}\ }\textbf
  {\bibinfo {volume} {52}},\ \bibinfo {pages} {81} (\bibinfo {year} {2020})},\
  \Eprint {http://arxiv.org/abs/2001.09793} {arXiv:2001.09793 [gr-qc]}
  \BibitemShut {NoStop}%
\bibitem [{\citenamefont {Janssen}\ \emph {et~al.}(2015)\citenamefont {Janssen}
  \emph {et~al.}}]{Janssen:2014dka}%
  \BibitemOpen
  \bibfield  {author} {\bibinfo {author} {\bibfnamefont {G.}~\bibnamefont
  {Janssen}} \emph {et~al.},\ }\href {\doibase 10.22323/1.215.0037} {\bibfield
  {journal} {\bibinfo  {journal} {PoS}\ }\textbf {\bibinfo {volume}
  {AASKA14}},\ \bibinfo {pages} {037} (\bibinfo {year} {2015})},\ \Eprint
  {http://arxiv.org/abs/1501.00127} {arXiv:1501.00127 [astro-ph.IM]}
  \BibitemShut {NoStop}%
\bibitem [{\citenamefont {Easther}\ and\ \citenamefont
  {Lim}(2006)}]{Easther:2006gt}%
  \BibitemOpen
  \bibfield  {author} {\bibinfo {author} {\bibfnamefont {R.}~\bibnamefont
  {Easther}}\ and\ \bibinfo {author} {\bibfnamefont {E.~A.}\ \bibnamefont
  {Lim}},\ }\href {\doibase 10.1088/1475-7516/2006/04/010} {\bibfield
  {journal} {\bibinfo  {journal} {JCAP}\ }\textbf {\bibinfo {volume} {04}},\
  \bibinfo {pages} {010} (\bibinfo {year} {2006})},\ \Eprint
  {http://arxiv.org/abs/astro-ph/0601617} {arXiv:astro-ph/0601617} \BibitemShut
  {NoStop}%
\bibitem [{\citenamefont {Ananda}\ \emph {et~al.}(2007)\citenamefont {Ananda},
  \citenamefont {Clarkson},\ and\ \citenamefont {Wands}}]{Ananda:2006af}%
  \BibitemOpen
  \bibfield  {author} {\bibinfo {author} {\bibfnamefont {K.~N.}\ \bibnamefont
  {Ananda}}, \bibinfo {author} {\bibfnamefont {C.}~\bibnamefont {Clarkson}}, \
  and\ \bibinfo {author} {\bibfnamefont {D.}~\bibnamefont {Wands}},\ }\href
  {\doibase 10.1103/PhysRevD.75.123518} {\bibfield  {journal} {\bibinfo
  {journal} {Phys. Rev.}\ }\textbf {\bibinfo {volume} {D75}},\ \bibinfo {pages}
  {123518} (\bibinfo {year} {2007})},\ \Eprint
  {http://arxiv.org/abs/gr-qc/0612013} {arXiv:gr-qc/0612013 [gr-qc]}
  \BibitemShut {NoStop}%
\bibitem [{\citenamefont {Baumann}\ \emph {et~al.}(2007)\citenamefont
  {Baumann}, \citenamefont {Steinhardt}, \citenamefont {Takahashi},\ and\
  \citenamefont {Ichiki}}]{Baumann:2007zm}%
  \BibitemOpen
  \bibfield  {author} {\bibinfo {author} {\bibfnamefont {D.}~\bibnamefont
  {Baumann}}, \bibinfo {author} {\bibfnamefont {P.~J.}\ \bibnamefont
  {Steinhardt}}, \bibinfo {author} {\bibfnamefont {K.}~\bibnamefont
  {Takahashi}}, \ and\ \bibinfo {author} {\bibfnamefont {K.}~\bibnamefont
  {Ichiki}},\ }\href {\doibase 10.1103/PhysRevD.76.084019} {\bibfield
  {journal} {\bibinfo  {journal} {Phys. Rev.}\ }\textbf {\bibinfo {volume}
  {D76}},\ \bibinfo {pages} {084019} (\bibinfo {year} {2007})},\ \Eprint
  {http://arxiv.org/abs/hep-th/0703290} {arXiv:hep-th/0703290 [hep-th]}
  \BibitemShut {NoStop}%
\bibitem [{\citenamefont {Saito}\ and\ \citenamefont
  {Yokoyama}(2009)}]{Saito:2008jc}%
  \BibitemOpen
  \bibfield  {author} {\bibinfo {author} {\bibfnamefont {R.}~\bibnamefont
  {Saito}}\ and\ \bibinfo {author} {\bibfnamefont {J.}~\bibnamefont
  {Yokoyama}},\ }\href {\doibase 10.1103/PhysRevLett.102.161101} {\bibfield
  {journal} {\bibinfo  {journal} {Phys. Rev. Lett.}\ }\textbf {\bibinfo
  {volume} {102}},\ \bibinfo {pages} {161101} (\bibinfo {year} {2009})},\
  \bibinfo {note} {[Erratum: Phys.Rev.Lett. 107, 069901 (2011)]},\ \Eprint
  {http://arxiv.org/abs/0812.4339} {arXiv:0812.4339 [astro-ph]} \BibitemShut
  {NoStop}%
\bibitem [{\citenamefont {Saito}\ and\ \citenamefont
  {Yokoyama}(2010)}]{Saito:2009jt}%
  \BibitemOpen
  \bibfield  {author} {\bibinfo {author} {\bibfnamefont {R.}~\bibnamefont
  {Saito}}\ and\ \bibinfo {author} {\bibfnamefont {J.}~\bibnamefont
  {Yokoyama}},\ }\href {\doibase 10.1143/PTP.126.351} {\bibfield  {journal}
  {\bibinfo  {journal} {Prog. Theor. Phys.}\ }\textbf {\bibinfo {volume}
  {123}},\ \bibinfo {pages} {867} (\bibinfo {year} {2010})},\ \bibinfo {note}
  {[Erratum: Prog.Theor.Phys. 126, 351--352 (2011)]},\ \Eprint
  {http://arxiv.org/abs/0912.5317} {arXiv:0912.5317 [astro-ph.CO]} \BibitemShut
  {NoStop}%
\bibitem [{\citenamefont {Kuroyanagi}\ \emph {et~al.}(2018)\citenamefont
  {Kuroyanagi}, \citenamefont {Chiba},\ and\ \citenamefont
  {Takahashi}}]{Kuroyanagi:2018csn}%
  \BibitemOpen
  \bibfield  {author} {\bibinfo {author} {\bibfnamefont {S.}~\bibnamefont
  {Kuroyanagi}}, \bibinfo {author} {\bibfnamefont {T.}~\bibnamefont {Chiba}}, \
  and\ \bibinfo {author} {\bibfnamefont {T.}~\bibnamefont {Takahashi}},\ }\href
  {\doibase 10.1088/1475-7516/2018/11/038} {\bibfield  {journal} {\bibinfo
  {journal} {JCAP}\ }\textbf {\bibinfo {volume} {11}},\ \bibinfo {pages} {038}
  (\bibinfo {year} {2018})},\ \Eprint {http://arxiv.org/abs/1807.00786}
  {arXiv:1807.00786 [astro-ph.CO]} \BibitemShut {NoStop}%
\bibitem [{\citenamefont {Espinosa}\ \emph {et~al.}(2018)\citenamefont
  {Espinosa}, \citenamefont {Racco},\ and\ \citenamefont
  {Riotto}}]{Espinosa:2018eve}%
  \BibitemOpen
  \bibfield  {author} {\bibinfo {author} {\bibfnamefont {J.~R.}\ \bibnamefont
  {Espinosa}}, \bibinfo {author} {\bibfnamefont {D.}~\bibnamefont {Racco}}, \
  and\ \bibinfo {author} {\bibfnamefont {A.}~\bibnamefont {Riotto}},\ }\href
  {\doibase 10.1088/1475-7516/2018/09/012} {\bibfield  {journal} {\bibinfo
  {journal} {JCAP}\ }\textbf {\bibinfo {volume} {09}},\ \bibinfo {pages} {012}
  (\bibinfo {year} {2018})},\ \Eprint {http://arxiv.org/abs/1804.07732}
  {arXiv:1804.07732 [hep-ph]} \BibitemShut {NoStop}%
\bibitem [{\citenamefont {Kohri}\ and\ \citenamefont
  {Terada}(2018)}]{Kohri:2018awv}%
  \BibitemOpen
  \bibfield  {author} {\bibinfo {author} {\bibfnamefont {K.}~\bibnamefont
  {Kohri}}\ and\ \bibinfo {author} {\bibfnamefont {T.}~\bibnamefont {Terada}},\
  }\href {\doibase 10.1103/PhysRevD.97.123532} {\bibfield  {journal} {\bibinfo
  {journal} {Phys. Rev. D}\ }\textbf {\bibinfo {volume} {97}},\ \bibinfo
  {pages} {123532} (\bibinfo {year} {2018})},\ \Eprint
  {http://arxiv.org/abs/1804.08577} {arXiv:1804.08577 [gr-qc]} \BibitemShut
  {NoStop}%
\bibitem [{\citenamefont {Inomata}\ \emph
  {et~al.}(2019{\natexlab{a}})\citenamefont {Inomata}, \citenamefont {Kohri},
  \citenamefont {Nakama},\ and\ \citenamefont {Terada}}]{Inomata:2019zqy}%
  \BibitemOpen
  \bibfield  {author} {\bibinfo {author} {\bibfnamefont {K.}~\bibnamefont
  {Inomata}}, \bibinfo {author} {\bibfnamefont {K.}~\bibnamefont {Kohri}},
  \bibinfo {author} {\bibfnamefont {T.}~\bibnamefont {Nakama}}, \ and\ \bibinfo
  {author} {\bibfnamefont {T.}~\bibnamefont {Terada}},\ }\href {\doibase
  10.1088/1475-7516/2019/10/071} {\bibfield  {journal} {\bibinfo  {journal}
  {JCAP}\ }\textbf {\bibinfo {volume} {10}},\ \bibinfo {pages} {071} (\bibinfo
  {year} {2019}{\natexlab{a}})},\ \Eprint {http://arxiv.org/abs/1904.12878}
  {arXiv:1904.12878 [astro-ph.CO]} \BibitemShut {NoStop}%
\bibitem [{\citenamefont {Inomata}\ \emph
  {et~al.}(2019{\natexlab{b}})\citenamefont {Inomata}, \citenamefont {Kohri},
  \citenamefont {Nakama},\ and\ \citenamefont {Terada}}]{Inomata:2019ivs}%
  \BibitemOpen
  \bibfield  {author} {\bibinfo {author} {\bibfnamefont {K.}~\bibnamefont
  {Inomata}}, \bibinfo {author} {\bibfnamefont {K.}~\bibnamefont {Kohri}},
  \bibinfo {author} {\bibfnamefont {T.}~\bibnamefont {Nakama}}, \ and\ \bibinfo
  {author} {\bibfnamefont {T.}~\bibnamefont {Terada}},\ }\href {\doibase
  10.1103/PhysRevD.100.043532} {\bibfield  {journal} {\bibinfo  {journal}
  {Phys. Rev. D}\ }\textbf {\bibinfo {volume} {100}},\ \bibinfo {pages}
  {043532} (\bibinfo {year} {2019}{\natexlab{b}})},\ \Eprint
  {http://arxiv.org/abs/1904.12879} {arXiv:1904.12879 [astro-ph.CO]}
  \BibitemShut {NoStop}%
\bibitem [{\citenamefont {Braglia}\ \emph {et~al.}(2020)\citenamefont
  {Braglia}, \citenamefont {Hazra}, \citenamefont {Finelli}, \citenamefont
  {Smoot}, \citenamefont {Sriramkumar},\ and\ \citenamefont
  {Starobinsky}}]{Braglia:2020eai}%
  \BibitemOpen
  \bibfield  {author} {\bibinfo {author} {\bibfnamefont {M.}~\bibnamefont
  {Braglia}}, \bibinfo {author} {\bibfnamefont {D.~K.}\ \bibnamefont {Hazra}},
  \bibinfo {author} {\bibfnamefont {F.}~\bibnamefont {Finelli}}, \bibinfo
  {author} {\bibfnamefont {G.~F.}\ \bibnamefont {Smoot}}, \bibinfo {author}
  {\bibfnamefont {L.}~\bibnamefont {Sriramkumar}}, \ and\ \bibinfo {author}
  {\bibfnamefont {A.~A.}\ \bibnamefont {Starobinsky}},\ }\href {\doibase
  10.1088/1475-7516/2020/08/001} {\bibfield  {journal} {\bibinfo  {journal}
  {JCAP}\ }\textbf {\bibinfo {volume} {08}},\ \bibinfo {pages} {001} (\bibinfo
  {year} {2020})},\ \Eprint {http://arxiv.org/abs/2005.02895} {arXiv:2005.02895
  [astro-ph.CO]} \BibitemShut {NoStop}%
\bibitem [{\citenamefont {Ragavendra}\ \emph
  {et~al.}(2021{\natexlab{a}})\citenamefont {Ragavendra}, \citenamefont {Saha},
  \citenamefont {Sriramkumar},\ and\ \citenamefont
  {Silk}}]{Ragavendra:2020sop}%
  \BibitemOpen
  \bibfield  {author} {\bibinfo {author} {\bibfnamefont {H.~V.}\ \bibnamefont
  {Ragavendra}}, \bibinfo {author} {\bibfnamefont {P.}~\bibnamefont {Saha}},
  \bibinfo {author} {\bibfnamefont {L.}~\bibnamefont {Sriramkumar}}, \ and\
  \bibinfo {author} {\bibfnamefont {J.}~\bibnamefont {Silk}},\ }\href {\doibase
  10.1103/PhysRevD.103.083510} {\bibfield  {journal} {\bibinfo  {journal}
  {Phys. Rev. D}\ }\textbf {\bibinfo {volume} {103}},\ \bibinfo {pages}
  {083510} (\bibinfo {year} {2021}{\natexlab{a}})},\ \Eprint
  {http://arxiv.org/abs/2008.12202} {arXiv:2008.12202 [astro-ph.CO]}
  \BibitemShut {NoStop}%
\bibitem [{\citenamefont {Ragavendra}\ \emph
  {et~al.}(2021{\natexlab{b}})\citenamefont {Ragavendra}, \citenamefont
  {Sriramkumar},\ and\ \citenamefont {Silk}}]{Ragavendra:2020vud}%
  \BibitemOpen
  \bibfield  {author} {\bibinfo {author} {\bibfnamefont {H.~V.}\ \bibnamefont
  {Ragavendra}}, \bibinfo {author} {\bibfnamefont {L.}~\bibnamefont
  {Sriramkumar}}, \ and\ \bibinfo {author} {\bibfnamefont {J.}~\bibnamefont
  {Silk}},\ }\href {\doibase 10.1088/1475-7516/2021/05/010} {\bibfield
  {journal} {\bibinfo  {journal} {JCAP}\ }\textbf {\bibinfo {volume} {05}},\
  \bibinfo {pages} {010} (\bibinfo {year} {2021}{\natexlab{b}})},\ \Eprint
  {http://arxiv.org/abs/2011.09938} {arXiv:2011.09938 [astro-ph.CO]}
  \BibitemShut {NoStop}%
\bibitem [{\citenamefont {Bhattacharya}\ \emph {et~al.}(2020)\citenamefont
  {Bhattacharya}, \citenamefont {Mohanty},\ and\ \citenamefont
  {Parashari}}]{Bhattacharya:2020lhc}%
  \BibitemOpen
  \bibfield  {author} {\bibinfo {author} {\bibfnamefont {S.}~\bibnamefont
  {Bhattacharya}}, \bibinfo {author} {\bibfnamefont {S.}~\bibnamefont
  {Mohanty}}, \ and\ \bibinfo {author} {\bibfnamefont {P.}~\bibnamefont
  {Parashari}},\ }\href@noop {} {\  (\bibinfo {year} {2020})},\ \Eprint
  {http://arxiv.org/abs/2010.05071} {arXiv:2010.05071 [astro-ph.CO]}
  \BibitemShut {NoStop}%
\bibitem [{\citenamefont {Akrami}\ \emph {et~al.}(2020)\citenamefont {Akrami}
  \emph {et~al.}}]{Akrami:2018odb}%
  \BibitemOpen
  \bibfield  {author} {\bibinfo {author} {\bibfnamefont {Y.}~\bibnamefont
  {Akrami}} \emph {et~al.} (\bibinfo {collaboration} {Planck}),\ }\href
  {\doibase 10.1051/0004-6361/201833887} {\bibfield  {journal} {\bibinfo
  {journal} {Astron. Astrophys.}\ }\textbf {\bibinfo {volume} {641}},\ \bibinfo
  {pages} {A10} (\bibinfo {year} {2020})},\ \Eprint
  {http://arxiv.org/abs/1807.06211} {arXiv:1807.06211 [astro-ph.CO]}
  \BibitemShut {NoStop}%
\bibitem [{\citenamefont {Aubourg}\ \emph {et~al.}(2015)\citenamefont {Aubourg}
  \emph {et~al.}}]{Aubourg:2014yra}%
  \BibitemOpen
  \bibfield  {author} {\bibinfo {author} {\bibfnamefont {E.}~\bibnamefont
  {Aubourg}} \emph {et~al.},\ }\href {\doibase 10.1103/PhysRevD.92.123516}
  {\bibfield  {journal} {\bibinfo  {journal} {Phys. Rev. D}\ }\textbf {\bibinfo
  {volume} {92}},\ \bibinfo {pages} {123516} (\bibinfo {year} {2015})},\
  \Eprint {http://arxiv.org/abs/1411.1074} {arXiv:1411.1074 [astro-ph.CO]}
  \BibitemShut {NoStop}%
\bibitem [{\citenamefont {V\'azquez}\ \emph {et~al.}(2018)\citenamefont
  {V\'azquez}, \citenamefont {Padilla},\ and\ \citenamefont
  {Matos}}]{Vazquez:2018qdg}%
  \BibitemOpen
  \bibfield  {author} {\bibinfo {author} {\bibfnamefont {J.~A.}\ \bibnamefont
  {V\'azquez}}, \bibinfo {author} {\bibfnamefont {L.~E.}\ \bibnamefont
  {Padilla}}, \ and\ \bibinfo {author} {\bibfnamefont {T.}~\bibnamefont
  {Matos}},\ }\href {\doibase 10.31349/RevMexFisE.17.73} {\  (\bibinfo {year}
  {2018}),\ 10.31349/RevMexFisE.17.73},\ \Eprint
  {http://arxiv.org/abs/1810.09934} {arXiv:1810.09934 [astro-ph.CO]}
  \BibitemShut {NoStop}%
\bibitem [{\citenamefont {Turner}\ \emph {et~al.}(1993)\citenamefont {Turner},
  \citenamefont {White},\ and\ \citenamefont {Lidsey}}]{Turner:1993vb}%
  \BibitemOpen
  \bibfield  {author} {\bibinfo {author} {\bibfnamefont {M.~S.}\ \bibnamefont
  {Turner}}, \bibinfo {author} {\bibfnamefont {M.~J.}\ \bibnamefont {White}}, \
  and\ \bibinfo {author} {\bibfnamefont {J.~E.}\ \bibnamefont {Lidsey}},\
  }\href {\doibase 10.1103/PhysRevD.48.4613} {\bibfield  {journal} {\bibinfo
  {journal} {Phys. Rev. D}\ }\textbf {\bibinfo {volume} {48}},\ \bibinfo
  {pages} {4613} (\bibinfo {year} {1993})},\ \Eprint
  {http://arxiv.org/abs/astro-ph/9306029} {arXiv:astro-ph/9306029} \BibitemShut
  {NoStop}%
\bibitem [{\citenamefont {Boyle}\ and\ \citenamefont
  {Steinhardt}(2008)}]{Boyle:2005se}%
  \BibitemOpen
  \bibfield  {author} {\bibinfo {author} {\bibfnamefont {L.~A.}\ \bibnamefont
  {Boyle}}\ and\ \bibinfo {author} {\bibfnamefont {P.~J.}\ \bibnamefont
  {Steinhardt}},\ }\href {\doibase 10.1103/PhysRevD.77.063504} {\bibfield
  {journal} {\bibinfo  {journal} {Phys. Rev. D}\ }\textbf {\bibinfo {volume}
  {77}},\ \bibinfo {pages} {063504} (\bibinfo {year} {2008})},\ \Eprint
  {http://arxiv.org/abs/astro-ph/0512014} {arXiv:astro-ph/0512014} \BibitemShut
  {NoStop}%
\bibitem [{\citenamefont {Sa}\ and\ \citenamefont
  {Henriques}(2008)}]{Sa:2007pc}%
  \BibitemOpen
  \bibfield  {author} {\bibinfo {author} {\bibfnamefont {P.~M.}\ \bibnamefont
  {Sa}}\ and\ \bibinfo {author} {\bibfnamefont {A.~B.}\ \bibnamefont
  {Henriques}},\ }\href {\doibase 10.1103/PhysRevD.77.064002} {\bibfield
  {journal} {\bibinfo  {journal} {Phys. Rev. D}\ }\textbf {\bibinfo {volume}
  {77}},\ \bibinfo {pages} {064002} (\bibinfo {year} {2008})},\ \Eprint
  {http://arxiv.org/abs/0712.2697} {arXiv:0712.2697 [astro-ph]} \BibitemShut
  {NoStop}%
\bibitem [{\citenamefont {Nakayama}\ \emph
  {et~al.}(2008{\natexlab{a}})\citenamefont {Nakayama}, \citenamefont {Saito},
  \citenamefont {Suwa},\ and\ \citenamefont {Yokoyama}}]{Nakayama:2008wy}%
  \BibitemOpen
  \bibfield  {author} {\bibinfo {author} {\bibfnamefont {K.}~\bibnamefont
  {Nakayama}}, \bibinfo {author} {\bibfnamefont {S.}~\bibnamefont {Saito}},
  \bibinfo {author} {\bibfnamefont {Y.}~\bibnamefont {Suwa}}, \ and\ \bibinfo
  {author} {\bibfnamefont {J.}~\bibnamefont {Yokoyama}},\ }\href {\doibase
  10.1088/1475-7516/2008/06/020} {\bibfield  {journal} {\bibinfo  {journal}
  {JCAP}\ }\textbf {\bibinfo {volume} {06}},\ \bibinfo {pages} {020} (\bibinfo
  {year} {2008}{\natexlab{a}})},\ \Eprint {http://arxiv.org/abs/0804.1827}
  {arXiv:0804.1827 [astro-ph]} \BibitemShut {NoStop}%
\bibitem [{\citenamefont {Nakayama}\ \emph
  {et~al.}(2008{\natexlab{b}})\citenamefont {Nakayama}, \citenamefont {Saito},
  \citenamefont {Suwa},\ and\ \citenamefont {Yokoyama}}]{Nakayama:2008ip}%
  \BibitemOpen
  \bibfield  {author} {\bibinfo {author} {\bibfnamefont {K.}~\bibnamefont
  {Nakayama}}, \bibinfo {author} {\bibfnamefont {S.}~\bibnamefont {Saito}},
  \bibinfo {author} {\bibfnamefont {Y.}~\bibnamefont {Suwa}}, \ and\ \bibinfo
  {author} {\bibfnamefont {J.}~\bibnamefont {Yokoyama}},\ }\href {\doibase
  10.1103/PhysRevD.77.124001} {\bibfield  {journal} {\bibinfo  {journal} {Phys.
  Rev. D}\ }\textbf {\bibinfo {volume} {77}},\ \bibinfo {pages} {124001}
  (\bibinfo {year} {2008}{\natexlab{b}})},\ \Eprint
  {http://arxiv.org/abs/0802.2452} {arXiv:0802.2452 [hep-ph]} \BibitemShut
  {NoStop}%
\bibitem [{\citenamefont {Sa}\ and\ \citenamefont
  {Henriques}(2010)}]{Sa:2010qw}%
  \BibitemOpen
  \bibfield  {author} {\bibinfo {author} {\bibfnamefont {P.~M.}\ \bibnamefont
  {Sa}}\ and\ \bibinfo {author} {\bibfnamefont {A.~B.}\ \bibnamefont
  {Henriques}},\ }\href {\doibase 10.1103/PhysRevD.81.124043} {\bibfield
  {journal} {\bibinfo  {journal} {Phys. Rev. D}\ }\textbf {\bibinfo {volume}
  {81}},\ \bibinfo {pages} {124043} (\bibinfo {year} {2010})},\ \Eprint
  {http://arxiv.org/abs/1003.4112} {arXiv:1003.4112 [gr-qc]} \BibitemShut
  {NoStop}%
\bibitem [{\citenamefont {Kuroyanagi}\ \emph {et~al.}(2011)\citenamefont
  {Kuroyanagi}, \citenamefont {Nakayama},\ and\ \citenamefont
  {Saito}}]{Kuroyanagi:2011fy}%
  \BibitemOpen
  \bibfield  {author} {\bibinfo {author} {\bibfnamefont {S.}~\bibnamefont
  {Kuroyanagi}}, \bibinfo {author} {\bibfnamefont {K.}~\bibnamefont
  {Nakayama}}, \ and\ \bibinfo {author} {\bibfnamefont {S.}~\bibnamefont
  {Saito}},\ }\href {\doibase 10.1103/PhysRevD.84.123513} {\bibfield  {journal}
  {\bibinfo  {journal} {Phys. Rev. D}\ }\textbf {\bibinfo {volume} {84}},\
  \bibinfo {pages} {123513} (\bibinfo {year} {2011})},\ \Eprint
  {http://arxiv.org/abs/1110.4169} {arXiv:1110.4169 [astro-ph.CO]} \BibitemShut
  {NoStop}%
\bibitem [{\citenamefont {Kuroyanagi}\ \emph {et~al.}(2015)\citenamefont
  {Kuroyanagi}, \citenamefont {Takahashi},\ and\ \citenamefont
  {Yokoyama}}]{Kuroyanagi:2014nba}%
  \BibitemOpen
  \bibfield  {author} {\bibinfo {author} {\bibfnamefont {S.}~\bibnamefont
  {Kuroyanagi}}, \bibinfo {author} {\bibfnamefont {T.}~\bibnamefont
  {Takahashi}}, \ and\ \bibinfo {author} {\bibfnamefont {S.}~\bibnamefont
  {Yokoyama}},\ }\href {\doibase 10.1088/1475-7516/2015/02/003} {\bibfield
  {journal} {\bibinfo  {journal} {JCAP}\ }\textbf {\bibinfo {volume} {02}},\
  \bibinfo {pages} {003} (\bibinfo {year} {2015})},\ \Eprint
  {http://arxiv.org/abs/1407.4785} {arXiv:1407.4785 [astro-ph.CO]} \BibitemShut
  {NoStop}%
\bibitem [{\citenamefont {Assadullahi}\ and\ \citenamefont
  {Wands}(2009)}]{Assadullahi:2009nf}%
  \BibitemOpen
  \bibfield  {author} {\bibinfo {author} {\bibfnamefont {H.}~\bibnamefont
  {Assadullahi}}\ and\ \bibinfo {author} {\bibfnamefont {D.}~\bibnamefont
  {Wands}},\ }\href {\doibase 10.1103/PhysRevD.79.083511} {\bibfield  {journal}
  {\bibinfo  {journal} {Phys. Rev. D}\ }\textbf {\bibinfo {volume} {79}},\
  \bibinfo {pages} {083511} (\bibinfo {year} {2009})},\ \Eprint
  {http://arxiv.org/abs/0901.0989} {arXiv:0901.0989 [astro-ph.CO]} \BibitemShut
  {NoStop}%
\bibitem [{\citenamefont {Nakayama}\ and\ \citenamefont
  {Yokoyama}(2010)}]{Nakayama:2009ce}%
  \BibitemOpen
  \bibfield  {author} {\bibinfo {author} {\bibfnamefont {K.}~\bibnamefont
  {Nakayama}}\ and\ \bibinfo {author} {\bibfnamefont {J.}~\bibnamefont
  {Yokoyama}},\ }\href {\doibase 10.1088/1475-7516/2010/01/010} {\bibfield
  {journal} {\bibinfo  {journal} {JCAP}\ }\textbf {\bibinfo {volume} {01}},\
  \bibinfo {pages} {010} (\bibinfo {year} {2010})},\ \Eprint
  {http://arxiv.org/abs/0910.0715} {arXiv:0910.0715 [astro-ph.CO]} \BibitemShut
  {NoStop}%
\bibitem [{\citenamefont {Durrer}\ and\ \citenamefont
  {Hasenkamp}(2011)}]{Durrer:2011bi}%
  \BibitemOpen
  \bibfield  {author} {\bibinfo {author} {\bibfnamefont {R.}~\bibnamefont
  {Durrer}}\ and\ \bibinfo {author} {\bibfnamefont {J.}~\bibnamefont
  {Hasenkamp}},\ }\href {\doibase 10.1103/PhysRevD.84.064027} {\bibfield
  {journal} {\bibinfo  {journal} {Phys. Rev. D}\ }\textbf {\bibinfo {volume}
  {84}},\ \bibinfo {pages} {064027} (\bibinfo {year} {2011})},\ \Eprint
  {http://arxiv.org/abs/1105.5283} {arXiv:1105.5283 [gr-qc]} \BibitemShut
  {NoStop}%
\bibitem [{\citenamefont {Alabidi}\ \emph {et~al.}(2013)\citenamefont
  {Alabidi}, \citenamefont {Kohri}, \citenamefont {Sasaki},\ and\ \citenamefont
  {Sendouda}}]{Alabidi:2013lya}%
  \BibitemOpen
  \bibfield  {author} {\bibinfo {author} {\bibfnamefont {L.}~\bibnamefont
  {Alabidi}}, \bibinfo {author} {\bibfnamefont {K.}~\bibnamefont {Kohri}},
  \bibinfo {author} {\bibfnamefont {M.}~\bibnamefont {Sasaki}}, \ and\ \bibinfo
  {author} {\bibfnamefont {Y.}~\bibnamefont {Sendouda}},\ }\href {\doibase
  10.1088/1475-7516/2013/05/033} {\bibfield  {journal} {\bibinfo  {journal}
  {JCAP}\ }\textbf {\bibinfo {volume} {05}},\ \bibinfo {pages} {033} (\bibinfo
  {year} {2013})},\ \Eprint {http://arxiv.org/abs/1303.4519} {arXiv:1303.4519
  [astro-ph.CO]} \BibitemShut {NoStop}%
\bibitem [{\citenamefont {D'Eramo}\ and\ \citenamefont
  {Schmitz}(2019)}]{DEramo:2019tit}%
  \BibitemOpen
  \bibfield  {author} {\bibinfo {author} {\bibfnamefont {F.}~\bibnamefont
  {D'Eramo}}\ and\ \bibinfo {author} {\bibfnamefont {K.}~\bibnamefont
  {Schmitz}},\ }\href {\doibase 10.1103/PhysRevResearch.1.013010} {\bibfield
  {journal} {\bibinfo  {journal} {Phys. Rev. Research.}\ }\textbf {\bibinfo
  {volume} {1}},\ \bibinfo {pages} {013010} (\bibinfo {year} {2019})},\ \Eprint
  {http://arxiv.org/abs/1904.07870} {arXiv:1904.07870 [hep-ph]} \BibitemShut
  {NoStop}%
\bibitem [{\citenamefont {Ricciardone}\ and\ \citenamefont
  {Tasinato}(2017)}]{Ricciardone:2016lym}%
  \BibitemOpen
  \bibfield  {author} {\bibinfo {author} {\bibfnamefont {A.}~\bibnamefont
  {Ricciardone}}\ and\ \bibinfo {author} {\bibfnamefont {G.}~\bibnamefont
  {Tasinato}},\ }\href {\doibase 10.1103/PhysRevD.96.023508} {\bibfield
  {journal} {\bibinfo  {journal} {Phys. Rev. D}\ }\textbf {\bibinfo {volume}
  {96}},\ \bibinfo {pages} {023508} (\bibinfo {year} {2017})},\ \Eprint
  {http://arxiv.org/abs/1611.04516} {arXiv:1611.04516 [astro-ph.CO]}
  \BibitemShut {NoStop}%
\bibitem [{\citenamefont {Koh}\ \emph {et~al.}(2018)\citenamefont {Koh},
  \citenamefont {Lee},\ and\ \citenamefont {Tumurtushaa}}]{Koh:2018qcy}%
  \BibitemOpen
  \bibfield  {author} {\bibinfo {author} {\bibfnamefont {S.}~\bibnamefont
  {Koh}}, \bibinfo {author} {\bibfnamefont {B.-H.}\ \bibnamefont {Lee}}, \ and\
  \bibinfo {author} {\bibfnamefont {G.}~\bibnamefont {Tumurtushaa}},\ }\href
  {\doibase 10.1103/PhysRevD.98.103511} {\bibfield  {journal} {\bibinfo
  {journal} {Phys. Rev. D}\ }\textbf {\bibinfo {volume} {98}},\ \bibinfo
  {pages} {103511} (\bibinfo {year} {2018})},\ \Eprint
  {http://arxiv.org/abs/1807.04424} {arXiv:1807.04424 [astro-ph.CO]}
  \BibitemShut {NoStop}%
\bibitem [{\citenamefont {Fujita}\ \emph {et~al.}(2019)\citenamefont {Fujita},
  \citenamefont {Kuroyanagi}, \citenamefont {Mizuno},\ and\ \citenamefont
  {Mukohyama}}]{Fujita:2018ehq}%
  \BibitemOpen
  \bibfield  {author} {\bibinfo {author} {\bibfnamefont {T.}~\bibnamefont
  {Fujita}}, \bibinfo {author} {\bibfnamefont {S.}~\bibnamefont {Kuroyanagi}},
  \bibinfo {author} {\bibfnamefont {S.}~\bibnamefont {Mizuno}}, \ and\ \bibinfo
  {author} {\bibfnamefont {S.}~\bibnamefont {Mukohyama}},\ }\href {\doibase
  10.1016/j.physletb.2018.12.025} {\bibfield  {journal} {\bibinfo  {journal}
  {Phys. Lett. B}\ }\textbf {\bibinfo {volume} {789}},\ \bibinfo {pages} {215}
  (\bibinfo {year} {2019})},\ \Eprint {http://arxiv.org/abs/1808.02381}
  {arXiv:1808.02381 [gr-qc]} \BibitemShut {NoStop}%
\bibitem [{\citenamefont {Bernal}\ and\ \citenamefont
  {Hajkarim}(2019)}]{Bernal:2019lpc}%
  \BibitemOpen
  \bibfield  {author} {\bibinfo {author} {\bibfnamefont {N.}~\bibnamefont
  {Bernal}}\ and\ \bibinfo {author} {\bibfnamefont {F.}~\bibnamefont
  {Hajkarim}},\ }\href {\doibase 10.1103/PhysRevD.100.063502} {\bibfield
  {journal} {\bibinfo  {journal} {Phys. Rev. D}\ }\textbf {\bibinfo {volume}
  {100}},\ \bibinfo {pages} {063502} (\bibinfo {year} {2019})},\ \Eprint
  {http://arxiv.org/abs/1905.10410} {arXiv:1905.10410 [astro-ph.CO]}
  \BibitemShut {NoStop}%
\bibitem [{\citenamefont {Bernal}\ \emph {et~al.}(2020)\citenamefont {Bernal},
  \citenamefont {Ghoshal}, \citenamefont {Hajkarim},\ and\ \citenamefont
  {Lambiase}}]{Bernal:2020ywq}%
  \BibitemOpen
  \bibfield  {author} {\bibinfo {author} {\bibfnamefont {N.}~\bibnamefont
  {Bernal}}, \bibinfo {author} {\bibfnamefont {A.}~\bibnamefont {Ghoshal}},
  \bibinfo {author} {\bibfnamefont {F.}~\bibnamefont {Hajkarim}}, \ and\
  \bibinfo {author} {\bibfnamefont {G.}~\bibnamefont {Lambiase}},\ }\href
  {\doibase 10.1088/1475-7516/2020/11/051} {\bibfield  {journal} {\bibinfo
  {journal} {JCAP}\ }\textbf {\bibinfo {volume} {11}},\ \bibinfo {pages} {051}
  (\bibinfo {year} {2020})},\ \Eprint {http://arxiv.org/abs/2008.04959}
  {arXiv:2008.04959 [gr-qc]} \BibitemShut {NoStop}%
\bibitem [{\citenamefont {Mishra}\ \emph {et~al.}(2021)\citenamefont {Mishra},
  \citenamefont {Sahni},\ and\ \citenamefont {Starobinsky}}]{Mishra:2021wkm}%
  \BibitemOpen
  \bibfield  {author} {\bibinfo {author} {\bibfnamefont {S.~S.}\ \bibnamefont
  {Mishra}}, \bibinfo {author} {\bibfnamefont {V.}~\bibnamefont {Sahni}}, \
  and\ \bibinfo {author} {\bibfnamefont {A.~A.}\ \bibnamefont {Starobinsky}},\
  }\href@noop {} {\  (\bibinfo {year} {2021})},\ \Eprint
  {http://arxiv.org/abs/2101.00271} {arXiv:2101.00271 [gr-qc]} \BibitemShut
  {NoStop}%
\bibitem [{\citenamefont {Weinberg}(2004)}]{Weinberg:2003ur}%
  \BibitemOpen
  \bibfield  {author} {\bibinfo {author} {\bibfnamefont {S.}~\bibnamefont
  {Weinberg}},\ }\href {\doibase 10.1103/PhysRevD.69.023503} {\bibfield
  {journal} {\bibinfo  {journal} {Phys. Rev. D}\ }\textbf {\bibinfo {volume}
  {69}},\ \bibinfo {pages} {023503} (\bibinfo {year} {2004})},\ \Eprint
  {http://arxiv.org/abs/astro-ph/0306304} {arXiv:astro-ph/0306304} \BibitemShut
  {NoStop}%
\bibitem [{\citenamefont {Mangilli}\ \emph {et~al.}(2008)\citenamefont
  {Mangilli}, \citenamefont {Bartolo}, \citenamefont {Matarrese},\ and\
  \citenamefont {Riotto}}]{Mangilli:2008bw}%
  \BibitemOpen
  \bibfield  {author} {\bibinfo {author} {\bibfnamefont {A.}~\bibnamefont
  {Mangilli}}, \bibinfo {author} {\bibfnamefont {N.}~\bibnamefont {Bartolo}},
  \bibinfo {author} {\bibfnamefont {S.}~\bibnamefont {Matarrese}}, \ and\
  \bibinfo {author} {\bibfnamefont {A.}~\bibnamefont {Riotto}},\ }\href
  {\doibase 10.1103/PhysRevD.78.083517} {\bibfield  {journal} {\bibinfo
  {journal} {Phys. Rev. D}\ }\textbf {\bibinfo {volume} {78}},\ \bibinfo
  {pages} {083517} (\bibinfo {year} {2008})},\ \Eprint
  {http://arxiv.org/abs/0805.3234} {arXiv:0805.3234 [astro-ph]} \BibitemShut
  {NoStop}%
\bibitem [{\citenamefont {Watanabe}\ and\ \citenamefont
  {Komatsu}(2006)}]{Watanabe:2006qe}%
  \BibitemOpen
  \bibfield  {author} {\bibinfo {author} {\bibfnamefont {Y.}~\bibnamefont
  {Watanabe}}\ and\ \bibinfo {author} {\bibfnamefont {E.}~\bibnamefont
  {Komatsu}},\ }\href {\doibase 10.1103/PhysRevD.73.123515} {\bibfield
  {journal} {\bibinfo  {journal} {Phys. Rev. D}\ }\textbf {\bibinfo {volume}
  {73}},\ \bibinfo {pages} {123515} (\bibinfo {year} {2006})},\ \Eprint
  {http://arxiv.org/abs/astro-ph/0604176} {arXiv:astro-ph/0604176} \BibitemShut
  {NoStop}%
\bibitem [{\citenamefont {Kuroyanagi}\ \emph {et~al.}(2009)\citenamefont
  {Kuroyanagi}, \citenamefont {Chiba},\ and\ \citenamefont
  {Sugiyama}}]{Kuroyanagi:2008ye}%
  \BibitemOpen
  \bibfield  {author} {\bibinfo {author} {\bibfnamefont {S.}~\bibnamefont
  {Kuroyanagi}}, \bibinfo {author} {\bibfnamefont {T.}~\bibnamefont {Chiba}}, \
  and\ \bibinfo {author} {\bibfnamefont {N.}~\bibnamefont {Sugiyama}},\ }\href
  {\doibase 10.1103/PhysRevD.79.103501} {\bibfield  {journal} {\bibinfo
  {journal} {Phys. Rev. D}\ }\textbf {\bibinfo {volume} {79}},\ \bibinfo
  {pages} {103501} (\bibinfo {year} {2009})},\ \Eprint
  {http://arxiv.org/abs/0804.3249} {arXiv:0804.3249 [astro-ph]} \BibitemShut
  {NoStop}%
\bibitem [{\citenamefont {Caldwell}\ \emph {et~al.}(2019)\citenamefont
  {Caldwell}, \citenamefont {Smith},\ and\ \citenamefont
  {Walker}}]{Caldwell:2018giq}%
  \BibitemOpen
  \bibfield  {author} {\bibinfo {author} {\bibfnamefont {R.~R.}\ \bibnamefont
  {Caldwell}}, \bibinfo {author} {\bibfnamefont {T.~L.}\ \bibnamefont {Smith}},
  \ and\ \bibinfo {author} {\bibfnamefont {D.~G.~E.}\ \bibnamefont {Walker}},\
  }\href {\doibase 10.1103/PhysRevD.100.043513} {\bibfield  {journal} {\bibinfo
   {journal} {Phys. Rev. D}\ }\textbf {\bibinfo {volume} {100}},\ \bibinfo
  {pages} {043513} (\bibinfo {year} {2019})},\ \Eprint
  {http://arxiv.org/abs/1812.07577} {arXiv:1812.07577 [astro-ph.CO]}
  \BibitemShut {NoStop}%
\bibitem [{\citenamefont {Martin}\ and\ \citenamefont
  {Ringeval}(2010)}]{Martin:2010kz}%
  \BibitemOpen
  \bibfield  {author} {\bibinfo {author} {\bibfnamefont {J.}~\bibnamefont
  {Martin}}\ and\ \bibinfo {author} {\bibfnamefont {C.}~\bibnamefont
  {Ringeval}},\ }\href {\doibase 10.1103/PhysRevD.82.023511} {\bibfield
  {journal} {\bibinfo  {journal} {Phys. Rev. D}\ }\textbf {\bibinfo {volume}
  {82}},\ \bibinfo {pages} {023511} (\bibinfo {year} {2010})},\ \Eprint
  {http://arxiv.org/abs/1004.5525} {arXiv:1004.5525 [astro-ph.CO]} \BibitemShut
  {NoStop}%
\bibitem [{\citenamefont {Dai}\ \emph {et~al.}(2014)\citenamefont {Dai},
  \citenamefont {Kamionkowski},\ and\ \citenamefont {Wang}}]{Dai:2014jja}%
  \BibitemOpen
  \bibfield  {author} {\bibinfo {author} {\bibfnamefont {L.}~\bibnamefont
  {Dai}}, \bibinfo {author} {\bibfnamefont {M.}~\bibnamefont {Kamionkowski}}, \
  and\ \bibinfo {author} {\bibfnamefont {J.}~\bibnamefont {Wang}},\ }\href
  {\doibase 10.1103/PhysRevLett.113.041302} {\bibfield  {journal} {\bibinfo
  {journal} {Phys. Rev. Lett.}\ }\textbf {\bibinfo {volume} {113}},\ \bibinfo
  {pages} {041302} (\bibinfo {year} {2014})},\ \Eprint
  {http://arxiv.org/abs/1404.6704} {arXiv:1404.6704 [astro-ph.CO]} \BibitemShut
  {NoStop}%
\bibitem [{\citenamefont {Cook}\ \emph {et~al.}(2015)\citenamefont {Cook},
  \citenamefont {Dimastrogiovanni}, \citenamefont {Easson},\ and\ \citenamefont
  {Krauss}}]{Cook:2015vqa}%
  \BibitemOpen
  \bibfield  {author} {\bibinfo {author} {\bibfnamefont {J.~L.}\ \bibnamefont
  {Cook}}, \bibinfo {author} {\bibfnamefont {E.}~\bibnamefont
  {Dimastrogiovanni}}, \bibinfo {author} {\bibfnamefont {D.~A.}\ \bibnamefont
  {Easson}}, \ and\ \bibinfo {author} {\bibfnamefont {L.~M.}\ \bibnamefont
  {Krauss}},\ }\href {\doibase 10.1088/1475-7516/2015/04/047} {\bibfield
  {journal} {\bibinfo  {journal} {JCAP}\ }\textbf {\bibinfo {volume} {04}},\
  \bibinfo {pages} {047} (\bibinfo {year} {2015})},\ \Eprint
  {http://arxiv.org/abs/1502.04673} {arXiv:1502.04673 [astro-ph.CO]}
  \BibitemShut {NoStop}%
\bibitem [{\citenamefont {Chung}\ \emph {et~al.}(1999)\citenamefont {Chung},
  \citenamefont {Kolb},\ and\ \citenamefont {Riotto}}]{Chung:1998rq}%
  \BibitemOpen
  \bibfield  {author} {\bibinfo {author} {\bibfnamefont {D.~J.~H.}\
  \bibnamefont {Chung}}, \bibinfo {author} {\bibfnamefont {E.~W.}\ \bibnamefont
  {Kolb}}, \ and\ \bibinfo {author} {\bibfnamefont {A.}~\bibnamefont
  {Riotto}},\ }\href {\doibase 10.1103/PhysRevD.60.063504} {\bibfield
  {journal} {\bibinfo  {journal} {Phys. Rev. D}\ }\textbf {\bibinfo {volume}
  {60}},\ \bibinfo {pages} {063504} (\bibinfo {year} {1999})},\ \Eprint
  {http://arxiv.org/abs/hep-ph/9809453} {arXiv:hep-ph/9809453} \BibitemShut
  {NoStop}%
\bibitem [{\citenamefont {Giudice}\ \emph {et~al.}(2001)\citenamefont
  {Giudice}, \citenamefont {Kolb},\ and\ \citenamefont
  {Riotto}}]{Giudice:2000ex}%
  \BibitemOpen
  \bibfield  {author} {\bibinfo {author} {\bibfnamefont {G.~F.}\ \bibnamefont
  {Giudice}}, \bibinfo {author} {\bibfnamefont {E.~W.}\ \bibnamefont {Kolb}}, \
  and\ \bibinfo {author} {\bibfnamefont {A.}~\bibnamefont {Riotto}},\ }\href
  {\doibase 10.1103/PhysRevD.64.023508} {\bibfield  {journal} {\bibinfo
  {journal} {Phys. Rev. D}\ }\textbf {\bibinfo {volume} {64}},\ \bibinfo
  {pages} {023508} (\bibinfo {year} {2001})},\ \Eprint
  {http://arxiv.org/abs/hep-ph/0005123} {arXiv:hep-ph/0005123} \BibitemShut
  {NoStop}%
\bibitem [{\citenamefont {Maity}\ and\ \citenamefont
  {Saha}(2018)}]{Maity:2018dgy}%
  \BibitemOpen
  \bibfield  {author} {\bibinfo {author} {\bibfnamefont {D.}~\bibnamefont
  {Maity}}\ and\ \bibinfo {author} {\bibfnamefont {P.}~\bibnamefont {Saha}},\
  }\href {\doibase 10.1103/PhysRevD.98.103525} {\bibfield  {journal} {\bibinfo
  {journal} {Phys. Rev. D}\ }\textbf {\bibinfo {volume} {98}},\ \bibinfo
  {pages} {103525} (\bibinfo {year} {2018})},\ \Eprint
  {http://arxiv.org/abs/1801.03059} {arXiv:1801.03059 [hep-ph]} \BibitemShut
  {NoStop}%
\bibitem [{\citenamefont {Haque}\ and\ \citenamefont
  {Maity}(2019)}]{Haque:2019prw}%
  \BibitemOpen
  \bibfield  {author} {\bibinfo {author} {\bibfnamefont {M.~R.}\ \bibnamefont
  {Haque}}\ and\ \bibinfo {author} {\bibfnamefont {D.}~\bibnamefont {Maity}},\
  }\href {\doibase 10.1103/PhysRevD.99.103534} {\bibfield  {journal} {\bibinfo
  {journal} {Phys. Rev. D}\ }\textbf {\bibinfo {volume} {99}},\ \bibinfo
  {pages} {103534} (\bibinfo {year} {2019})},\ \Eprint
  {http://arxiv.org/abs/1902.09491} {arXiv:1902.09491 [hep-th]} \BibitemShut
  {NoStop}%
\bibitem [{\citenamefont {Haque}\ \emph {et~al.}(2020)\citenamefont {Haque},
  \citenamefont {Maity},\ and\ \citenamefont {Saha}}]{Haque:2020zco}%
  \BibitemOpen
  \bibfield  {author} {\bibinfo {author} {\bibfnamefont {M.~R.}\ \bibnamefont
  {Haque}}, \bibinfo {author} {\bibfnamefont {D.}~\bibnamefont {Maity}}, \ and\
  \bibinfo {author} {\bibfnamefont {P.}~\bibnamefont {Saha}},\ }\href {\doibase
  10.1103/PhysRevD.102.083534} {\bibfield  {journal} {\bibinfo  {journal}
  {Phys. Rev. D}\ }\textbf {\bibinfo {volume} {102}},\ \bibinfo {pages}
  {083534} (\bibinfo {year} {2020})},\ \Eprint
  {http://arxiv.org/abs/2009.02794} {arXiv:2009.02794 [hep-th]} \BibitemShut
  {NoStop}%
\bibitem [{\citenamefont {Arzoumanian}\ \emph {et~al.}(2020)\citenamefont
  {Arzoumanian} \emph {et~al.}}]{Arzoumanian:2020vkk}%
  \BibitemOpen
  \bibfield  {author} {\bibinfo {author} {\bibfnamefont {Z.}~\bibnamefont
  {Arzoumanian}} \emph {et~al.} (\bibinfo {collaboration} {NANOGrav}),\ }\href
  {\doibase 10.3847/2041-8213/abd401} {\bibfield  {journal} {\bibinfo
  {journal} {Astrophys. J. Lett.}\ }\textbf {\bibinfo {volume} {905}},\
  \bibinfo {pages} {L34} (\bibinfo {year} {2020})},\ \Eprint
  {http://arxiv.org/abs/2009.04496} {arXiv:2009.04496 [astro-ph.HE]}
  \BibitemShut {NoStop}%
\bibitem [{\citenamefont {Pol}\ \emph {et~al.}(2020)\citenamefont {Pol} \emph
  {et~al.}}]{Pol:2020igl}%
  \BibitemOpen
  \bibfield  {author} {\bibinfo {author} {\bibfnamefont {N.~S.}\ \bibnamefont
  {Pol}} \emph {et~al.} (\bibinfo {collaboration} {NANOGrav}),\ }\href@noop {}
  {\  (\bibinfo {year} {2020})},\ \Eprint {http://arxiv.org/abs/2010.11950}
  {arXiv:2010.11950 [astro-ph.HE]} \BibitemShut {NoStop}%
\bibitem [{\citenamefont {Kuroyanagi}\ \emph {et~al.}(2021)\citenamefont
  {Kuroyanagi}, \citenamefont {Takahashi},\ and\ \citenamefont
  {Yokoyama}}]{Kuroyanagi:2020sfw}%
  \BibitemOpen
  \bibfield  {author} {\bibinfo {author} {\bibfnamefont {S.}~\bibnamefont
  {Kuroyanagi}}, \bibinfo {author} {\bibfnamefont {T.}~\bibnamefont
  {Takahashi}}, \ and\ \bibinfo {author} {\bibfnamefont {S.}~\bibnamefont
  {Yokoyama}},\ }\href {\doibase 10.1088/1475-7516/2021/01/071} {\bibfield
  {journal} {\bibinfo  {journal} {JCAP}\ }\textbf {\bibinfo {volume} {01}},\
  \bibinfo {pages} {071} (\bibinfo {year} {2021})},\ \Eprint
  {http://arxiv.org/abs/2011.03323} {arXiv:2011.03323 [astro-ph.CO]}
  \BibitemShut {NoStop}%
\bibitem [{\citenamefont {Inomata}\ \emph {et~al.}(2020)\citenamefont
  {Inomata}, \citenamefont {Kawasaki}, \citenamefont {Mukaida},\ and\
  \citenamefont {Yanagida}}]{Inomata:2020xad}%
  \BibitemOpen
  \bibfield  {author} {\bibinfo {author} {\bibfnamefont {K.}~\bibnamefont
  {Inomata}}, \bibinfo {author} {\bibfnamefont {M.}~\bibnamefont {Kawasaki}},
  \bibinfo {author} {\bibfnamefont {K.}~\bibnamefont {Mukaida}}, \ and\
  \bibinfo {author} {\bibfnamefont {T.~T.}\ \bibnamefont {Yanagida}},\
  }\href@noop {} {\  (\bibinfo {year} {2020})},\ \Eprint
  {http://arxiv.org/abs/2011.01270} {arXiv:2011.01270 [astro-ph.CO]}
  \BibitemShut {NoStop}%
\bibitem [{\citenamefont {Tahara}\ and\ \citenamefont
  {Kobayashi}(2020)}]{Tahara:2020fmn}%
  \BibitemOpen
  \bibfield  {author} {\bibinfo {author} {\bibfnamefont {H.~W.~H.}\
  \bibnamefont {Tahara}}\ and\ \bibinfo {author} {\bibfnamefont
  {T.}~\bibnamefont {Kobayashi}},\ }\href {\doibase
  10.1103/PhysRevD.102.123533} {\bibfield  {journal} {\bibinfo  {journal}
  {Phys. Rev. D}\ }\textbf {\bibinfo {volume} {102}},\ \bibinfo {pages}
  {123533} (\bibinfo {year} {2020})},\ \Eprint
  {http://arxiv.org/abs/2011.01605} {arXiv:2011.01605 [gr-qc]} \BibitemShut
  {NoStop}%
\bibitem [{\citenamefont {Kitajima}\ \emph {et~al.}(2020)\citenamefont
  {Kitajima}, \citenamefont {Soda},\ and\ \citenamefont
  {Urakawa}}]{Kitajima:2020rpm}%
  \BibitemOpen
  \bibfield  {author} {\bibinfo {author} {\bibfnamefont {N.}~\bibnamefont
  {Kitajima}}, \bibinfo {author} {\bibfnamefont {J.}~\bibnamefont {Soda}}, \
  and\ \bibinfo {author} {\bibfnamefont {Y.}~\bibnamefont {Urakawa}},\
  }\href@noop {} {\  (\bibinfo {year} {2020})},\ \Eprint
  {http://arxiv.org/abs/2010.10990} {arXiv:2010.10990 [astro-ph.CO]}
  \BibitemShut {NoStop}%
\bibitem [{\citenamefont {Dom\`enech}\ and\ \citenamefont
  {Pi}(2020)}]{Domenech:2020ers}%
  \BibitemOpen
  \bibfield  {author} {\bibinfo {author} {\bibfnamefont {G.}~\bibnamefont
  {Dom\`enech}}\ and\ \bibinfo {author} {\bibfnamefont {S.}~\bibnamefont
  {Pi}},\ }\href@noop {} {\  (\bibinfo {year} {2020})},\ \Eprint
  {http://arxiv.org/abs/2010.03976} {arXiv:2010.03976 [astro-ph.CO]}
  \BibitemShut {NoStop}%
\bibitem [{\citenamefont {Li}\ \emph {et~al.}(2021)\citenamefont {Li},
  \citenamefont {Ye},\ and\ \citenamefont {Piao}}]{Li:2020cjj}%
  \BibitemOpen
  \bibfield  {author} {\bibinfo {author} {\bibfnamefont {H.-H.}\ \bibnamefont
  {Li}}, \bibinfo {author} {\bibfnamefont {G.}~\bibnamefont {Ye}}, \ and\
  \bibinfo {author} {\bibfnamefont {Y.-S.}\ \bibnamefont {Piao}},\ }\href
  {\doibase 10.1016/j.physletb.2021.136211} {\bibfield  {journal} {\bibinfo
  {journal} {Phys. Lett. B}\ }\textbf {\bibinfo {volume} {816}},\ \bibinfo
  {pages} {136211} (\bibinfo {year} {2021})},\ \Eprint
  {http://arxiv.org/abs/2009.14663} {arXiv:2009.14663 [astro-ph.CO]}
  \BibitemShut {NoStop}%
\bibitem [{\citenamefont {Bian}\ \emph {et~al.}(2021)\citenamefont {Bian},
  \citenamefont {Cai}, \citenamefont {Liu}, \citenamefont {Yang},\ and\
  \citenamefont {Zhou}}]{Bian:2020bps}%
  \BibitemOpen
  \bibfield  {author} {\bibinfo {author} {\bibfnamefont {L.}~\bibnamefont
  {Bian}}, \bibinfo {author} {\bibfnamefont {R.-G.}\ \bibnamefont {Cai}},
  \bibinfo {author} {\bibfnamefont {J.}~\bibnamefont {Liu}}, \bibinfo {author}
  {\bibfnamefont {X.-Y.}\ \bibnamefont {Yang}}, \ and\ \bibinfo {author}
  {\bibfnamefont {R.}~\bibnamefont {Zhou}},\ }\href {\doibase
  10.1103/PhysRevD.103.L081301} {\bibfield  {journal} {\bibinfo  {journal}
  {Phys. Rev. D}\ }\textbf {\bibinfo {volume} {103}},\ \bibinfo {pages}
  {L081301} (\bibinfo {year} {2021})},\ \Eprint
  {http://arxiv.org/abs/2009.13893} {arXiv:2009.13893 [astro-ph.CO]}
  \BibitemShut {NoStop}%
\bibitem [{\citenamefont {Blasi}\ \emph {et~al.}(2021)\citenamefont {Blasi},
  \citenamefont {Brdar},\ and\ \citenamefont {Schmitz}}]{Blasi:2020mfx}%
  \BibitemOpen
  \bibfield  {author} {\bibinfo {author} {\bibfnamefont {S.}~\bibnamefont
  {Blasi}}, \bibinfo {author} {\bibfnamefont {V.}~\bibnamefont {Brdar}}, \ and\
  \bibinfo {author} {\bibfnamefont {K.}~\bibnamefont {Schmitz}},\ }\href
  {\doibase 10.1103/PhysRevLett.126.041305} {\bibfield  {journal} {\bibinfo
  {journal} {Phys. Rev. Lett.}\ }\textbf {\bibinfo {volume} {126}},\ \bibinfo
  {pages} {041305} (\bibinfo {year} {2021})},\ \Eprint
  {http://arxiv.org/abs/2009.06607} {arXiv:2009.06607 [astro-ph.CO]}
  \BibitemShut {NoStop}%
\bibitem [{\citenamefont {De~Luca}\ \emph {et~al.}(2021)\citenamefont
  {De~Luca}, \citenamefont {Franciolini},\ and\ \citenamefont
  {Riotto}}]{DeLuca:2020agl}%
  \BibitemOpen
  \bibfield  {author} {\bibinfo {author} {\bibfnamefont {V.}~\bibnamefont
  {De~Luca}}, \bibinfo {author} {\bibfnamefont {G.}~\bibnamefont
  {Franciolini}}, \ and\ \bibinfo {author} {\bibfnamefont {A.}~\bibnamefont
  {Riotto}},\ }\href {\doibase 10.1103/PhysRevLett.126.041303} {\bibfield
  {journal} {\bibinfo  {journal} {Phys. Rev. Lett.}\ }\textbf {\bibinfo
  {volume} {126}},\ \bibinfo {pages} {041303} (\bibinfo {year} {2021})},\
  \Eprint {http://arxiv.org/abs/2009.08268} {arXiv:2009.08268 [astro-ph.CO]}
  \BibitemShut {NoStop}%
\bibitem [{\citenamefont {Vaskonen}\ and\ \citenamefont
  {Veerm\"ae}(2021)}]{Vaskonen:2020lbd}%
  \BibitemOpen
  \bibfield  {author} {\bibinfo {author} {\bibfnamefont {V.}~\bibnamefont
  {Vaskonen}}\ and\ \bibinfo {author} {\bibfnamefont {H.}~\bibnamefont
  {Veerm\"ae}},\ }\href {\doibase 10.1103/PhysRevLett.126.051303} {\bibfield
  {journal} {\bibinfo  {journal} {Phys. Rev. Lett.}\ }\textbf {\bibinfo
  {volume} {126}},\ \bibinfo {pages} {051303} (\bibinfo {year} {2021})},\
  \Eprint {http://arxiv.org/abs/2009.07832} {arXiv:2009.07832 [astro-ph.CO]}
  \BibitemShut {NoStop}%
\bibitem [{\citenamefont {Ellis}\ and\ \citenamefont
  {Lewicki}(2021)}]{Ellis:2020ena}%
  \BibitemOpen
  \bibfield  {author} {\bibinfo {author} {\bibfnamefont {J.}~\bibnamefont
  {Ellis}}\ and\ \bibinfo {author} {\bibfnamefont {M.}~\bibnamefont
  {Lewicki}},\ }\href {\doibase 10.1103/PhysRevLett.126.041304} {\bibfield
  {journal} {\bibinfo  {journal} {Phys. Rev. Lett.}\ }\textbf {\bibinfo
  {volume} {126}},\ \bibinfo {pages} {041304} (\bibinfo {year} {2021})},\
  \Eprint {http://arxiv.org/abs/2009.06555} {arXiv:2009.06555 [astro-ph.CO]}
  \BibitemShut {NoStop}%
\bibitem [{\citenamefont {Buchmuller}\ \emph {et~al.}(2020)\citenamefont
  {Buchmuller}, \citenamefont {Domcke},\ and\ \citenamefont
  {Schmitz}}]{Buchmuller:2020lbh}%
  \BibitemOpen
  \bibfield  {author} {\bibinfo {author} {\bibfnamefont {W.}~\bibnamefont
  {Buchmuller}}, \bibinfo {author} {\bibfnamefont {V.}~\bibnamefont {Domcke}},
  \ and\ \bibinfo {author} {\bibfnamefont {K.}~\bibnamefont {Schmitz}},\ }\href
  {\doibase 10.1016/j.physletb.2020.135914} {\bibfield  {journal} {\bibinfo
  {journal} {Phys. Lett. B}\ }\textbf {\bibinfo {volume} {811}},\ \bibinfo
  {pages} {135914} (\bibinfo {year} {2020})},\ \Eprint
  {http://arxiv.org/abs/2009.10649} {arXiv:2009.10649 [astro-ph.CO]}
  \BibitemShut {NoStop}%
\bibitem [{\citenamefont {Kohri}\ and\ \citenamefont
  {Terada}(2021)}]{Kohri:2020qqd}%
  \BibitemOpen
  \bibfield  {author} {\bibinfo {author} {\bibfnamefont {K.}~\bibnamefont
  {Kohri}}\ and\ \bibinfo {author} {\bibfnamefont {T.}~\bibnamefont {Terada}},\
  }\href {\doibase 10.1016/j.physletb.2020.136040} {\bibfield  {journal}
  {\bibinfo  {journal} {Phys. Lett. B}\ }\textbf {\bibinfo {volume} {813}},\
  \bibinfo {pages} {136040} (\bibinfo {year} {2021})},\ \Eprint
  {http://arxiv.org/abs/2009.11853} {arXiv:2009.11853 [astro-ph.CO]}
  \BibitemShut {NoStop}%
\bibitem [{\citenamefont {Vagnozzi}(2021)}]{Vagnozzi:2020gtf}%
  \BibitemOpen
  \bibfield  {author} {\bibinfo {author} {\bibfnamefont {S.}~\bibnamefont
  {Vagnozzi}},\ }\href {\doibase 10.1093/mnrasl/slaa203} {\bibfield  {journal}
  {\bibinfo  {journal} {Mon. Not. Roy. Astron. Soc.}\ }\textbf {\bibinfo
  {volume} {502}},\ \bibinfo {pages} {L11} (\bibinfo {year} {2021})},\ \Eprint
  {http://arxiv.org/abs/2009.13432} {arXiv:2009.13432 [astro-ph.CO]}
  \BibitemShut {NoStop}%
\bibitem [{\citenamefont {Maggiore}(2000)}]{Maggiore:1999vm}%
  \BibitemOpen
  \bibfield  {author} {\bibinfo {author} {\bibfnamefont {M.}~\bibnamefont
  {Maggiore}},\ }\href {\doibase 10.1016/S0370-1573(99)00102-7} {\bibfield
  {journal} {\bibinfo  {journal} {Phys. Rept.}\ }\textbf {\bibinfo {volume}
  {331}},\ \bibinfo {pages} {283} (\bibinfo {year} {2000})},\ \Eprint
  {http://arxiv.org/abs/gr-qc/9909001} {arXiv:gr-qc/9909001} \BibitemShut
  {NoStop}%
\bibitem [{\citenamefont {Kallosh}\ and\ \citenamefont
  {Linde}(2013)}]{Kallosh:2013hoa}%
  \BibitemOpen
  \bibfield  {author} {\bibinfo {author} {\bibfnamefont {R.}~\bibnamefont
  {Kallosh}}\ and\ \bibinfo {author} {\bibfnamefont {A.}~\bibnamefont
  {Linde}},\ }\href {\doibase 10.1088/1475-7516/2013/07/002} {\bibfield
  {journal} {\bibinfo  {journal} {JCAP}\ }\textbf {\bibinfo {volume} {07}},\
  \bibinfo {pages} {002} (\bibinfo {year} {2013})},\ \Eprint
  {http://arxiv.org/abs/1306.5220} {arXiv:1306.5220 [hep-th]} \BibitemShut
  {NoStop}%
\bibitem [{\citenamefont {Kallosh}\ \emph {et~al.}(2013)\citenamefont
  {Kallosh}, \citenamefont {Linde},\ and\ \citenamefont
  {Roest}}]{Kallosh:2013yoa}%
  \BibitemOpen
  \bibfield  {author} {\bibinfo {author} {\bibfnamefont {R.}~\bibnamefont
  {Kallosh}}, \bibinfo {author} {\bibfnamefont {A.}~\bibnamefont {Linde}}, \
  and\ \bibinfo {author} {\bibfnamefont {D.}~\bibnamefont {Roest}},\ }\href
  {\doibase 10.1007/JHEP11(2013)198} {\bibfield  {journal} {\bibinfo  {journal}
  {JHEP}\ }\textbf {\bibinfo {volume} {11}},\ \bibinfo {pages} {198} (\bibinfo
  {year} {2013})},\ \Eprint {http://arxiv.org/abs/1311.0472} {arXiv:1311.0472
  [hep-th]} \BibitemShut {NoStop}%
\bibitem [{\citenamefont {Starobinsky}(1980)}]{Starobinsky:1980te}%
  \BibitemOpen
  \bibfield  {author} {\bibinfo {author} {\bibfnamefont {A.~A.}\ \bibnamefont
  {Starobinsky}},\ }\href {\doibase 10.1016/0370-2693(80)90670-X} {\bibfield
  {journal} {\bibinfo  {journal} {Phys. Lett. B}\ }\textbf {\bibinfo {volume}
  {91}},\ \bibinfo {pages} {99} (\bibinfo {year} {1980})}\BibitemShut {NoStop}%
\bibitem [{\citenamefont {Bezrukov}\ and\ \citenamefont
  {Shaposhnikov}(2008)}]{Bezrukov:2007ep}%
  \BibitemOpen
  \bibfield  {author} {\bibinfo {author} {\bibfnamefont {F.~L.}\ \bibnamefont
  {Bezrukov}}\ and\ \bibinfo {author} {\bibfnamefont {M.}~\bibnamefont
  {Shaposhnikov}},\ }\href {\doibase 10.1016/j.physletb.2007.11.072} {\bibfield
   {journal} {\bibinfo  {journal} {Phys. Lett. B}\ }\textbf {\bibinfo {volume}
  {659}},\ \bibinfo {pages} {703} (\bibinfo {year} {2008})},\ \Eprint
  {http://arxiv.org/abs/0710.3755} {arXiv:0710.3755 [hep-th]} \BibitemShut
  {NoStop}%
\bibitem [{\citenamefont {Drewes}\ \emph {et~al.}(2017)\citenamefont {Drewes},
  \citenamefont {Kang},\ and\ \citenamefont {Mun}}]{Drewes:2017fmn}%
  \BibitemOpen
  \bibfield  {author} {\bibinfo {author} {\bibfnamefont {M.}~\bibnamefont
  {Drewes}}, \bibinfo {author} {\bibfnamefont {J.~U.}\ \bibnamefont {Kang}}, \
  and\ \bibinfo {author} {\bibfnamefont {U.~R.}\ \bibnamefont {Mun}},\ }\href
  {\doibase 10.1007/JHEP11(2017)072} {\bibfield  {journal} {\bibinfo  {journal}
  {JHEP}\ }\textbf {\bibinfo {volume} {11}},\ \bibinfo {pages} {072} (\bibinfo
  {year} {2017})},\ \Eprint {http://arxiv.org/abs/1708.01197} {arXiv:1708.01197
  [astro-ph.CO]} \BibitemShut {NoStop}%
\bibitem [{\citenamefont {Podolsky}\ \emph {et~al.}(2006)\citenamefont
  {Podolsky}, \citenamefont {Felder}, \citenamefont {Kofman},\ and\
  \citenamefont {Peloso}}]{Podolsky:2005bw}%
  \BibitemOpen
  \bibfield  {author} {\bibinfo {author} {\bibfnamefont {D.~I.}\ \bibnamefont
  {Podolsky}}, \bibinfo {author} {\bibfnamefont {G.~N.}\ \bibnamefont
  {Felder}}, \bibinfo {author} {\bibfnamefont {L.}~\bibnamefont {Kofman}}, \
  and\ \bibinfo {author} {\bibfnamefont {M.}~\bibnamefont {Peloso}},\ }\href
  {\doibase 10.1103/PhysRevD.73.023501} {\bibfield  {journal} {\bibinfo
  {journal} {Phys. Rev. D}\ }\textbf {\bibinfo {volume} {73}},\ \bibinfo
  {pages} {023501} (\bibinfo {year} {2006})},\ \Eprint
  {http://arxiv.org/abs/hep-ph/0507096} {arXiv:hep-ph/0507096} \BibitemShut
  {NoStop}%
\bibitem [{\citenamefont {Figueroa}\ and\ \citenamefont
  {Torrenti}(2017)}]{Figueroa:2016wxr}%
  \BibitemOpen
  \bibfield  {author} {\bibinfo {author} {\bibfnamefont {D.~G.}\ \bibnamefont
  {Figueroa}}\ and\ \bibinfo {author} {\bibfnamefont {F.}~\bibnamefont
  {Torrenti}},\ }\href {\doibase 10.1088/1475-7516/2017/02/001} {\bibfield
  {journal} {\bibinfo  {journal} {JCAP}\ }\textbf {\bibinfo {volume} {02}},\
  \bibinfo {pages} {001} (\bibinfo {year} {2017})},\ \Eprint
  {http://arxiv.org/abs/1609.05197} {arXiv:1609.05197 [astro-ph.CO]}
  \BibitemShut {NoStop}%
\bibitem [{\citenamefont {Maity}\ and\ \citenamefont
  {Saha}(2019)}]{Maity:2018qhi}%
  \BibitemOpen
  \bibfield  {author} {\bibinfo {author} {\bibfnamefont {D.}~\bibnamefont
  {Maity}}\ and\ \bibinfo {author} {\bibfnamefont {P.}~\bibnamefont {Saha}},\
  }\href {\doibase 10.1088/1475-7516/2019/07/018} {\bibfield  {journal}
  {\bibinfo  {journal} {JCAP}\ }\textbf {\bibinfo {volume} {07}},\ \bibinfo
  {pages} {018} (\bibinfo {year} {2019})},\ \Eprint
  {http://arxiv.org/abs/1811.11173} {arXiv:1811.11173 [astro-ph.CO]}
  \BibitemShut {NoStop}%
\bibitem [{\citenamefont {Mukhanov}(2005)}]{Mukhanov:2005sc}%
  \BibitemOpen
  \bibfield  {author} {\bibinfo {author} {\bibfnamefont {V.}~\bibnamefont
  {Mukhanov}},\ }\href@noop {} {\emph {\bibinfo {title} {{Physical Foundations
  of Cosmology}}}}\ (\bibinfo  {publisher} {Cambridge University Press},\
  \bibinfo {address} {Oxford},\ \bibinfo {year} {2005})\BibitemShut {NoStop}%
\bibitem [{\citenamefont {Haque}\ \emph {et~al.}(2021)\citenamefont {Haque},
  \citenamefont {Maity},\ and\ \citenamefont {Pal}}]{Haque:2020bip}%
  \BibitemOpen
  \bibfield  {author} {\bibinfo {author} {\bibfnamefont {M.~R.}\ \bibnamefont
  {Haque}}, \bibinfo {author} {\bibfnamefont {D.}~\bibnamefont {Maity}}, \ and\
  \bibinfo {author} {\bibfnamefont {S.}~\bibnamefont {Pal}},\ }\href {\doibase
  10.1103/PhysRevD.103.103540} {\bibfield  {journal} {\bibinfo  {journal}
  {Phys. Rev. D}\ }\textbf {\bibinfo {volume} {103}},\ \bibinfo {pages}
  {103540} (\bibinfo {year} {2021})},\ \Eprint
  {http://arxiv.org/abs/2012.10859} {arXiv:2012.10859 [hep-th]} \BibitemShut
  {NoStop}%
\bibitem [{\citenamefont {Gradshteyn}\ and\ \citenamefont
  {Ryzhik}(2007)}]{gradshteyn2007}%
  \BibitemOpen
  \bibfield  {author} {\bibinfo {author} {\bibfnamefont {I.~S.}\ \bibnamefont
  {Gradshteyn}}\ and\ \bibinfo {author} {\bibfnamefont {I.~M.}\ \bibnamefont
  {Ryzhik}},\ }\href@noop {} {\emph {\bibinfo {title} {Table of integrals,
  series, and products}}},\ \bibinfo {edition} {seventh}\ ed.\ (\bibinfo
  {publisher} {Elsevier/Academic Press, Amsterdam},\ \bibinfo {year} {2007})\
  pp.\ \bibinfo {pages} {xlviii+1171},\ \bibinfo {note} {translated from the
  Russian, Translation edited and with a preface by Alan Jeffrey and Daniel
  Zwillinger, With one CD-ROM (Windows, Macintosh and UNIX)}\BibitemShut
  {NoStop}%
\bibitem [{\citenamefont {Moore}\ \emph {et~al.}(2015)\citenamefont {Moore},
  \citenamefont {Cole},\ and\ \citenamefont {Berry}}]{Moore:2014lga}%
  \BibitemOpen
  \bibfield  {author} {\bibinfo {author} {\bibfnamefont {C.}~\bibnamefont
  {Moore}}, \bibinfo {author} {\bibfnamefont {R.}~\bibnamefont {Cole}}, \ and\
  \bibinfo {author} {\bibfnamefont {C.}~\bibnamefont {Berry}},\ }\href
  {\doibase 10.1088/0264-9381/32/1/015014} {\bibfield  {journal} {\bibinfo
  {journal} {Class. Quant. Grav.}\ }\textbf {\bibinfo {volume} {32}},\ \bibinfo
  {pages} {015014} (\bibinfo {year} {2015})},\ \Eprint
  {http://arxiv.org/abs/1408.0740} {arXiv:1408.0740 [gr-qc]} \BibitemShut
  {NoStop}%
\bibitem [{\citenamefont {Pagano}\ \emph {et~al.}(2016)\citenamefont {Pagano},
  \citenamefont {Salvati},\ and\ \citenamefont {Melchiorri}}]{Pagano:2015hma}%
  \BibitemOpen
  \bibfield  {author} {\bibinfo {author} {\bibfnamefont {L.}~\bibnamefont
  {Pagano}}, \bibinfo {author} {\bibfnamefont {L.}~\bibnamefont {Salvati}}, \
  and\ \bibinfo {author} {\bibfnamefont {A.}~\bibnamefont {Melchiorri}},\
  }\href {\doibase 10.1016/j.physletb.2016.07.078} {\bibfield  {journal}
  {\bibinfo  {journal} {Phys. Lett. B}\ }\textbf {\bibinfo {volume} {760}},\
  \bibinfo {pages} {823} (\bibinfo {year} {2016})},\ \Eprint
  {http://arxiv.org/abs/1508.02393} {arXiv:1508.02393 [astro-ph.CO]}
  \BibitemShut {NoStop}%
\bibitem [{\citenamefont {Hazra}\ \emph {et~al.}(2013)\citenamefont {Hazra},
  \citenamefont {Sriramkumar},\ and\ \citenamefont {Martin}}]{Hazra:2012yn}%
  \BibitemOpen
  \bibfield  {author} {\bibinfo {author} {\bibfnamefont {D.~K.}\ \bibnamefont
  {Hazra}}, \bibinfo {author} {\bibfnamefont {L.}~\bibnamefont {Sriramkumar}},
  \ and\ \bibinfo {author} {\bibfnamefont {J.}~\bibnamefont {Martin}},\ }\href
  {\doibase 10.1088/1475-7516/2013/05/026} {\bibfield  {journal} {\bibinfo
  {journal} {JCAP}\ }\textbf {\bibinfo {volume} {05}},\ \bibinfo {pages} {026}
  (\bibinfo {year} {2013})},\ \Eprint {http://arxiv.org/abs/1201.0926}
  {arXiv:1201.0926 [astro-ph.CO]} \BibitemShut {NoStop}%
\bibitem [{\citenamefont {Kurkela}\ and\ \citenamefont
  {Moore}(2011)}]{Kurkela:2011ti}%
  \BibitemOpen
  \bibfield  {author} {\bibinfo {author} {\bibfnamefont {A.}~\bibnamefont
  {Kurkela}}\ and\ \bibinfo {author} {\bibfnamefont {G.~D.}\ \bibnamefont
  {Moore}},\ }\href {\doibase 10.1007/JHEP12(2011)044} {\bibfield  {journal}
  {\bibinfo  {journal} {JHEP}\ }\textbf {\bibinfo {volume} {12}},\ \bibinfo
  {pages} {044} (\bibinfo {year} {2011})},\ \Eprint
  {http://arxiv.org/abs/1107.5050} {arXiv:1107.5050 [hep-ph]} \BibitemShut
  {NoStop}%
\bibitem [{\citenamefont {Harigaya}\ and\ \citenamefont
  {Mukaida}(2014)}]{Harigaya:2013vwa}%
  \BibitemOpen
  \bibfield  {author} {\bibinfo {author} {\bibfnamefont {K.}~\bibnamefont
  {Harigaya}}\ and\ \bibinfo {author} {\bibfnamefont {K.}~\bibnamefont
  {Mukaida}},\ }\href {\doibase 10.1007/JHEP05(2014)006} {\bibfield  {journal}
  {\bibinfo  {journal} {JHEP}\ }\textbf {\bibinfo {volume} {05}},\ \bibinfo
  {pages} {006} (\bibinfo {year} {2014})},\ \Eprint
  {http://arxiv.org/abs/1312.3097} {arXiv:1312.3097 [hep-ph]} \BibitemShut
  {NoStop}%
\bibitem [{\citenamefont {Ellis}\ \emph {et~al.}(1987)\citenamefont {Ellis},
  \citenamefont {Enqvist}, \citenamefont {Nanopoulos},\ and\ \citenamefont
  {Olive}}]{Ellis:1987rw}%
  \BibitemOpen
  \bibfield  {author} {\bibinfo {author} {\bibfnamefont {J.~R.}\ \bibnamefont
  {Ellis}}, \bibinfo {author} {\bibfnamefont {K.}~\bibnamefont {Enqvist}},
  \bibinfo {author} {\bibfnamefont {D.~V.}\ \bibnamefont {Nanopoulos}}, \ and\
  \bibinfo {author} {\bibfnamefont {K.~A.}\ \bibnamefont {Olive}},\ }\href
  {\doibase 10.1016/0370-2693(87)90620-4} {\bibfield  {journal} {\bibinfo
  {journal} {Phys. Lett. B}\ }\textbf {\bibinfo {volume} {191}},\ \bibinfo
  {pages} {343} (\bibinfo {year} {1987})}\BibitemShut {NoStop}%
\bibitem [{\citenamefont {McDonald}(2000)}]{McDonald:1999hd}%
  \BibitemOpen
  \bibfield  {author} {\bibinfo {author} {\bibfnamefont {J.}~\bibnamefont
  {McDonald}},\ }\href {\doibase 10.1103/PhysRevD.61.083513} {\bibfield
  {journal} {\bibinfo  {journal} {Phys. Rev. D}\ }\textbf {\bibinfo {volume}
  {61}},\ \bibinfo {pages} {083513} (\bibinfo {year} {2000})},\ \Eprint
  {http://arxiv.org/abs/hep-ph/9909467} {arXiv:hep-ph/9909467} \BibitemShut
  {NoStop}%
\bibitem [{\citenamefont {Allahverdi}(2000)}]{Allahverdi:2000ss}%
  \BibitemOpen
  \bibfield  {author} {\bibinfo {author} {\bibfnamefont {R.}~\bibnamefont
  {Allahverdi}},\ }\href {\doibase 10.1103/PhysRevD.62.063509} {\bibfield
  {journal} {\bibinfo  {journal} {Phys. Rev. D}\ }\textbf {\bibinfo {volume}
  {62}},\ \bibinfo {pages} {063509} (\bibinfo {year} {2000})},\ \Eprint
  {http://arxiv.org/abs/hep-ph/0004035} {arXiv:hep-ph/0004035} \BibitemShut
  {NoStop}%
\bibitem [{\citenamefont {Kawasaki}\ and\ \citenamefont
  {Takahashi}(2005)}]{Kawasaki:2004rx}%
  \BibitemOpen
  \bibfield  {author} {\bibinfo {author} {\bibfnamefont {M.}~\bibnamefont
  {Kawasaki}}\ and\ \bibinfo {author} {\bibfnamefont {F.}~\bibnamefont
  {Takahashi}},\ }\href {\doibase 10.1016/j.physletb.2005.05.022} {\bibfield
  {journal} {\bibinfo  {journal} {Phys. Lett. B}\ }\textbf {\bibinfo {volume}
  {618}},\ \bibinfo {pages} {1} (\bibinfo {year} {2005})},\ \Eprint
  {http://arxiv.org/abs/hep-ph/0410158} {arXiv:hep-ph/0410158} \BibitemShut
  {NoStop}%
\bibitem [{\citenamefont {Kuroyanagi}\ \emph {et~al.}(2013)\citenamefont
  {Kuroyanagi}, \citenamefont {Ringeval},\ and\ \citenamefont
  {Takahashi}}]{Kuroyanagi:2013ns}%
  \BibitemOpen
  \bibfield  {author} {\bibinfo {author} {\bibfnamefont {S.}~\bibnamefont
  {Kuroyanagi}}, \bibinfo {author} {\bibfnamefont {C.}~\bibnamefont
  {Ringeval}}, \ and\ \bibinfo {author} {\bibfnamefont {T.}~\bibnamefont
  {Takahashi}},\ }\href {\doibase 10.1103/PhysRevD.87.083502} {\bibfield
  {journal} {\bibinfo  {journal} {Phys. Rev. D}\ }\textbf {\bibinfo {volume}
  {87}},\ \bibinfo {pages} {083502} (\bibinfo {year} {2013})},\ \Eprint
  {http://arxiv.org/abs/1301.1778} {arXiv:1301.1778 [astro-ph.CO]} \BibitemShut
  {NoStop}%
\bibitem [{\citenamefont {Hattori}\ \emph {et~al.}(2015)\citenamefont
  {Hattori}, \citenamefont {Kobayashi}, \citenamefont {Omoto},\ and\
  \citenamefont {Seto}}]{Hattori:2015xla}%
  \BibitemOpen
  \bibfield  {author} {\bibinfo {author} {\bibfnamefont {H.}~\bibnamefont
  {Hattori}}, \bibinfo {author} {\bibfnamefont {T.}~\bibnamefont {Kobayashi}},
  \bibinfo {author} {\bibfnamefont {N.}~\bibnamefont {Omoto}}, \ and\ \bibinfo
  {author} {\bibfnamefont {O.}~\bibnamefont {Seto}},\ }\href {\doibase
  10.1103/PhysRevD.92.103518} {\bibfield  {journal} {\bibinfo  {journal} {Phys.
  Rev. D}\ }\textbf {\bibinfo {volume} {92}},\ \bibinfo {pages} {103518}
  (\bibinfo {year} {2015})},\ \Eprint {http://arxiv.org/abs/1510.03595}
  {arXiv:1510.03595 [hep-ph]} \BibitemShut {NoStop}%
\bibitem [{\citenamefont {Banerjee}\ and\ \citenamefont
  {Paul}(2017)}]{Banerjee:2017lxi}%
  \BibitemOpen
  \bibfield  {author} {\bibinfo {author} {\bibfnamefont {N.}~\bibnamefont
  {Banerjee}}\ and\ \bibinfo {author} {\bibfnamefont {T.}~\bibnamefont
  {Paul}},\ }\href {\doibase 10.1140/epjc/s10052-017-5256-0} {\bibfield
  {journal} {\bibinfo  {journal} {Eur. Phys. J. C}\ }\textbf {\bibinfo {volume}
  {77}},\ \bibinfo {pages} {672} (\bibinfo {year} {2017})},\ \Eprint
  {http://arxiv.org/abs/1706.05964} {arXiv:1706.05964 [hep-th]} \BibitemShut
  {NoStop}%
\bibitem [{\citenamefont {Elizalde}\ \emph {et~al.}(2019)\citenamefont
  {Elizalde}, \citenamefont {Odintsov}, \citenamefont {Paul},\ and\
  \citenamefont {S\'aez-Chill\'on~G\'omez}}]{Elizalde:2018rmz}%
  \BibitemOpen
  \bibfield  {author} {\bibinfo {author} {\bibfnamefont {E.}~\bibnamefont
  {Elizalde}}, \bibinfo {author} {\bibfnamefont {S.~D.}\ \bibnamefont
  {Odintsov}}, \bibinfo {author} {\bibfnamefont {T.}~\bibnamefont {Paul}}, \
  and\ \bibinfo {author} {\bibfnamefont {D.}~\bibnamefont
  {S\'aez-Chill\'on~G\'omez}},\ }\href {\doibase 10.1103/PhysRevD.99.063506}
  {\bibfield  {journal} {\bibinfo  {journal} {Phys. Rev. D}\ }\textbf {\bibinfo
  {volume} {99}},\ \bibinfo {pages} {063506} (\bibinfo {year} {2019})},\
  \Eprint {http://arxiv.org/abs/1811.02960} {arXiv:1811.02960 [gr-qc]}
  \BibitemShut {NoStop}%
\bibitem [{\citenamefont {Unnikrishnan}\ \emph {et~al.}(2012)\citenamefont
  {Unnikrishnan}, \citenamefont {Sahni},\ and\ \citenamefont
  {Toporensky}}]{Unnikrishnan:2012zu}%
  \BibitemOpen
  \bibfield  {author} {\bibinfo {author} {\bibfnamefont {S.}~\bibnamefont
  {Unnikrishnan}}, \bibinfo {author} {\bibfnamefont {V.}~\bibnamefont {Sahni}},
  \ and\ \bibinfo {author} {\bibfnamefont {A.}~\bibnamefont {Toporensky}},\
  }\href {\doibase 10.1088/1475-7516/2012/08/018} {\bibfield  {journal}
  {\bibinfo  {journal} {JCAP}\ }\textbf {\bibinfo {volume} {08}},\ \bibinfo
  {pages} {018} (\bibinfo {year} {2012})},\ \Eprint
  {http://arxiv.org/abs/1205.0786} {arXiv:1205.0786 [astro-ph.CO]} \BibitemShut
  {NoStop}%
\bibitem [{\citenamefont {Yokoyama}(2021)}]{Yokoyama:2021hsa}%
  \BibitemOpen
  \bibfield  {author} {\bibinfo {author} {\bibfnamefont {J.}~\bibnamefont
  {Yokoyama}},\ }\href@noop {} {\  (\bibinfo {year} {2021})},\ \Eprint
  {http://arxiv.org/abs/2105.07629} {arXiv:2105.07629 [gr-qc]} \BibitemShut
  {NoStop}%
\end{thebibliography}%
\end{document}